\pdfoutput=1
\documentclass[a4paper,11pt,english]{article}

\usepackage[bindingoffset=0.3cm,textheight=22.5cm,hdivide={2.7cm,*,2.7cm}, vdivide={*,22cm,*}]{geometry}
\usepackage{cite}
\usepackage{amsmath,amsfonts,amssymb,babel,slashed,color,empheq}
\usepackage{hyperref}

\makeatletter

\newcommand{\dd}{\mathrm{d}}

\newcommand{\del}{\partial}

\def\nn{\nonumber} 
\def\obar{\overline}
\numberwithin{equation}{section}
\def\a{\alpha}  \def\b{\beta}
 \def\g{\gamma} \def\G{\Gamma}
 \def\d{\delta} 

\def\l{\lambda} \def\L{\Lambda}  \def\m{\mu}
    \def\r{\rho}
\def\s{\sigma}  \def\t{\tau}

\def\cA{{\cal A}}  \def\cC{{\cal C}} 
\def\cD{{\cal D}}   
\def\cG{{\cal G}} \def\cH{{\cal H}} \def\cI{{\cal I}} 
 \def\cK{{\cal K}}  
\def\cM{{\cal M}}   
   
\def\cS{{\cal S}}

\def\R{{\mathbb R}} \def\C{{\mathbb C}} 

 \def\one{\mbox{1 \kern-.59em {\rm l}}}

\def\mmu{\mathfrak{u}}

\def\mso{\mathfrak{so}}

\newcommand{\Tr}{\mathrm{Tr}}

\newcommand{\End}{\mathrm{End}}

\def\Tr{\mbox{Tr}}

\newcommand{\im}{\mathrm{i}}

\newcommand{\diag}{\rm diag}

 \allowdisplaybreaks[3]

\begin{document}

%%%%%%%%%%%%%%%%%%%%%%%%%%%%%%%%%%%%%%%%%%%%%%%%%%%%%%%%%%%%%%%5%%%%%%%%%%%%%%%%%%%%%%%%%%

\parindent=0cm

\renewcommand{\title}[1]{\vspace{10mm}\noindent{\Large{\bf#1}}\vspace{8mm}} 
\newcommand{\authors}[1]{\noindent{\large #1}\vspace{5mm}}
\newcommand{\address}[1]{{\itshape #1\vspace{2mm}}}

%\thispagestyle{hepth}

%%%% --- TITLE PAGE --- %%%%
\begin{titlepage}
\begin{flushright}
 UWThPh-2019-27
%  \\
%  \textcolor{red}{\today}
\end{flushright}
\begin{center}
\title{ {\Large Scalar modes and the linearized Schwarzschild solution \\[1ex] on a
quantized FLRW space-time in Yang-Mills matrix models}  }

\vskip 3mm

\authors{Harold C.\ Steinacker${}^\ddagger$}

\vskip 3mm

 \address{ 
${}^\ddagger${\it Faculty of Physics, University of Vienna\\
Boltzmanngasse 5, A-1090 Vienna, Austria  }  \\
Email: {\tt harold.steinacker@univie.ac.at}  
  }

\bigskip

\vskip 1.4cm

%%%% --- ABSTRACT --- %%%%
\textbf{Abstract}
\vskip 3mm

\begin{minipage}{14cm}%

We study scalar perturbations of a recently found
3+1-dimensional FLRW quantum space-time solution in Yang-Mills matrix models.
In particular, the linearized Schwarzschild metric is obtained as a solution. 
It arises from a quasi-static  would-be massive graviton mode,
and  slowly decreases during the cosmic expansion. 
Along with the propagating graviton modes,
this strongly suggests that 3+1 dimensional (quantum) gravity 
emerges from the IKKT matrix model on this background.
%without requiring an Einstein-Hilbert action or compactification.
For the dynamical scalar modes, non-linear effects must be taken into account.
We argue that they lead to non-Ricci-flat metric perturbations with very long wavelengths,
which would be perceived as dark matter from the GR point of view.

\end{minipage}

\end{center}

\end{titlepage}

\tableofcontents
%
%%%%%%%%%%%%%%%%%%%%%%%%%%%%%%%%%%%%%%%%%%%%%%%%%%%%%%%%%%
%
\section{Introduction} 

The starting point of this paper is a 
recent solution of the IKKT-type matrix models with mass term \cite{Sperling:2019xar}, 
which is naturally interpreted as 3+1-dimensional cosmological FLRW quantum space-time.
It was  shown that the fluctuation modes around this background 
include spin-2 metric fluctuations, as well as a truncated tower of higher-spin modes which are 
organized in a higher-spin gauge theory. The 2 standard Ricci-flat massless 
graviton modes were found, as well as some additional 
vector-like and scalar modes whose significance was not fully clarified.

The aim of the present paper is to study in more detail the metric perturbations, 
and in particular to 
see if and how the (linearized) Schwarzschild solution can be obtained.
 We will indeed find such a solution, which
 is realized in the scalar sector of the linearized perturbation modes exhibited in \cite{Sperling:2019xar}. 
This means that the model has a good chance to satisfy the precision solar system tests of gravity.
We will also elaborate and discuss in some detail the extra scalar mode, which is not present in GR.
This seems to provide a natural candidate for apparent dark matter.

Since the notorious problems in attempts to quantize gravity 
arise primarily from the Einstein-Hilbert action, is is very desirable to find another 
framework for gravity, which is more suitable for quantization.
String theory provides such a framework, but the traditional approach using compactifications 
leads to a host of issues, notably lack of predictivity.
This suggests to use matrix models as a starting point, and in particular
the IKKT or IIB model \cite{Ishibashi:1996xs}, which was originally proposed as a constructive definition of string theory. Remarkably,
numerical studies in this non-perturbative formulation  provide evidence  
\cite{Kim:2011cr,Nishimura:2019qal,Aoki:2019tby} that 
3+1-dimensional configurations arise at the non-perturbative level, 
tentatively interpreted as expanding universe.
However, this requires a new mechanism for gravity on 3+1-dimensional
non-commutative backgrounds 
as in \cite{Sperling:2019xar}, 
which does not rely on compactification.
The present paper provides further evidence and insights for this mechanism.

The (linearized) Schwarzschild metric is clearly the benchmark for any
viable theory of gravity. 
There has been considerable effort to find noncommutative analogs of the Schwarzschild metric from various approaches,
leading to a number of proposals 
\cite{Schupp:2009pt,Blaschke:2010ye,Ohl:2009pv,Chaichian:2007dr} and references therein, cf. also \cite{Nicolini:2008aj}; 
however, none is truly satisfactory.
The proposals are typically obtained by some  
ad-hoc modification of the classical solution, without any intrinsic role of noncommutativity,
which is put in by hand. 
In contrast, %the matrix model framework is inherently non-commutative, and
the quantum structure 
(or its semi-classical limit) plays a central role in the present framework. 
Our solution is a deformation of the noncommutative background  which respects an exact $SO(3)$
rotation symmetry, even though there are only finitely many d.o.f. per unit volume.
The solution has a good asymptotics at large distances, 
allowing superpositions corresponding to arbitrary mass distributions.
In fact we obtain generic quasi-static Ricci-flat linearized perturbations,
which complement the Ricci-flat propagating gravitons 
 found in \cite{Sperling:2019xar}.

This  realization of the (linearized) Schwarzschild solution 
is remarkable and may seem surprising, because the action is of Yang-Mills type, and
no Einstein-Hilbert-like action is required\footnote{It is well-known that gravity 
can be obtained from a Yang-Mills-type action by imposing constraints, 
cf. \cite{MacDowell:1977jt,Chamseddine:2002fd,Manolakos:2019fle}. 
However this essentially amounts to a reformulation of classical GR,
and the usual problems are expected to arise upon quantization.
In contrast, we do not impose any constraints on the Yang-Mills action. Nevertheless, 
quantum effects are expected  to induce an
Einstein-Hilbert-like term, 
as discussed in \cite{Sperling:2019xar}. This may play an important role here as well,
but we focus on the classical mechanism. Another interesting possibility was proposed in \cite{Hanada:2005vr}, which has some 
similarities to the present mechanism but leads to many additional fields, possibly including ghosts.}.
This  means that the theory has a good chance to survive upon quantization, which is 
naturally defined via integration over the space of matrices.
The IKKT model is indeed well suited for quantization, and quite clearly free of ghosts and 
other obvious pathologies.  
It is background-independent 
in the sense that it has a large class of solutions with different geometries,
and defines a gauge theory for fluctuations on any background.

The price to pay is a considerable complexity of the resulting theory. 
As explained in \cite{Steinacker:2016vgf,Sperling:2019xar} the background leads to a higher-spin gauge theory, with a truncated tower of higher 
spin modes, and many similarities with (but also distinctions from) Vasiliev theory 
\cite{Vasiliev:1990en}. 
Since space-time itself is part of the background solution, it is not unreasonable to expect 
Ricci-flat deformations, cf. \cite{Rivelles:2002ez,Yang:2006dk,Steinacker:2010rh}. 
However, Lorentz invariance is very tricky on noncommutative backgrounds. In the present case
 the space-like isometries $SO(3,1)$  of the $k=-1$ FLRW space-time are manifest, but invariance under (local) boosts is not. 
Nevertheless, the propagation of  all physical modes is governed by the same effective metric.  
In particular, the concept of spin has to be used with caution, and would-be spin $s$ modes decompose further 
into sectors governed by the space-like $SO(3,1)$ isometry.
The tensor fields are accordingly characterized by the transformation  under the local $SO(3)$ stabilizer group,
and the term ``scalar modes'' is understood in this sense throughout the paper.
However this  complication is in fact helpful to identify physical degrees of freedom
 in the physical sector, and to understand the absence of ghosts.

Let us describe  the new results in some details. 
We focus on the scalar fluctuation mode which was found in  \cite{Sperling:2019xar}, and elaborate the associated metric fluctuations.
The main result is that there is a preferred ``quasi-static'' vacuum solution
which leads  to the linearized Schwarzschild metric on the 
FLRW background.
This strongly suggests that a near-realistic gravity  emerges on the background, 
however only the vacuum solution is considered here.
Quasi-static 
means that the solution is static on local scales at late times, but slowly decays on cosmic scales, in a specific way.
This is a somewhat unexpected result, whose significance is not entirely clear.
The quasi-static solution is singled out because all other solutions lead to a large diffeo term, which makes the linearized treatment 
problematic. Hence the Schwarzschild solution is the ``cleanest'' case, while 
the generic dynamical scalar modes require non-linear considerations 
somewhat reminiscent of the Vainshtein mechanism \cite{Vainshtein:1972sx}. We  offer a heuristic way to 
understand them, which points to the intriguing -  albeit quite speculative - possibility that 
these non-Ricci-flat  scalar modes might provide a geometrical explanation for   
dark matter at galactic scales.

%%H new V2
It may seem strange to start with a curved cosmological background rather than flat Minkowski 
space, since the Schwarzschild solution is basically a local structure. 
The reason is that no flat counterpart of the underlying 
quantum-spacetime with the required structure is known. We will thus largely neglect the 
 contributions of the Schwarzschild solution at the cosmic scales.

Along the way, we also find the missing $4^{th}$ off-shell scalar fluctuation mode, which was missing in \cite{Sperling:2019xar}.
Thus all 10 off-shell metric fluctuation modes for the most general 
 metric fluctuations are realized, and
the model is certainly rich enough for a realistic theory of gravity. That theory
would clearly deviate from GR at cosmic scale, since the FLRW background solution is 
not Ricci flat, but requires no stabilization by matter (or energy) and no fine-tuning.

Finally, it should be stressed that even though the model is 
intrinsically noncommutative, it should be viewed 
in the spirit of almost-local and almost-classical field theory. 
Space-time arises as a condensation of matrices 
rather than some non-local holographic image, with dynamical local fluctuations
described by an effective field theory.

The paper is  meant to be as self-contained and compact as possible. 
We start with a lightning introduction to the $\cM^{3,1}_n$ space-time under consideration,
and elaborate only the specific modes and aspects needed to obtain the Schwarzschild solution.
For some results we have to refer to \cite{Sperling:2019xar}, but the essential new computations are mostly spelled out.
For the skeptical reader, some of the missing steps may be 
uncovered from the file in the arXive.

% 
%%%%%%%%%%%%%%%%%%%%%%%%%%%%%%%%%%%%%%%%%%%%%%%
%%%%%%%%%%%%%%%%%%%%%%%%%%%%%%%%%%%%%%%%%%%%%%%
%

%
%
%
\section{Quantum FLRW space-time   \texorpdfstring{$\cM^{3,1}_n$}{M(3,1)}}
\label{sec:projection-Lorentzian}

The quantum space-time under consideration is based on a particular 
representation $\cH_n$  of $SO(4,2)$, which is a lowest weight unitary irrep 
in the short discrete series 
known as \emph{minireps} or \emph{doubletons} \cite{Mack:1975je,Fernando:2009fq}.
Those are the unique irreps which remain irreducible under 
the restriction to $SO(4,1) \subset SO(4,2)$. 
We denote the generators in this representation by
$\cM^{ab}$, which are Hermitian operators satisfying
\begin{align}
  [\cM_{ab},\cM_{cd}] &=\im \left(\eta_{ac}\cM_{bd} - \eta_{ad}\cM_{bc} - 
\eta_{bc}\cM_{ad} + \eta_{bd}\cM_{ac}\right) \ 
 \label{M-M-relations-noncompact}
\end{align}
where $\eta^{ab} = \diag(-1,1,1,1,1,-1)$ is the invariant metric of $SO(4,2)$.
We then define 
\begin{alignat}{2}
 X^\mu &\coloneqq r\cM^{\mu 5} ,  \qquad X^4 :=  r\cM^{4 5}  \nn\\
 T^\mu &\coloneqq R^{-1} \cM^{\mu 4}        \qquad \mu,\nu = 0,\ldots,3 \ .
\end{alignat}  
Then the $X^a$ transform as vector operators under $SO(4,1)$, while the $T^\mu$ are  vector operators
under $SO(3,1) \subset SO(4,1)$.
The  $SO(3,1)$-invariant fuzzy or quantum space-time $\cM^{3,1}_n$ is then defined through the algebra of functions $\phi(X^\mu)$
generated by the $X^\mu$ for $\mu=0,1,2,3$.
Here $r$ is a microscopic length scale related to the internal quantum structure, while
$R$ is a macroscopic scale as specified in \refeq{X-T-relations-1}.
The commutation relations \eqref{M-M-relations-noncompact} imply
 \begin{subequations}
 \label{basic-CR-H4}
 \begin{align}
  [X^\mu,X^\nu] &= -  \im\, r^2\cM^{\mu\nu}  \eqqcolon \im \Theta^{\mu\nu} \,,
  \label{X-X-CR}\\
   [T^\mu,X^\nu] &=  \im \frac{1}{R}\eta^{\mu\nu} X^4 \,,  \label{T-X-CR}\\
[T^\mu, T^\nu] &=  -\frac{\im}{r^2 R^2} \Theta^{\mu  \nu} \, , \label{T-T-CR} \\
  [T^\mu,X^4] &=  - \im \frac{1}{R} X^\mu \,,\\
 [X^\mu,X^4] &=  - \im r^2 R\,  T^\mu  \, ,
\end{align}
\end{subequations}
and the irreducibility of $\cH_n$ under $SO(4,1)$
implies the relations \cite{Sperling:2019xar}
 \begin{subequations}
  \label{basic-constraints}
 \begin{align}
 X_{\mu} X^{\mu}  &= -R^2 -  X^4 X^4, \
  \qquad R^2 = \frac{r^2}{4}(n^2-4)  \label{X-T-relations-1}\\
 T_{\mu} T^{\mu}  &= \frac 1{r^2} + \frac 1{r^2 R^{2}}\, X^4 X^4,  \\
 X_\mu T^\mu + T^\mu X_\mu  &= 0 \ .
 \label{X-T-relations}
\end{align} 
\end{subequations}
There are some extra constraints involving $\Theta^{\mu\nu}$,
which will only be given in the semi-classical version below.
Unless otherwise stated, indices will be raised and lowered with $\eta^{ab}$ or $\eta^{\mu\nu}$.
Apart from the extra constraints,
the construction is quite close to that of Snyder 
\cite{Snyder:1946qz} and Yang \cite{Yang:1947ud}.

\begin{figure}
%\begin{center}
\hspace{2cm} \includegraphics[width=0.5\textwidth]{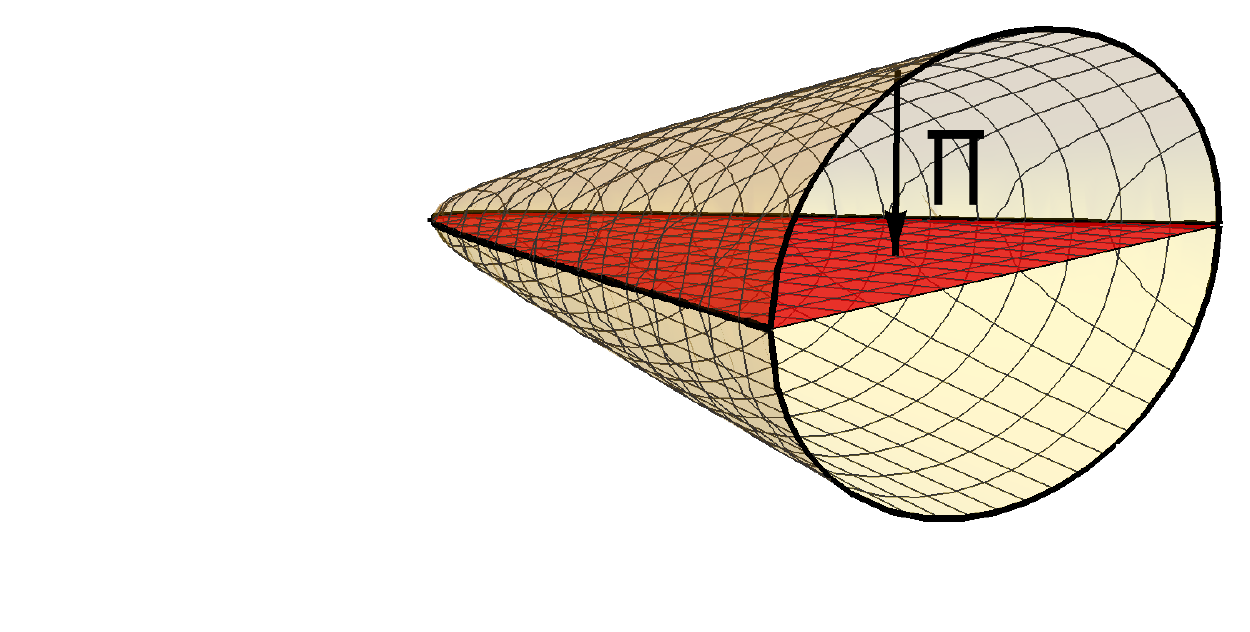}
 \caption{Sketch of the projection $\Pi$ from $H^4$ to $\cM^{3,1}$  with 
Minkowski signature.}
%\end{center}
 \label{fig:projection}
\end{figure}

The proper interpretation of this structure is not obvious a priori, due to the 
extra generators $T^\mmu$ and $\Theta^{\mu\nu}$. 
These cannot  be dropped, because the full algebra  $\End(\cH_n)$ is generated by
the $X^\mu$ alone.
A proper geometrical understanding is obtained by considering all the generators 
$\cM_{ab}$ of $\mso(4,2)$. 
As explained in \cite{Steinacker:2017bhb,Sperling:2017dts,Medina:2002pc},
these are naturally viewed as quantized embedding 
functions of a coadjoint orbit $m^{ab}:\ \C P^{1,2} \hookrightarrow \mso(4,2) \cong\R^{15}$.
Here  $\C P^{1,2}$ is a 6-dimensional noncompact analog of $\C P^3$, which is 
singled out by the constraints satisfied by $m^{ab}$. 
Hence the full  algebra  $\End(\cH_n)$ can be interpreted as a quantized 
algebra of functions on  $\C P^{1,2}$, dubbed fuzzy $\C P^{1,2}_n$.
Furthermore,  $\C P^{1,2}$ is naturally a $S^2$ bundle over $H^4$,
which is defined by the $X^a$ satisfying \eqref{X-T-relations-1}.
Hence the space $\cM^{3,1}$ generated by the $X^\mu \sim x^\mu$, $\mu = 0,...,3$ 
can be viewed as projection of $H^4\subset \R^{4,1}$ to $\R^{3,1}$ along $X^4$,
as sketched in figure \ref{fig:projection}.
This is the space-time of interest here, which is
covariant under $SO(3,1)$. 
For similar covariant quantum spaces see e.g. 
\cite{Grosse:1996mz,Heckman:2014xha,Sperling:2017dts,Ramgoolam:2001zx,Hasebe:2012mz,Buric:2017yes,Steinacker:2017vqw}.

%
%%%%%%%%%%%%%%%%%%%%%%%%%%%%%%%%%%%%%%%%%%%%%%%
%
\subsection{Semi-classical structure of \texorpdfstring{$\cM^{3,1}$}{M(3,1)}}

We will mostly restrict ourselves to the semi-classical limit 
$n \to \infty$
of the above 
space, working with commutative functions of $x^\mu\sim X^\mu$ and $t^\mu \sim T^\mu$, but keeping 
the Poisson or symplectic structure  $[.,.] \sim i \{.,.\}$ encoded in $\theta^{\mu\nu}$.
The constraints \eqref{basic-constraints} etc. imply the following relations 
\begin{subequations}
\label{geometry-H-M}
\begin{align}
 x_\mu x^\mu &= -R^2 - x_4^2 = -R^2 \cosh^2(\eta) \, , 
 \qquad R \sim \frac{r}{2}n   \label{radial-constraint}\\
 t_{\mu} t^{\mu}  &=  r^{-2}\, \cosh^2(\eta) \,, \\
 t_\mu x^\mu &= 0, \ \\
 t_\mu \theta^{\mu\a} &= - \sinh(\eta) x^\a , \\
 x_\mu \theta^{\mu\a} &= - r^2 R^2 \sinh(\eta) t^\a , \label{x-theta-contract}\\
 \eta_{\mu\nu}\theta^{\mu\a} \theta^{\nu\b} &= R^2 r^2 \eta^{\a\b} - R^2 r^4 
t^\a t^\b - r^2 x^\a x^\b  
% 
% \theta^{\mu\nu}\theta_{\mu\nu} &= 2R^2r^2\big(2-\cosh^2(\eta)\big)
\end{align}
\end{subequations}
where $\mu,\a = 0,\ldots ,3$.
Here $\eta$ is a global time coordinate defined by 
\begin{align}
 x^4 = R \sinh(\eta) \ ,
 \label{x4-eta-def}
\end{align}
which will be related to the scale parameter  of the universe \eqref{a-eta}.
Clearly
the $x^\mu:\, \cM^{3,1}\hookrightarrow \R^{3,1}$ can be viewed as Cartesian  coordinate functions. Similarly,
the $t^\mu$  describe the $S^2$ fiber over $\cM^{3,1}$
as discussed above.
On the other hand, the relation \eqref{T-X-CR} implies that
the derivations
\begin{align}
 -i[T^\mu,.] \sim \{t^\mu,.\}  \ = \sinh(\eta) \del_\mu
\end{align}
act as momentum generators on $\cM^{3,1}$,
leading to  the useful relation 
\begin{align}
 \del_\mu \phi = \b\{t_\mu,\phi\}, \qquad \b = \frac{1}{\sinh(\eta)}
 \label{del-t-rel}
\end{align}
for $\phi = \phi(x)$.
In particular, a $SO(3,1)$-invariant matrix d'Alembertian can be defined as 
\begin{align}
 \Box := [T^\mu,[T_\mu,.]] \  \sim \ -\{t^\mu,\{t_\mu,.\}\}  \ .
 \label{Box-def}
\end{align}
It acts on any  $\phi \in  \End(\cH)$, and
will play a  central role throughout this paper.
We also define a globally defined time-like vector field 
\begin{align}
 \t := x^\mu \del_\mu  .
 \label{tau-def}
\end{align}
To get some insight into the $\theta^{\mu\nu}$,  fix some
reference point $\xi$  on  $\cM^{3,1}$, which using $SO(3,1)$ invariance
can be chosen as
\begin{align}
 \xi = (x^0,0,0,0) , \qquad x^0 = R\cosh(\eta) \ .
\end{align}
%It corresponds to some $\hat\xi = (x^0,0,0,0,x^4)$ on  $H^4$.
Then \eqref{x-theta-contract}
provides a relation between the $t^{\mu}$ and the  $\theta^{\mu\nu}$ generators,
\begin{align}
 t^\mu = -\frac{1}{R r^2 x^4}\, x_\nu\theta^{\nu\mu} 
  \ &\stackrel{\xi}{=} \ -\frac{1}{R r^2}\, \frac{1}{\tanh(\eta)}\theta^{0\mu} \ , \
\, \qquad t^0  \stackrel{\xi}{=} \  0 \ .
\end{align}
Conversely,  the self-duality relation on $H^4_n$ \cite{Sperling:2018xrm}
\begin{align}
 \epsilon_{abcde} \theta^{ab} x^{c} &= 2R \theta_{de}  
 \label{SD-H-class}
\end{align}
relates the space-like and the time-like components of $\theta^{\mu\nu}$ 
 on $\cM^{3,1}$,
%  , leading to 
% \begin{align}
% \label{SD-H-ref_point}
% t^i  & \stackrel{\xi}{=} \ \frac 1{nr^3} \cosh(\eta)\epsilon^{i jk} \theta^{jk}  , 
% \qquad t^0  \stackrel{\xi}{=} \  0 \, .
% \end{align}
%Note that here the quantum number $n$ enters. 
and an explicit expression of $\theta^{\mu\nu}$ 
 in terms of  $t^\mu$ can be derived  \cite{Sperling:2019xar}
\begin{align}
 \theta^{\mu\nu} &= \ c(x^\mu t^\nu - x^\nu t^\mu) + b \epsilon^{\mu\nu\a\b} x_\a 
t_\b   
 \label{theta-P-relation} \\
\text{with} \qquad 
c &=  r^2 \frac{\sinh(\eta)}{\cosh^2(\eta)} 
 \qquad \text{and} \qquad
 b = \frac{r^2}{\cosh^2(\eta)} \,.
\end{align}

\paragraph{Hyperbolic coordinates.}

Now consider the adapted hyperbolic coordinates
\begin{align}
 \begin{pmatrix}
  x^0 \\ x^1 \\ x^2 \\x^3 
 \end{pmatrix}
= R \cosh(\eta) 
\begin{pmatrix}
\cosh(\chi) \\
\sinh(\chi)\sin(\theta) \cos(\varphi) \\
\sinh(\chi)\sin(\theta) \sin(\varphi) \\
\sinh(\chi)\cos(\theta)
\end{pmatrix} \ .
\label{hyperbolic-coords}
\end{align}
We will see that $\eta$ measures the cosmic time, cf. \eqref{x4-eta-def}, while
the  space-like distance from the origin on each time slice $H^3$ is measured by $\chi$.
Noting that
\begin{align}
 \frac{x_\mu x_\nu}{R^2\cosh^2(\eta)}\dd x^\mu \dd x^\nu &=R^2 \sinh^2(\eta) d\eta^2 \ 
 \label{metric-sphericalcoords-trafo}
\end{align}
which follows from \refeq{radial-constraint}, we obtain
the induced (flat) metric of $\R^{3,1}$ in these coordinates 
\begin{align}
ds^2_g &=  \eta_{\mu\nu}\dd x^\mu \dd x^\nu = R^2\Big(-\sinh^2(\eta)d\eta^2 
 + \cosh^2(\eta)d\Sigma^2\Big) 
 \label{ind-metric-explicit}
 \end{align}
 where $d\Sigma^2$ is the metric on the 
unit hyperboloid $H^3$,
 \begin{align}
 d\Sigma^2 &=  d\chi^2 + \sinh^2(\chi)d\Omega^2,  \qquad 
 d\Omega^2 = d\theta^2 + \sin^2(\theta)d\varphi^2 \ .
 \label{ds-induced}
\end{align}
However, the effective metric % in the present framework 
is a different one, which is also $SO(3,1)$ invariant but not flat.

\subsection{Effective metric and d'Alembertian}
\label{sec:metric}

In the matrix model framework considered below, the 
effective metric on the background $\cM^{3,1}$   under consideration is given by
\cite{Sperling:2019xar}
\begin{align}
   G^{\mu\nu} &= \a\, \g^{\mu\nu} \ = \ \sinh^{-1}(\eta) \eta^{\mu\nu}
   \qquad 
   \qquad \a = \sqrt{\frac 1{\tilde\rho^2|\g^{\mu\nu}|}} 
             =  \sinh^{-3}(\eta) \  \nn\\
    \g^{\a\b} &= \eta_{\mu\nu}\theta^{\mu\a}\theta^{\nu\b} 
  = \sinh^2(\eta) \eta^{\a\b} \ .
   \label{eff-metric-G}
\end{align}
This is an $SO(3,1)$-invariant FLRW metric with signature $(-+++)$. 
Here $\tilde\rho^2$ is an irrelevant constant which adjusts the dimensions.
There are several ways to obtain this metric. One is by rewriting the 
kinetic term in covariant form \cite{Sperling:2019xar,Steinacker:2010rh}
\begin{align}
S[\phi] =   \Tr [T^\mu,\phi][T_\mu,\phi] 
\sim \int d⁴ x\,\sqrt{|G|}G^{\mu\nu}\del_\mu\phi \del_\nu \phi \ ,
\label{scalar-action-metric}
\end{align}
and another way is given below  by showing \eqref{Box-deldel}.
Using \eqref{ind-metric-explicit}, this metric can be written as
\begin{align}
 \dd s^2_G = G_{\mu\nu} \dd x^\mu \dd x^\nu 
   &= -R^2 \sinh^3(\eta) \dd \eta^2 + R^2\sinh(\eta) \cosh^2(\eta)\, \dd \Sigma^2 \ \nn\\
   &= -\dd t^2 + a^2(t)\dd\Sigma^2 \, 
   \label{eff-metric-FRW}
\end{align}
and we can read off the cosmic scale parameter $a(t)$  
\begin{align}
a(t)^2 &=  R^2\sinh(\eta) \cosh^2(\eta) \ \stackrel {t\to\infty}{\sim}  \  R^2\sinh^3(\eta) ,  \label{a-eta}\\
\dd t &=  R \sinh(\eta)^{\frac{3}{2}} \dd\eta \ .
\end{align}
Hence $a(t) \sim \frac 32 t$ for late times, and the Hubble rate is decreasing as $\frac{\dot a}{a} \sim a^{-5/3}$.
This is related to the  time-like vector field $\t$ \eqref{tau-def} via
\begin{align}
 \frac{\del}{\del\eta} &= \tanh(\eta) \t , \qquad 
 \frac{\del}{\del t}
  % &= \frac 1{R \sinh(\eta)^{\frac{3}{2}}}\tanh(\eta) \t 
  = \frac 1R \frac{1}{\sqrt{\sinh(\eta)\cosh(\eta)}} \t 
  \ \stackrel {t\to\infty}{\sim} \ \frac 1R \b \t \ . 
  \label{tau-eta}
\end{align}
As a consistency check,
it is shown in appendix \ref{sec:diffeo-FRW}  
that the covariant d'Alembertian $\Box_G$ of a scalar field 
% \begin{align}
%  \nabla^\a \del_\a \phi &=  (G^{\a\b}\del_\a \del_\b  - \Gamma^\g \del_\g)\phi \nn\\
% %   &=    G^{\a\b}\big(\del_\a \del_\b  + \frac{1}{2x_4^2} (\d^\g_{\b} \eta_{\a\r}x^\r  
% %       + \d^{\g}_{\a}\eta_{\b\r}x^\r  - \eta_{\a\b}x^\g)   \del_\g\big)\phi \nn\\
% %   &=   \b \big(\eta^{\a\b}\del_\a \del_\b  + \frac{1}{x_4^2} (x^\b  \del_\b 
% %   - \frac 12 \eta^{\a\b} \eta_{\a\b}(x^\g  \del_\g))  \big)\phi\nn\\
%   &=   \b \big(\eta^{\a\b}\del_\a \del_\b  - \frac{1}{x_4^2} x^\b \del_\b  \big)\phi
%   \label{Box-explicit}
% \end{align}
is indeed given by $\Box$ up to a factor \cite{Steinacker:2010rh},
\begin{align}
 -\Box \phi &= \eta^{\a\b}\{t_\a,\{t_\b,\phi\}\} 
  = \eta^{\a\b} \b^{-1} (\del_\a \b^{-1}\del_\b \phi) \nn\\
  &= \b^{-2} \big(  \eta^{\a\b} \del_\a \del_\b 
  -  \frac{1}{x_4^2} x^\b \del_\b \big) \phi   \nn\\
  &= \b^{-3} \nabla^\a \del_\a \phi = \b^{-3} \Box_G
\label{Box-deldel}
\end{align}
where $\nabla$ is the covariant derivative w.r.t. $G_{\mu\nu}$.
In particular, we note the useful formula    
\begin{align}
 \del^\a\del_\a \phi = \b^2\big(-\Box + \frac 1{R^2}\t\big)\phi \ .
 \label{deldel-Box-relation}
\end{align}
We would like to decompose $\Box$ into time derivatives $\t$ and 
the space-like  Laplacian $\Delta^{(3)}$ on $H^3$ 
\begin{align}
  -\Delta^{(3)} \phi &=  \nabla_\mu^{(3)}\nabla^{(3)\mu}\phi =
  \del_\mu( P_\perp^{\mu\nu}\del_\nu\phi) 
  %=   \b\{t_\mu, P_\perp^{\mu\nu}\b\{t_\nu,\phi\}\} 
%  &=   \b\{t_\mu, \big(\eta^{\mu\nu} + \frac{x^\mu x^\nu}{R^2\cosh^2(\eta)}\big)\b\{t_\nu,\phi\}\} \nn\\
%  &=   \b\eta^{\mu\nu}\{t_\mu, \b\{t_\nu,\phi\}\} 
%    +  \b\{t_\mu, \frac{x^\mu }{R^2\cosh^2(\eta)}\b x^\nu\{t_\nu,\phi\}\} \nn\\
%  &=   \b\eta^{\mu\nu}\{t_\mu, \b\}\{t_\nu,\phi\} - \b^2 \Box \phi
%    +\frac{4\b}{R^2\cosh^2(\eta)}\t\phi
%    +  \b x^\mu \{t_\mu, \frac{1}{R^2\cosh^2(\eta)}\t\phi\} \nn\\
%  &=  - \b^2 \Box \phi + \frac{\b^3}{R^2} x^\mu\{t_\nu,\phi\}
%    +\frac{4}{R^2\cosh^2(\eta)}\t\phi
%    + \frac{1}{R^2} \t (\frac{1}{\cosh^2(\eta)}\t\phi) \nn\\
% &=  - \b^2 \Box \phi + \frac{\b^2}{R^2} \t \phi
%   +\frac{1}{R^2\cosh^2(\eta)} (2 + \t)\t\phi 
\label{Laplacian-H3-def}
\end{align}
using the time-like and space-like projectors
\begin{align}
 P_\t^{\mu\nu} &:=  \frac{1}{x_\a x^\a} x^\mu x^\nu, \qquad 
 P_\perp^{\mu\nu} := \eta^{\mu\nu} - P_\t^{\mu\nu} \ .
 \label{H3-projector}
\end{align}
After some calculations using \eqref{del-t-rel} and the formulas in section 
\ref{sec:useful-formulas}, one obtains
\begin{align}
 \boxed{
\Box\phi = \Big( \b^{-2}\Delta^{(3)} + \frac{1}{R^2} \t 
   +\frac{\sinh^2(\eta)}{R^2\cosh^2(\eta)} (2 + \t)\t\Big)\phi
 }
 \label{Box-Laplace-tau}
\end{align}
for scalar fields $\phi(x)$.
This can be checked e.g. for $\phi = x^\a$.
% \begin{align}
%   & - \b^2 \Box x^\a + \frac{\b^2}{R^2} \t x^\a
%    +\frac{1}{R^2\cosh^2(\eta)} \Big(2 + \t\Big)\t\phi  x^\a   \nn\\
%  &= - \frac{\b^2}{R^2} x^\a + \frac{\b^2}{R^2} x^\a
%    +\frac{3}{R^2\cosh^2(\eta)}  x^\a   \nn\\
%  &= \frac{3}{R^2\cosh^2(\eta)} x^\a  
% \end{align}
% which is correct. Can also be checked for $\phi = x^4$. 
% Then $\Box x^4 = \frac{4}{R^2} x^4$ and $\Delta^{(3)}x^4 = 0$ and using \eqref{tau-x4}
On the other hand we can use  
the above hyperbolic coordinates \eqref{hyperbolic-coords}, where
\begin{align}
G_{\mu\nu} &= R^2\sinh(\eta)\diag \Big(-\sinh^2(\eta),\cosh^2(\eta),
\cosh^2(\eta)\sinh^2(\chi),\cosh^2(\eta)\sinh^2(\chi)\sin^2(\theta)\Big)
 %\nn\\
 % |G_{\mu\nu}| &= R^8\sinh^{6}(\eta)\cosh^{6}(\eta)\sinh^4(\chi)\sin^2(\theta)
\end{align}
so that
\begin{align}
 \Box_G &= -\frac{1}{\sqrt{|G_{\mu\nu}|}}\del_\mu\big(\sqrt{|G_{\mu\nu}|}\, G^{\mu\nu}\del_\nu\big) \nn\\
  &=  \frac{1}{R^2\sinh^3(\eta)\cosh^3(\eta)} 
 \del_\eta\big(\cosh^3(\eta)\del_\eta \phi\big)
  + \frac{1}{\sinh(\eta)}\Delta^{(3)} \phi \ .
 \label{G-Box-relation}
\end{align}
This reduces indeed to \eqref{Box-Laplace-tau} using $\Box = \b^{-3}\Box_G$
and \eqref{tau-eta}.
The Laplacian  $\Delta^{(3)}$  \eqref{Laplacian-H3-def} 
on the space-like $H^3$ reduces
for rotationally invariant functions $\phi(\chi)$  to
\begin{align}
 \Delta^{(3)} \phi(\chi)
  &= - \frac{1}{R^2\cosh^2(\eta)} \frac 1{\sinh^2(\chi)}
  \del_\chi\big(\sinh^2(\chi)\del_\chi\phi\big) \ .
  \label{Delta-3-H}
\end{align}

\subsection{Higher spin sectors and filtration}
\label{sec:higher-spin}

Due to the extra generators $t^\mu$, the full algebra of functions decomposes
into sectors $\cC^s$ which correspond to spin $s$ harmonics on the $S^2$ fiber:
\begin{align}
    \End(\cH_n) =\cC = \cC^0 \oplus \cC^1 \oplus \ldots \oplus  \cC^n\qquad \text{with} 
\quad 
  \cS^2|_{\cC^s} = 2s(s+1)  \,
  \label{EndH-Cs-decomposition}
\end{align}
Here $\cS^2 = \frac 12\sum_{a,b< 5} [\cM_{ab},[\cM^{ab},\cdot]] 
  + r^{-2} [X_a,[X^a,\cdot]]$ can be viewed as a spin operator\footnote{Since
  local Lorentz invariance is not manifest, the usual notion of spin cannot be used, 
  and $\cS^2$ is a substitute.} on $H^4_n$
  \cite{Sperling:2018xrm}, which commutes with $\Box$.
  In the semi-classical limit, the  $\cC^s$  are modules over $\cC^0$,
  and can be realized explicitly in terms of totally symmetric traceless space-like 
rank $s$ tensor 
fields on $\cM^{3,1}$
\begin{align}
 \phi^{(s)} = \phi_{\mu_1 ... \mu_s}(x) t^{\mu_1} ... t^{\mu_s} , 
 \qquad \phi_{\mu_1 ... \mu_s} x^{\mu_i} = 0
 \label{Cs-explicit}
\end{align}
due to  \eqref{geometry-H-M}.
The underlying $\mso(4,2)$ structure provides an $SO(3,1)$ -invariant derivation
\begin{align}
 D\phi &:= \{x^4,\phi\} \ 
  = r^2 R^2 \frac{1}{x^4} t^\mu \{t_\mu,\phi\}   
  = -\frac{1}{x^4}x_\mu\{x^\mu,\phi\} \nn\\
   &= r^2 R\, t^{\mu_1}\ldots t^{\mu_s} t^\mu \, \nabla^{(3)}_\mu\phi_{\mu_1\ldots\mu_s}(x) 
 \label{D-properties}
\end{align}
where  $\nabla^{(3)}$ is the covariant derivative along the 
space-like $H^3 \subset \cM^{3,1}$.
Hence $D$ relates the different spin sectors in \eqref{EndH-Cs-decomposition}:
\begin{align}
  D = D^- + D^+: \ \cC^{s} \ &\to \cC^{s-1} \oplus \cC^{s+1}, \qquad
    D^\pm \phi^{(s)} = [D\phi^{(s)}]_{s\pm 1} \ 
\end{align}
where $[.]_{s}$ denotes the projection to $\cC^{s}$ defined through 
\eqref{EndH-Cs-decomposition}.
For example, $D x^\mu = r^2 R\, t^\mu$ and $D t^\mu =  R^{-1}\, x^\mu$. 
This allows to define a further refinement \cite{Sperling:2019xar}
\begin{align}
 \cC^{(s,k)} \coloneqq \cK^{(s,k)} /  \cK^{(s,k-1)} , \qquad 
 \cK^{(s,k)} = \ker (D^-)^{k+1} \subset \cC^{s} \ .
 \label{C-sk-def}
\end{align}
Then 
\begin{align}
  D^\pm: \quad \cC^{(s,k)} &\to \cC^{(s-1,k-1)}  \ .
  \label{D-refined}
\end{align}
In particular,  $\cC^{(s,0)} \subset \cC^{s}$ is the space of divergence-free traceless
space-like rank $s$ tensor fields on $\cM^{3,1}$,
while $D^+D\phi^{(0)} = [t^\mu t^\nu]_2 \nabla^{(3)}_\mu\del_\nu \phi^{(0)} \in \cC^{(2,2)} \subset  \cC^2$ encodes the 
traceless second derivatives of the scalar field $\phi^{(0)}$.
These will play an important role below. Finally, 
$\t$ is extended to $\cC^s$ via \cite{Sperling:2019xar}
\begin{align}
 \sinh(\eta)(\t + s)\phi^{(s)}   = x^\mu\{t_\mu,\phi^{(s)} \} \ ,
 \label{tau-relns}
\end{align}
which gives \eqref{D-tau-relation}.

\paragraph{Averaging.}

We will  need some explicit formulas for the projection $[.]_0$ to $\cC^0$:
\begin{align}
 [t^\mu t^\nu]_0 &\eqqcolon \frac{\cosh^2(\eta)}{3r^2} P_\perp^{\mu\nu} \,,
   \label{kappa-average}  
\end{align}
in terms or the projector $P_\perp$ \eqref{H3-projector}
on the time-slices $H^3$. This can be viewed as an averaging over $S^2$.
Explicitly, one finds \cite{Sperling:2019xar}
\begin{subequations}
\label{averaging-relns}
\begin{align}
\left[t^{\a}  \theta^{\mu\nu}\right]_{0} 
  &= \frac{1}{3} \Big(\sinh(\eta) ( \eta^{\a\nu} x^\mu - \eta^{\a\mu} x^\nu)  +  x_\b \varepsilon^{\b 4\a\mu \nu}  \Big)\,, \\
 [t^{\mu_1} \ldots t^{\mu_4}]_0 &= 
  \frac 35 \big([t^{\mu_1}t^{\mu_2}][t^{\mu_3} t^{\mu_4}]_0 
   + [t^{\mu_1}t^{\mu_3}][t^{\mu_2} t^{\mu_4}]_0  + 
[t^{\mu_1}t^{\mu_4}][t^{\mu_2} t^{\mu_3}]_0\big) \,.  \nn\\
%%%
[t^{\a} t^\b t^\g ]_{1} 
  &=  \frac 35 \Big([t^{\a} t^\b]_{0} t^\g + t^{\a} [t^\b t^\g]_{0} 
   +  t^\b[t^{\a} t^\g ]_{0}  \Big) \ .
   \label{average-3}
\end{align}
\end{subequations}
As an application, one can derive the following formula 
\begin{align}
 \{x^\mu,\{x_\mu,\phi\}\}_0 
 % &= \{x_\mu,\theta^{\mu\a}\del_\a\phi\}_0  \nn\\
%   &= 3 r^2 x^\a\del_\a\phi + [\theta^{\mu\a}\theta^{\mu\b}]_0\del_\b\del_\a\phi \nn\\
%   &= 3 r^2 \t\phi + \frac{r^2 R^2}{3}
%   \big(P^{\mu\mu}P^{\a\b} - P^{\mu\a}P^{\mu\b}\big)\del_\b\del_\a\phi \nn\\
%   &= 3 r^2 \t\phi + \frac{r^2 R^2}{3}
%   \Big(\big(4 -\cosh^2(\eta)\big)  
%   \big(\eta^{\a\b} +\frac{1}{R^2}x^\a x^\b\big)  \nn\\
%   &\quad - \big(\eta^{\mu\a} +\frac{1}{R^2}x^\mu x^\a\big)  
%   \big(\eta^{\mu\b} +\frac{1}{R^2}x^\mu x^\b\big)\Big)\del_\b\del_\a\phi \nn\\
%   &= 3 r^2 \t\phi + \frac{r^2 R^2}{3}
%   \Big(\big(4 -\cosh^2(\eta)\big)  
%   \big(\eta^{\a\b} +\frac{1}{R^2}x^\a x^\b\big)  \nn\\
%   &\quad - \eta^{\a\b} - \frac{2}{R^2} x^\a x^\b  
%    + \frac{1}{R^2}\cosh^2(\eta)  x^\a x^\b\Big)\del_\b\del_\a\phi \nn\\
%   &= 3 r^2 \t\phi +
%   \frac {r^2}3 \Big(R^2\big(3 -\cosh^2(\eta)\big) \eta^{\a\b} 
%  +  2 x^\a x^\b  \Big)\del_\b\del_\a\phi\nn\\
%   &= 
%   \frac {r^2}3 \Big(R^2\big(3 -\cosh^2(\eta)\big)  \del^\a\del_\a
%  +  (2\t  + 7)\t\Big)\phi  \nn\\
&=  \frac{r^2R^2}3(3 -\cosh^2(\eta))\b^2(-\Box+\frac{1}{R^2}\t)\phi 
 + \frac {r^2}3 (2\t + 7)\t \phi\, 
 \label{dAlembertian-x}
\end{align}
for $\phi\in\cC^0$.
This could be  another natural d'Alembertian on $\cM^{3,1}$  which exhibits a transition 
from a Euclidean to a Minkowski era, as discussed in \cite{Steinacker:2017bhb}.
However in this paper the effective 
d'Alembertian will be $\Box$, which respects the spin sectors $\cC^s$ 
\eqref{EndH-Cs-decomposition}.

%%%%%%%%%%%%%%%%%%%%%%%%%%%%%%%%%%%%%%%%%%%%%%%%%%%
%%%%%%%%%%%%%%%%%%%%%%%%%%%%%%%%%%%%%%%%%%%%%%%%%%%
%
\section{Matrix model and higher-spin gauge theory}
\label{sec:fluctuations}

Now we return to the  noncommutative setting, and 
define a dynamical model for the fuzzy $\cM^{3,1}$ space-time under consideration.
We consider a Yang-Mills matrix model  with  mass term,
\begin{align}
 S[Y] &= \frac 1{g^2}\Tr \Big([Y^\mu,Y^\nu][Y^{\mu'},Y^{\nu'}] \eta_{\mu\mu'} \eta_{\nu\nu'} \, 
  +\frac{6}{R^2} Y^\mu Y^\nu \eta_{\mu\nu} \Big) \ . 
 \label{bosonic-action}
\end{align}
This includes in particular the  IKKT or IIB matrix model \cite{Ishibashi:1996xs} with mass term,
which is best suited for quantization because maximal supersymmetry protects from UV/IR mixing \cite{Minwalla:1999px}.
As observed in \cite{Sperling:2019xar}, $\cM^{3,1}$ is indeed a solution of this model\footnote{any other 
positive mass parameter in \eqref{bosonic-action} would of course just result 
in a trivial rescaling. For negative mass parameter, $Y^\mu \sim X^\mu$ would be a solution \cite{Steinacker:2017bhb}, 
but the fluctuations are more difficult to analyze.}, through
\begin{align}
 Y^\mu = T^\mu \ .
\end{align}
Now consider  tangential
deformations of the above background solution, i.e.
\begin{align}
 Y^\mu = T^\mu  + \cA^\mu \ , 
\end{align}
where $\cA^\mu \in \End(\cH_n) \otimes \C^4$ is an arbitrary (Hermitian) fluctuation.
The  Yang-Mills action \eqref{bosonic-action}  can be expanded  as
\begin{align}
 S[Y] = S[T]  +  S_2[\cA] + O(\cA^3) \ ,  
 \end{align}
 and the quadratic fluctuations are  governed by  
 \begin{align}
S_2[\cA] = -\frac{2}{g^2} \,\Tr \left( \cA_\mu 
\Big(\cD^2 -\frac{3}{R^2}\Big) \cA^\mu + \cG\left(\cA\right)^2 \right) .
\label{eff-S-expand}
\end{align}
Here 
\begin{align}
\cD^2 \cA =  \left(\Box  - 2\cI \right)\cA  
\label{vector-Laplacian}
\end{align}
is the  vector d'Alembertian, which involves  the scalar matrix d'Alembertian
$\Box \sim \a^{-1} \Box_G$   on the $\cM^{3,1}$ background 
\eqref{Box-def},   \eqref{Box-deldel} as discussed before,
 and the intertwiner
\begin{align}
 \cI (\cA)^\mu = -\im [[ Y^\mu, Y^\nu],\cA_\nu] =  \frac{\im}{r^2 R^2} 
[\Theta^{\mu\nu},\cA_\nu] 
 \eqqcolon -\frac{1}{r^2 R^2}\tilde\cI (\cA)^\mu \ 
\end{align}
using \eqref{T-T-CR}.
As usual in Yang-Mills theories, $\cA$ transforms under gauge transformations as
\begin{align}
 \d_\L\cA = -i[T^\mu  + \cA^\mu,\L] \sim \{t^\mu,\L\}  + \{\cA^\mu,\L\}
\end{align}
for any $\L\in\cC$,
and the scalar ghost mode
 \begin{align}
\cG(\cA) = -\im [T^\mu,\cA_\mu] \sim \{t^\mu,\cA_\mu\}  ,  
 \label{gaugefix-intertwiner}
 \end{align}
should be removed to get a meaningful theory.
This can be achieved by adding a gauge-fixing term $-\cG(\cA)^2$ to the action
as well as the corresponding Faddeev-Popov (or BRST) ghost. Then the quadratic 
action becomes 
\begin{align}
 S_2[\cA] + S_{g.f} + S_{ghost} &= -\frac{2}{g^2}\Tr\, 
\left( \cA_\mu \Big(\cD^2  -\frac{3}{R^2} \Big) \cA^\mu + 2 \obar{c} \Box 
c \right) \ 
\label{eff-S-gaugefixed}
\end{align}
where $c$ denotes the fermionic BRST ghost; see e.g.\ \cite{Blaschke:2011qu} 
for more details.
% 

%%%%%%%%%%%%%%%%%%%%%%%%%%%%%%%%%%%%%%%%%%%%% 
% 
\section{Fluctuation modes}

All indices will be raised and lowered with $\eta^{\mu\nu}$ in this section.
We should expand the vector modes into higher spin modes according to 
\eqref{EndH-Cs-decomposition}, \eqref{Cs-explicit}
\begin{align}
 \cA^\mu &= A^{\mu}(x) + A^{\mu}_\a(x)\, t^\a +   A^{\mu}_{\a\b}(x)\, t^\a t^\b + \ldots 
  \ \in \ \cC^0   \oplus  \cC^1  \oplus  \cC^2  \oplus \ \ldots
 \label{A-M31-spins}
\end{align}
However these are neither irreducible nor eigenmodes of $\cD^2$. 
In \cite{Sperling:2019xar}, three series of
spin $s$ eigenmodes $\cA_\mu$ were found of the form
\begin{align}
\label{A2-mode-ansatz}
 \boxed{ \
\begin{aligned}
 \cA_\mu^{(g)}[\phi^{(s)}] &= \{t_\mu,\phi^{(s)}\}  \quad \in \cC^{s}\,,
\\
 \cA_\mu^{(+)}[\phi^{(s)}] &= \{x_\mu,\phi^{(s)}\}|_{\cC^{s+1}} \ \equiv  
\{x_\mu,\phi^{(s)}\}_+ \quad \in \cC^{s+1} \,, \\
 \cA_\mu^{(-)}[\phi^{(s)}] &= \{x_\mu,\phi^{(s)}\}|_{\cC^{s-1}} \  \equiv  
\{x_\mu,\phi^{(s)}\}_- \quad \in \cC^{s-1} \,
 \end{aligned}
 }
\end{align}
for any $\phi^{(s)}\in\cC^s$, which satisfy
\begin{align}
\cD^2 \cA_\mu^{(g)}[\phi] &= 
\cA_\mu^{(g)}\left[\left(\Box+\frac{3}{R^2}\right) 
\phi\right] \,,
 \label{puregauge-D2}  \\
 %%%
 \cD^2 \cA_\mu^{(+)}[\phi^{(s)}]  
    &= \cA_\mu^{(+)}\left[\left(\Box + \frac{2s+5}{R^2} 
\right)\phi^{(s)}\right]
    \label{D2-A2p-eigenvalues} \,, \\
 \cD^2 \cA_\mu^{(-)}[\phi^{(s)}] 
   &= \cA_\mu^{(-)}\left[\left(\Box + 
\frac{-2s+3}{R^2}\right)\phi^{(s)}\right] \, .
   \label{D2-A2m-eigenvalues}
\end{align}
We provide in appendix \ref{sec:ladder-ops} a simple new derivation for the 
last two relations.
Hence diagonalizing $\cD^2$ is reduced  to diagonalizing $\Box$ on $\cC^s$, 
and we have the on-shell modes $\big(\cD^2  -\frac{3}{R^2} \big) \cA = 0$ for
\begin{align}
 \cA^{(+)}[\phi^{(s)}]  \qquad & \text{for } \  \ \left(\Box + 
\frac{2s+2}{R^2}\right) \phi^{(s)} = 0 \,,\\
 \cA^{(-)}[\phi^{(s)}] \qquad& \text{for } \quad  \ \left(\Box + \frac{-2s}{R^2} 
\right)\phi^{(s)} = 0 \, ,   \\
\cA^{(g)}[\phi^{(s)}] \qquad &\text{for } \qquad \qquad  \ \Box \phi^{(s)} = 0 \ .
  \label{on-shell-B2pm}
\end{align}
Of course $\cA^{(g)}$ is a pure gauge mode and hence unphysical.
Furthermore, the following gauge fixing identities\footnote{As a check, 
consider e.g. $\cA^{(-)}[\phi^{(1,0)}]$. It satisfies 
$\{t_\mu,\cA^\mu\} = 0 = x_\mu\cA^\mu$ due to \eqref{D-properties}, and \eqref{div-A-full} gives
$\nabla_\mu \cA^\mu = 0$ and $\tilde\cI(\cA^\mu) = r^2 \cA^\mu$, 
consistent with (A.33) in \cite{Sperling:2019xar}.} were shown in \cite{Sperling:2019xar}
\begin{align}
\{t^{\mu},\cA_\mu^{(+)}[\phi^{(s)}]\} &= \frac{s+3}{R} D^+\phi^{(s)} \,,\\
\{t^{\mu},\cA_\mu^{(-)}[\phi^{(s)}]\} &= \frac{-s + 2}{R} D^-\phi^{(s)} \, .
 \label{A2-gaugefix}
\end{align}
In particular
for $s=2$, $\cA_\mu^{(-)}[\phi^{(2)}]$ is 
already gauge fixed\footnote{For $s\neq 2$ some linear combinations of 
$\cA_\mu^{(+)}$ and $\cA_\mu^{(-)}$ must be taken to obtain a gauge-fixed physical
solution. However, this is  not our concern here.}. 
This will lead to the physical spin 2 metric fluctuations.
According to the discussion in section \ref{sec:higher-spin}, they decompose into the 
modes $\cA_\mu^{(-)}[\phi^{(2,0)}], \cA_\mu^{(-)}[D\phi^{(1,0)}]$ and 
$\cA_\mu^{(-)}[D^+D\phi^{(0)}]$, which we will denote
-- in slight abuse of language -- as 
helicity 2, 1 and 0 sectors of the would-be massive spin 2 modes, respectively.
We will focus on the physical helicity 0 or scalar mode, with  on-shell condition
\begin{align}
\boxed{ \
 \cA_\mu^{(-)}[D^+D\phi], \qquad \big(\Box + \frac{2}{R^2}\big)\phi = 0, \qquad \phi \in \cC^0
 \ }
 \label{onshell-0}
\end{align}
due to  \eqref{Box-x4-relations}.
However, one series of spin $s$ (off-shell) eigenmodes $\cA_\mu$ 
of $\cD^2$  is still missing, and was not known 
up to now. We will find  the missing scalar mode in section \ref{sec:time-mode}, in terms of
\begin{align}
 \cA_\mu^{(\t)}[\phi^{(s)}] = x_\mu \phi^{(s)} \ .
 \label{A4-mode-ansatz}
\end{align}
That ansatz was also considered in \cite{Sperling:2019xar}, where it was shown to satisfy
\begin{align}
  \cD^2 \cA_\mu^{(\t)}[\phi^{(s)}] &= 
\cA^{(\t)}_\mu\left[\left(\Box+\frac{7}{R^2}\right)\phi^{(s)}\right] + 2\eth_\mu \phi ^{(s)} 
\,   \label{A4-eom}   \\
 \{t^\mu,\cA_\mu^{(\t)}[\phi^{(s)}]\} &= \sinh(\eta)\big(4 +s + \t \big) \phi^{(s)} \ .
 \label{gaugefix-timelike}
\end{align}
Here $\eth_\mu$ will be defined in \eqref{eth-tbracket-0}.
We will show in the following that $\cA^{(-)}[D^+D\phi]$ provides the on-shell 
mode leading to the linearized Schwarzschild metric. Moreover, an ansatz based on 
$\cA_\mu^{(\t)}$ will give solutions which are 
equivalent on-shell, but not off-shell.

\subsection{Scalar $\cA^{(-)}[D^+D\phi]$ mode}

We  need the explicit form of $\cA^{(-)}[D^+D\phi]$. This is quite tedious to work out
and delegated to the appendix \ref{sec:eval-A-DD},
where we provide an exact expression  in \eqref{A-DDphi-1}.
This simplifies considerably using the on-shell condition 
$(\Box + \frac{2}{R^2})\phi = 0$ \eqref{onshell-0}, leading to 
\begin{align}
\boxed{ \ 
  \cA_\mu^{(-)}[D^+D^+\phi] 
  = \frac {2r^4}5  \Big(\b (t^\mu  + x^\mu t^\a  \del_\a) 
   - \frac 1{3r^2} \theta^{\m\g}\del_\g (\t+4 + \b^2)\Big)(\t+2)\phi  + \{t_\mu,\L\} \ 
    \ }
   \label{A-DDphi-explicit}
\end{align}
with $\L$ given in \eqref{gaugeparam-schwarzschild}.
This is a reasonable perturbation of the background $Y^\mu = t^\mu$, 
as long as $\phi$ remains bounded.
Remarkably, \eqref{A-DDphi-explicit} can be rewritten
via $\theta^{\m\g}\del_\g \phi = \cA^{(+)\mu}[\phi]$ as
\begin{align}
  \cA_\mu^{(-)}[D^+D^+\phi]   &=   \frac 25 \frac{r^2}{R} \Big(D(x^\mu \phi') 
   - \frac R{3}  \cA^{(+)\mu}[(\t+4+\b^2)(\t+2)\phi]\Big)
  + \{t^\mu,\L \} \nn\\
 %  &=  \frac 25 \frac{r^2}{R} D\Big(\cA^{(\t)\mu}[\b\phi'] 
 %  - \frac{R^2}{3}  \{t^\mu,(\sinh(\eta)(\t+5) +2\b)\b\phi']\Big) \ 
 %  + \{t^\mu,\L' \} \nn\\
   &= \frac 25 \frac{r^2}{R}  \cA_\mu^{(S)}[\phi']  
   +  \{t^\mu, \L' \} 
  \label{ADDphi-onshell-specialform}
\end{align}
where $\cA_\mu^{(S)}[\phi']$ is the new mode defined in \eqref{A-S-def}, with
\begin{align}
 \phi' = \b(\t+2)\phi, \qquad 
 \L' = \L  +  \frac 2{15}r^2 R D(\t+4+\b^2)(\t+2)\phi \ .
 \label{phi-prime}
\end{align}
To see this, the identities
\begin{align}
 \b(t^\mu  +  x^\mu t^\a  \del_\a)(\t+2)\phi
   &= \frac{1}{r^2R} D(x^\mu \phi')  \nn\\
 (\t+4+\b^2)(\t+2)\phi &= (\sinh(\eta)(\t+5)+2\b)\phi'
 \end{align}
 and the on-shell equations
 \begin{align}
 \big(\Box + \frac{2}{R^2}(3+\t-\b^2)\big)\phi' &= 0  \label{eom-phi'}\\
 \Box \L' &= 0 \ 
\end{align} 
are needed, which
can be checked using the results of section \ref{sec:useful-formulas}.
The last form implies that $\cA_\mu^{(-)}[D^+D^+\phi]$ 
differs from $\cA_\mu^{(S)}[\phi']$ by an on-shell pure gauge mode.
This means that even though these  are distinct off-shell modes, they 
become degenerate on-shell, so that there is only one physical scalar graviton mode.
This is  essential for a ghost-free theory.

Strictly speaking the form \eqref{A-DDphi-explicit} 
collapses for $\t=-2$. 
However, its expression in terms of $\phi'$ -- or alternatively 
the form \eqref{ADDphi-onshell-specialform} -- makes sense also in the limit
$\t\to -2$. This is important, because $\t=-2$ gives precisely the 
Ricci-flat quasi-Schwarzschild solution, as discussed in section  \ref{sec:A-DD-metric}.

For completeness
we also provide the explicit form of 
the  pure gauge field $\cA^{(g)}$ corresponding to  \eqref{gaugeparam-schwarzschild} 
\begin{align}
 \cA^{(g)\mu}[\L] &= \{t_\mu,\L\} 
 % = -\frac 25  r^4 R^2\{t_\mu,t^\a \del_\a(\t+3)\phi \} \nn\\
%   &= -\frac 25  r^4 R^2\Big(\{t_\mu,t^\a\} \del_\a(\t+3)\phi \ 
%      + t^\a \{t_\mu,\del_\a(\t+3)\phi \}\Big) \nn\\
%   &= -\frac 25  r^4 R^2\Big(-\frac{1}{r^2 R^2}\theta^{\mu\a} \del_\a(\t+3)\phi \ 
%   + \sinh(\eta) t^\a \del_\mu\del_\a(\t+3)\phi \Big)  \nn\\
  = \frac 25  r^2 \Big(\theta^{\mu\a} \del_\a\ 
   -R\sinh(\eta) D\del^\mu\Big)(\t+3)\phi  \ .
   \label{A-puregauge-L}
\end{align}

\paragraph{Gauge fixing.}

A non-trivial consistency check of \eqref{A-DDphi-explicit} is obtained by 
verifying that it satisfies the gauge-fixing constraint.
For the pure gauge contribution, this is 
\begin{align}
 \{t^\mu,\{t_\mu,\L\}\} &= 
  \frac 25  r^2 R \Box D(\t+3) \phi
 %= \frac 25  r^2 R D(\Box+\frac{2}{R})((\t+3) \phi) 
 % &= \frac 25  r^2 R D(\Box+\frac{2}{R})\t \phi
  = \frac {4r^2}{5R} D \b^2(2+\t) \phi
\end{align}
using \eqref{box-tau-relation-1}.
Together with the relations \eqref{gauge-fixing-relations-scalar},
 one verifies indeed $\{t^\mu, \cA_\mu^{(-)}[D^+D^+\phi]\} =0$.
% \begin{align}
% \{t^\mu, \cA_\mu^{(-)}[D^+D^+\phi]\} 
% %   &= \frac{2}{5}r^4 \{t_\mu,  \b t^\mu (2 + \t) \phi \}
% %    + \frac{2}{5} r^4 \{t_\mu, \b x^\mu t^\a  \del_\a  (2 + \t)\phi \} \nn\\
% %  &\quad 
% %   - \frac2{15} r^2\{t_\mu, \theta^{\nu\a} \del_\a (\t+4+\b^2)(\t+2) \phi\}
% %   +  \{t^\mu,\{t_\mu,\L\}\} \nn\\
% %  &=  \frac{2}{5} r^4 t^\a  \del_\a \Big(
% %  1 +  (\t+3-\b^2) - (\t+4+\b^2) +2\b^2
% %  \Big)(2 + \t)\phi \nn\\
%   &= 0 \ .
% \end{align}

\subsection{Time-like  scalar mode $\tilde\cA_\mu^{(\t)}$}
\label{sec:time-mode}

In this section we will show that a refined ansatz involving $\cA^{(\t)}[\phi]$ 
provides a further scalar eigenmode of $\cD^2$. This will also  
provide the missing $10^{th}$ degree of freedom for 
the off-shell metric fluctuations. While this is not essential to understand the Schwarzschild solution,
 it provides further insights.

First we recall the relation \eqref{A4-eom}, which involves the derivation 
\begin{align}
  \eth^\mu \phi &= -\frac{1}{r^2 R^2}\theta^{\mu b}\{x_b,\phi\} 
   = \{t^\mu,\b\phi\} 
  +  \frac 1{R^2} x^\mu \Big(-\b^2 + \t \Big)\phi ,
   \label{eth-tbracket-0}
\end{align}
for $b=0,...,4$ and $\phi\in\cC^0$. 
The second form is obtained noting that 
\begin{align}
 % \eth^\mu x^\nu &= \eta^{\mu\nu} + \frac 1{R^2} x^\mu x^\nu      \nn\\
 \eth^\mu \phi &= \del^\mu\phi +  \frac 1{R^2} x^\mu \t\phi \qquad \mbox{for}
     \qquad \phi\in\cC^0   \ ,
 \end{align}
 and rewriting the first term using 
$\del^\mu\phi  = \{t^\mu,\b\phi\} - \frac 1{R^2} x^\mu \ \b^2\phi$.
Hence \eqref{A4-eom} can be written as 
\begin{align}
  \cD^2 \cA_\mu^{(\t)}[\phi] 
 % &= \cA^{(\t)}_\mu\left[\left(\Box+\frac{7}{R^2}\right)\phi\right] 
 %+ 2 \{t^\mu,\b\phi\} 
 %+ \frac 2{R^2} x^\mu \Big(-\b^2 + \t \Big)\phi  \nn\\
 %%
 &= \cA^{(\t)}_\mu\left[
 \left(\Box + \frac 1{R^2}\Big(- 2\b^2 + 2\t +7 \Big)\right)\phi\right] 
 + 2 \{t^\mu,\b\phi\} \ .
 \label{D2-timelike-rel}
\end{align}
Since the last term is a pure gauge mode, this provides a new eigenmode of $\cD^2$:

\paragraph{Scalar time-like $\cC^0$ mode.}

Combining  the above with  %$\cD^2\{t^\mu,\tilde\phi\} = \{t^\mu,(\Box+\frac{3}{R^2})\tilde\phi\}$
\eqref{puregauge-D2}, the ansatz
\begin{align}
\boxed{  \ \ 
 \tilde\cA_\mu^{(\t)}[\phi] = \cA_\mu^{(\t)}[\phi] + \{t^\mu,\tilde\phi\} \ \ 
 }
 \label{tilde-A-t-0}
 \end{align}
 leads to new scalar eigenmode of $\cD^2$ 
 \begin{align}
 \cD^2 \tilde\cA_\mu^{(\t)}[\phi] 
 % &= \cA^{(\t)}_\mu\left[
 %\left(\Box + \frac 1{R^2}\Big(-2\b^2 + 2\t +7 \Big)\right)\phi\right] 
 %+ \{t^\mu,2\b\phi + (\Box+\frac{3}{R^2})\tilde\phi\} \nn\\
 &= \l \tilde\cA_\mu^{(\t)}[\phi]
\end{align}
provided
\begin{align}
\Big(\Box + \frac 1{R^2}\big(- 2\b^2 + 2\t +7 \big)\Big)\phi
  &= \l \phi\nn\\
   (\Box+\frac{3}{R^2})\tilde\phi + 2\b\phi  &= \l \tilde\phi \ .
 \label{timelikemode-0}
\end{align}
The first equation can be solved, and 
has propagating solutions $\phi$. Then $\tilde\phi$
is determined by the second equation, up to solutions of 
$(\Box+\frac{3}{R^2}-\l)\tilde\phi = 0$.
This $4^{th}$ eigenmode is needed e.g. for the off-shell propagator. 
In particular, $\tilde\cA_\mu^{(\t)}[\phi]$ is on-shell, $(\cD^2  -\frac{3}{R^2} \Big) \tilde\cA^{(\t)} = 0$ for 
\begin{align}
\Big(\Box + \frac 2{R^2}\big(2+\t  -\b^2 \big)\Big)\phi &= 0 \nn\\
  \Box \tilde\phi +  2\b\phi  &= 0  \ .
 \label{timelikemode-0a}
\end{align}
However the gauge fixing condition for this  mode is very restrictive on-shell, 
\begin{align}
 \{t_\mu,\tilde\cA_\mu^{(\t)}[\phi] \} 
 &=  \{t_\mu,\cA_\mu^{(\t)}[\phi] \} - \Box\tilde\phi  
 =  \big(\sinh(\eta)(4 + \t )  + 2\b\big) \phi 
 \label{gaugefix-At-0}
\end{align}
or
\begin{align}
  \t \phi  &=  -(4 + 2\b^2) \phi  \ ,
\end{align}
which means that $\phi$ is decaying in time with a fixed rate.
Hence these modes are ``frozen'' 
rather than propagating, which is good because they would otherwise be ghosts. 
We will see that these $\cA \in\cC^0$ modes do not contribute to the linearized metric fluctuations.

\paragraph{Scalar time-like $\cC^1$ mode.}

Based on the above mode and using the ladder property \eqref{D2-D+-relation}, we can similarly find a 
new eigenmode $\cA \in \cC^1$ 
with the ansatz
\begin{align}
 D\tilde\cA_\mu^{(\t)}[\phi]
  &= D\big(\cA_\mu^{(\t)}[\phi] + \{t^\mu,\tilde\phi\} \big)  \
  = r^2 R t^\mu \phi + x^\mu D\phi 
    + \frac{1}{R}\{x^\mu,\tilde\phi\} + \{t^\mu,D\tilde\phi\} \ .
    \label{D-tilde-A-t-1}
\end{align}
This is an eigenmode of $\cD^2$ provided $\tilde\cA_\mu^{(\t)}$ is an eigenmode,  with shifted eigenvalue 
\begin{align}
 \cD^2 (D\tilde\cA_\mu^{(\t)}[\phi])
  &= D(\cD^2 + \frac{2}{R^2}) \tilde\cA_\mu^{(\t)}[\phi] \ .
\end{align}
In particular, $D^+\tilde\cA_\mu^{(\t)}[\phi]$ is on-shell if 
$(\cD^2 -\frac{1}{R^2})\tilde\cA_\mu^{(\t)}[\phi] = 0$,
which means by \eqref{timelikemode-0}
\begin{subequations}
\label{onshell-tildephi}
\begin{align}
 \Big(\Box + \frac 2{R^2}(-\b^2 + \t + 3)\Big)\phi
  &= 0 \label{onshell-tildephi-1} \\
  (\Box + \frac{2}{R^2})\tilde\phi + 2 \b\phi  &= 0 \ .
  \label{onshell-tildephi-2}
\end{align}
\end{subequations}
This provides the missing $4^{th}$ scalar eigenmode in $\cC^1$.
The  gauge-fixing condition is
\begin{align}
  \{t^\mu,D\tilde\cA_\mu^{(\t)}[\phi] \} 
 &=  \{t^\mu, r^2 R t_\mu \phi + x_\mu D\phi \} + \{t^\mu,D \{t_\mu,\tilde\phi\}\}  \nn\\
 % &=  r^2 R \{t^\mu, t_\mu \phi\} + \{t^\mu,x_\mu D\phi \} 
 % + \frac 1R \{t^\mu,\{x_\mu,\tilde\phi\}\} + \{t^\mu,\{t_\mu,D\tilde\phi\}\}    \nn\\
  &=  r^2 R t_\mu \{t^\mu, \phi\} + 4\sinh(\eta) D\phi +  x_\mu \{t^\mu,D\phi \} 
  + \frac 1R \{t^\mu,\{x_\mu,\tilde\phi\}\} - \Box D\tilde\phi      \nn\\
  &=  \sinh(\eta) D\phi + 4\sinh(\eta) D\phi +  \sinh(\eta) (\t+1) D\phi 
  + \frac 3{R^2} D\tilde\phi -  D(\Box+\frac{2}{R^2})\tilde\phi      \nn\\
 % &=  \sinh(\eta) (\t+6) D\phi -  D(\Box - \frac{1}{R^2})\tilde\phi   \nn\\
 % &=  \Big(\sinh(\eta) (\t+6) + 2\b\Big)D\phi +  \frac{3}{R^2}D\tilde\phi  \nn\\
  &=  D\Big(\big(\sinh(\eta) (\t+5) +  2\b\big)\phi 
    +  \frac{3}{R^2}\tilde\phi \Big)
\end{align}
using \eqref{gaugefix-At-0}, \eqref{tau-relns}, \eqref{D-tau-relation}, \eqref{D-properties}
and  the on-shell equations \eqref{onshell-tildephi}.
This implies
\begin{align}
 \big(\sinh(\eta) (\t+5) +  2 \b\big)\phi 
    +  \frac{3}{R^2}\tilde\phi = f(x^4) \ .
 \label{gaugefix-t-relation}
\end{align}
For now we set $f=0$. Then 
\begin{align}
 \tilde\phi = -\frac{R^2}3\big(\sinh(\eta) (\t+5) +  2 \b\big)\phi \ ,
 \label{tilde-phi-sol}
\end{align}
and one can verify that the equations of motion  \eqref{onshell-tildephi-2} for $\tilde\phi$
indeed follow from those of $\phi$,
%\begin{align}
% 0 &=  -R^2(\Box+\frac{2}{R^2}) \big(\sinh(\eta) (\t+5) +  2 \b\big)\phi + 6 \b\phi 
%   &= - R^2\Box\big(\sinh(\eta) (\t+5) +  2 \b\big)\phi 
%    - 2\big(\sinh(\eta) (\t+5) +  2 \b\big)\phi+ 6 \b\phi \nn\\
%    &= - R^2\Box\sinh(\eta) (\t+5)\phi  - 2 R^2\Box \b\phi 
%    - 2\sinh(\eta) (\t+5)\phi + 2 \b\phi \nn\\
%    &= - R^2\sinh(\eta)(\Box+\frac{2}{R^2}(\t+2)) (\t+5)\phi 
%     - 2 R^2(\b\Box-\frac{2}{R^2}(\t+2)\b)\phi 
%    - 2\sinh(\eta) (\t+5)\phi + 2 \b\phi \nn\\
%    &= - R^2\sinh(\eta)\Box(\t+5)\phi - 2\sinh(\eta)(\t+2)(\t+5)\phi 
%     - 2 \b R^2\Box \phi + 4\b(\t+2)\phi 
%      -4(1+\b^2)\b\phi \nn\\
%   &\quad - 2\sinh(\eta) (\t+5)\phi + 2 \b\phi \nn\\
%    &= - R^2\sinh(\eta)\Box(\t+5)\phi 
%       - 2 \b R^2\Box \phi 
%     - 2\sinh(\eta)(\t+3)(\t+5)\phi 
%          -4(1+\b^2)\b\phi
%      + 2\b(2\t+5)\phi     \nn\\
%    &= - R^2\sinh(\eta)\big((\t+5)\Box\phi
%       +2\b^2(-\Box+\frac{1}{R^2}\t)\phi\big) 
%       - 2 \b R^2\Box \phi 
%     - 2\sinh(\eta)(\t+3)(\t+5)\phi   \nn\\
%  &\quad        -4(1+\b^2)\b\phi
%      + 2\b(2\t+5)\phi   \nn\\
%    &= - R^2\sinh(\eta) (\t+5)\Box\phi
%     - 2\b\t\phi
%     - 2\sinh(\eta)(\t+3)(\t+5)\phi  
%          -4(1+\b^2)\b\phi
%      + 2\b(2\t+5)\phi \nn\\
%    &= 2\sinh(\eta) (\t+5)(-\b^2 + \t + 3)\phi
%     - 2\sinh(\eta)(\t+3)(\t+5)\phi  
%          -4(1+\b^2)\b\phi
%      + 2\b(2\t+5)\phi - 2\b\t\phi \nn\\
%    &= -2\sinh(\eta) (\t+5)\b^2\phi  + 10\b\phi
%       -4(1+\b^2)\b\phi + 2\b\t\phi 
% &= 0
%\end{align}
using the relations in section \ref{sec:useful-formulas}.
% Hence
% \begin{align}
%  0 &=  - (\t+5)\b^2\phi  + 5\b^2\phi - 2(1+\b^2)\b^2\phi + \b^2\t\phi \nn\\
%   &= \big(-\t \b^2  -2(1+\b^2)\b^2  + \b^2\t \big)\phi  \nn\\
%  &=  \big( 2(1 +\b^2)\b^2  - 2(1+\b^2)\b^2  \big)\phi  \nn\\
%   &= 0
%  \end{align} 
This means that the gauge-fixing condition leading to \eqref{tilde-phi-sol}
is consistent with the equations of motion, and we have found a physical propagating mode of the form 
\begin{align}
\boxed{ \
 \cA_\mu^{(S)}[\phi]
  := D\Big(\cA_\mu^{(\t)}[\phi] -\frac{R^2}3 \{t^\mu, \big(\sinh(\eta) (\t+5) +  2 \b\big)\phi \} \Big) 
  \  }
  \label{A-S-def}
\end{align}
with $\phi$ satisfying  \eqref{onshell-tildephi-1}.
On-shell, this coincides precisely with the on-shell eigenmode
$\cA^{(-)}[D^+D\phi]$  \eqref{ADDphi-onshell-specialform}, although off-shell (hence in the propagator) they are distinct modes.
In the quasi-static case $\t=-2$, this will give the linearized Schwarzschild metric.

\section{Scalar metric fluctuation modes}
\label{sec:graviton}

In this section, we elaborate the  metric fluctuations arising from the above scalar modes.
The effective metric for functions of $\cM^{3,1}$ on a perturbed background $Y = T + \cA$ can be extracted from the kinetic 
term  in \eqref{scalar-action-metric}, which defines the 
bi-derivation
\begin{align}
\begin{aligned}
 \g:\quad \cC\times \cC  \ &\to  \quad \cC  \\
  (\phi,\phi') &\mapsto \{Y^\a,\phi\}\{Y_\a,\phi'\}
  \label{metric-full}
  \end{aligned}
\end{align}
up to a conformal factor as discussed in section \ref{sec:metric}.
Specializing to $\phi=x^\mu, \phi' = x^\nu$ we obtain  the coordinate form
\begin{align}
\g_Y^{\mu\nu} &=  \obar\g^{\mu\nu} + \d_\cA \g^{\mu\nu} + [\{\cA^\a,x^\mu\}\{\cA_\a,x^\nu\}]_0
%  =  \obar\g^{\mu\nu} + \sinh(\eta)h^{\mu\nu} + \sinh^2(\eta)h^{\mu\a}\eta_{\a\b}h^{\nu\b}
\label{gamma-nonlinear}
\end{align}
 where the linearized contribution is given by
\begin{align}
\begin{aligned}
 \d_\cA \g^{\mu\nu} &\coloneqq 
  [\{t^\a,x^\mu\}\{\cA_\a,x^\nu\}]_0 + (\mu \leftrightarrow \nu) 
  = \sinh(\eta) \{\cA_\mu,x^\nu\}_0 + (\mu \leftrightarrow \nu) \ .
  \label{gravitons-H1}
  \end{aligned}
\end{align}
The projection on $\cC^0$ ensures that this is the metric for functions on $\cM^{3,1}$.
We will focus on the linearized contribution in $\cA$ in the following.
To evaluate this explicitly,
it is convenient to consider the following rescaled graviton mode: 
\begin{align}
  h^{\mu\nu}[\cA] &\coloneqq   \{\cA^\mu,x^\nu\}_0 + (\mu \leftrightarrow \nu) \ , 
   \qquad h[\cA] = 2\{\cA^\mu,x_\mu\}_0 \ .
 \label{tilde-H-def}
\end{align}
Clearly only $\cA \in \cC^1$ can contribute to $ h^{\mu\nu}[\cA]$.
% We will need 
% \begin{align}
%  \{t_\mu,h^{\mu\nu}\} 
%   &=  \{t_\mu, \{\cA^\mu,x^\nu\}_-\} + \{t_\mu, \{\cA^\nu,x^\mu\}_-\} \nn\\
%   &=  \{\{t_\mu, \cA^\mu\},x^\nu\}_-
%   +\{\cA^\nu,\sinh(\eta)\}_-
%   - \{t_\mu, \{x^\mu,\cA^\nu\}_-\} 
%   \nn\\
%   &=  \{\{t_\mu, \cA^\mu\},x^\nu\}_-
%   - \frac{2}{R} D^- \cA^\nu
% \end{align}
% and
% \begin{align}
%  \{t_\mu,h\} &= 2\{t_\mu,\{\cA^\a,x_\a\}_0\}  
%   = 2\{\{t_\mu,\cA^\a\},x_\a\}_0 - \frac 2R D^-\cA^\mu
% \end{align}
% as well as
% \begin{align}
%  x_\mu h^{\mu\nu}
%   &= x_\mu \{\cA^\mu,x^\nu\}_- +  x_\mu \{\cA^\nu,x^\mu\}_-  \nn\\
%   &= \{ x_\mu \cA^\mu,x^\nu\}_- - [\cA_\mu\theta^{\mu\nu}]_0 + x^4 D^-\cA^\nu  \ .
% \end{align}
Taking into account the conformal factor as identified in section \ref{sec:metric},
the effective metric $G^{\mu\nu}$ \eqref{eff-metric-G} is  
\begin{align}
 G^{\mu\nu} &= \obar G^{\mu\nu} + \d G^{\mu\nu}  \nn\\
 &= \a\left[\g^{\mu\nu} +  \d_\cA \g^{\mu\nu} 
  -\frac{1}{2} \eta^{\mu\nu}\,\left(\eta_{\a\b}\ \d_\cA \g^{\a\b}\right)\right]   \nn\\ 
 \d G^{\mu\nu} &= \b^2\big(h^{\mu\nu} -\frac{1}{2} \eta^{\mu\nu}\,h\big) \ .
\label{eq:def_phys_graviton}
\end{align}
Here $\obar G^{\mu\nu} = \a\g^{\mu\nu} = \b\eta^{\mu\nu}$ \eqref{eff-metric-G}
is the effective background metric, 
$\a = \b^3$ is the conformal factor arising from the fixed symplectic measure on $\C P^{1,2}$, and $\b=\sinh(\eta)^{-1}$ \eqref{del-t-rel}.
Equivalently,     
\begin{align}
G_{\mu\nu} &=\obar G_{\mu\nu} - \d G_{\mu\nu} ,  \nn\\
 \d G_{\mu\nu} &= \obar G_{\mu\mu'} \obar G_{\nu\nu'} \d G^{\mu'\nu'} 
 = h_{\mu\nu} -\frac{1}{2} \eta_{\mu\nu}\,h
\end{align}
where $h_{\a\b} = \eta_{\a\a'}\eta_{\b\b'}h^{\a'\b'}$.
One has to be very careful in rising and lowering indices, because there are different metrics in the game.
The indices of the effective metric $G$ will always be raised and lowered with the effective background metric $\obar G^{\mu\nu}$, while 
the indices of $h^{\mu\nu}$ and most other tensorial objects will be raised and lowered with $\eta^{\mu\nu}$.
In case of ambiguity, we will typically spell this out.
With this convention, we can write
the fluctuations of the effective background effective metric \eqref{eff-metric-FRW} as
\begin{align}
 (G_{\mu\nu} - \d G_{\mu\nu})  \dd x^\mu \dd x^\nu 
   &= -\dd t^2 + a^2(t)\dd\Sigma^2 
    - (h_{\mu\nu} -\frac{1}{2} \eta_{\mu\nu}\,h) \dd x^\mu \dd x^\nu \ . 
    \label{perturbed-metric-cosm}
\end{align}

\subsection{Linearized Ricci tensor}
\label{sec:lin-curv}

To understand the significance of the metric modes, we consider the
linearized Ricci tensor 
\begin{align}
2\d R_{(\rm lin)}^{\mu\nu}[G] 
 &= -\nabla^\a \nabla_\a \d G^{\mu \nu}  
 + \nabla^{\mu} \nabla_\r \d G^{\nu\r} + \nabla^{\nu} \nabla_\r \d  G^{\mu\r} 
  - \nabla^\mu\nabla^\nu \d G  
 \label{gmunu-lin}
\end{align}
for a metric fluctuation $\d G^{\mu\nu} = \b^2 \tilde h^{\mu\nu}$ with
\begin{align}
  \tilde h^{\mu\nu} &= h^{\mu\nu} - \frac 12 \eta^{\mu\nu} h,
 \qquad \tilde h = - h
\end{align}
around the background 
 $\obar G^{\mu\nu} = \b \eta^{\mu\nu}$.
 For simplicity, we will neglect contributions of the order of the cosmic background
 curvature. Then we can replace $\nabla$ by $\del$ in Cartesian coordinates, and
\begin{align}
2R_{(\rm lin)}^{\mu\nu}[G] \
 &\stackrel{\eta\to\infty}{\approx} \ 
  \b^2\Big(-\del^\a\del_\a  \tilde h^{\mu\nu}
   + \del^\mu\del_\r  \tilde h^{\r\nu}
   + \del^\nu\del_\r  \tilde h^{\r\mu}
   - \del^\mu\del^\nu  \tilde h\Big) \nn\\
  & \ \ =  \quad  \b^2\Big(-\del^\a\del_\a  (h^{\mu\nu} -\frac 12 \eta^{\mu\nu} h)
   + \del^\mu\del_\r  h^{\r\nu} 
   + \del^\nu\del_\r h^{\r\mu} \Big) 
   \label{Ricci-lin-approx-1}
\end{align}
 neglecting the $\del\b$ terms at late times $\eta\to\infty$, because \eqref{x-t-beta-relations}
\begin{align}
 \b^{-1}\del_\mu\b %&= \frac{\b}{x_4^2} \eta_{\mu\nu }x^\nu 
   =  \frac{\b^2}{R} G_{\mu\nu } \frac{x^\nu}{x_4} \ = O(\b^2) .
\end{align}
Now we can use the intertwiner relation (6.25) in \cite{Sperling:2019xar}
\begin{align}
 \Big(\Box + \frac{2}{R^2r^2}\tilde\cI\Big) h^{\mu\nu}[\cA]
  &=  h_{\mu\nu}[\cD^2\cA]   
  +  \frac{2}{R^2}\Big(3 h^{\mu\nu}[\cA] -  \eta^{\mu \nu} h[\cA]\Big) \, 
  \label{Box-h-relation}
\end{align}
and the on-shell relation $(\cD^2-\frac{3}{R^2})\cA = 0$.
We should also drop the contribution from $\tilde\cI$ in the same approximation, because 
 \begin{align}
 \Box \phi &\sim -\sinh^2(\eta)\del^\a\del_\a \phi \ \gg  \ \frac 1{R^2 r^2}\tilde\cI(h^{\mu\nu}) 
  \ \sim \frac{x}{R^2}\del h^{\mu\nu}
 \label{M-estimate}
  \end{align}
 for  $\del \gg \frac 1{x_4}$,
  using  \eqref{deldel-Box-relation} and $\{\theta^{\mu\a},\phi\} = r^2(x^\mu\del^\nu - x^\nu \del^\mu)\phi$ \cite{Sperling:2019xar}.
Therefore \eqref{Box-h-relation} reduces on-shell to
\begin{align}
  \del^\a\del_\a  h^{\mu\nu}
 &\approx - \frac{1}{x_4^2}\left(9 h^{\mu\nu} - 2 \eta^{\mu \nu} h\right) 
  % \  \ll \ \del\del h^{\mu\nu}
%\del^\a\del_\a h[\cA] &\approx - \frac{1}{x_4^2} h[\cA] \, 
\end{align}
which is negligible  at late times compared to the terms involving second derivatives $\del\del h^{\mu\nu}$
 in \refeq{Ricci-lin-approx-1},
and similarly for the trace.
This means that 
the linearized Ricci tensor reduces on-shell to
\begin{align}
 2R_{(\rm lin)}^{\mu\nu}[G^{\a\b}] \
 &=  \ \b^2\Big(  \del^\mu\del_\r  h^{\r\nu} 
   + \del^\nu\del_\r h^{\r\mu} \ +  O(\frac{\del h^{\mu\nu}}{x_4})  \Big)
   \label{Ricci-tensor-onshell}
\end{align}
 on scales much shorter than the cosmic curvature scale, or
for late times i.e. large $\eta$.

\subsection{Pure gauge modes}
Now consider the metric fluctuation corresponding to the pure gauge fields $\cA^{(g)}[\phi]$, where 
$\phi = \phi^{(1)}$  is a  spin 1 field. This has the form (cf. \cite{Sperling:2019xar})
\begin{subequations}
\label{eq:prop_pure_gauge}
\begin{align}
  h^{\mu\nu}_{(g)}[\phi] &\coloneqq
  h^{\mu\nu}[\cA^{(g)}]  
  = -\{t^\mu,\cA^{(-)\nu}[\phi]\} + (\mu \leftrightarrow \nu)  
  + \frac{1}{3}  h^{(g)} \eta^{\mu\nu} \,,
  \label{pure-gauge-metric} \\
h_{(g)}[\phi] &\coloneqq  \eta_{\mu\nu} h_{(g)}^{\mu\nu}[\phi]
  = \frac 6R D^-\phi\, 
 = 6 \{t_\mu,\cA^{(-)\mu}[\phi]\} \, .
 \label{pure-gauge-H}
\end{align}
\end{subequations}
It is not hard to show the following formulas 
\begin{align}
 \{t_\mu,  h^{\mu\nu}_{(g)}[\phi] \} 
 \ &= \ -\{\Box \phi,x^\nu\}_- - \frac 2R\, D^-\{t^\nu,\phi\} \,,
 \label{t-h-g-relation} \\
 x_\nu x_\mu  h^{\mu\nu}_{(g)}[\phi] 
%   &= -2 x_\nu  x_\mu\{x^\mu,\{t^\nu,\phi\}\}   
%   = -2  x_\mu\{x^\mu,x_\nu \{t^\nu,\phi\}\} + 2  x_\mu\theta^{\mu\nu} \{t_\nu,\phi\}  \nn\\
%  &= 2 x^4 D^-(\sinh(\eta)(\t+1) \phi) - 2R \sinh^2(\eta) D^-(\phi)  \nn\\
  \ &= \ 2R \sinh^2(\eta) D^-\t\phi \ 
  \label{xxh-puregauge-id}
\end{align}
using \eqref{D-properties} cf. \cite{Sperling:2019xar},
and in particular
\begin{align}
 \{t_\mu,  h^{\mu\nu}_{(g)}[\phi^{(1,0)}] \} 
 \ &= - \frac 2{R^2}\, \cA^{(-)\nu}[\phi] \qquad \mbox{for}\ \ \Box\phi^{(1,0)} = 0 \ \nn\\
  x_\mu h^{\mu\nu}_{(g)}[\phi^{(1,0)}] 
 % &= - x_\mu \{t^\mu,\cA^{(-)\nu}[\phi]\} - x_\mu \{t^\nu,\cA^{(-)\mu}[\phi]\}  \nn\\
  &= - \sinh(\eta)(\t-1)\cA^{(-)\nu}[\phi] \ .
 \label{div-puregauge-spin1}
\end{align}
Taking into account the  conformal factor  \eqref{eq:def_phys_graviton}, 
the pure gauge contribution to the effective metric is 
\begin{align}
 \d G^{\mu\nu}_{(g)} &= \b^2\big(h^{\mu\nu} -\frac{1}{2} \eta^{\mu\nu}\,h\big) \nn\\
 %  &= \sinh^{-2}\Big(-\{t^\mu,\cA^{\nu(-)}\} + (\mu \leftrightarrow \nu)  
 % + (\frac{1}{3}-\frac{1}{2}) \eta^{\mu\nu}\,h \Big)   \nn\\
   &= \b^2\Big(-\{t^\mu,\cA^{\nu}\}  -\{t^\nu,\cA^{\mu}\}  
  - \eta^{\mu\nu}\,\{t_\a,\cA^\a \} \Big)   \nn\\
   &= -\del^\mu\cA^{\nu}  -\del^\nu\cA^{\mu} - G^{\mu\nu}\,(\del_\a\cA^\a)    
  \label{eff-graviton-gauge}
\end{align}
where   $\cA^{\a} = \cA^{(-)\a}[\phi]$ and $\del^\mu = G^{\mu\nu}\del_\nu$.
This formula is valid in  Cartesian coordinates, and we must be 
very careful with using upper indices, e.g.
 $\{t^\mu,\phi\} = \sinh(\eta)\eta^{\mu\nu}\del_\nu \phi 
  = \sinh^{2}(\eta) G^{\mu\nu}\del_\nu \phi$.

\paragraph{Relation with diffeomorphisms.}

We can rewrite these pure gauge modes as  diffeomorphism modes  by
comparing with \eqref{puregauge-grav-covar}  on the present FLRW background. This gives 
\begin{align}
  \d G^{\mu\nu}_{(g)} 
   %&=  -\del^\mu\cA^{\nu}  -\del^\nu\cA^{\mu} - G^{\mu\nu}\,(\del_\a\cA^\a)  \nn\\
    &= \del^\mu\xi^{\nu} + \del^\nu\xi^{\mu} - \frac{1}{x_4^2} G^{\mu\nu}\, x \cdot \xi \nn\\
   &=  \nabla^\mu\xi^\nu + \nabla^\nu \xi^\mu, \qquad \xi^\mu = - \cA^\mu 
 \label{puregauge-diffeo-rel}
\end{align}
using
\begin{align}
 x_\a \cA^\a &= \eta_{\a\b} x^\a\{x^\b,\phi\}_- = -x^4 D^- \phi  \nn\\
 \sinh \del_\a \cA^\a &= \{t_\a,\{x^\a,\phi\}_-\} = \frac{1}{R} D^- \phi  \nn\\
 \del_\a \cA^\a &= - \frac{1}{x_4^2} x \cdot \cA  \ 
\end{align}
where $\cA^{\a} = \cA^{\a(-)}[\phi]$,  using the notation $x\cdot\cA \equiv \eta_{\a\b}x^\a\cA^\b$. 
Hence the pure gauge metric modes in the present framework can be identified with 
diffeomorphisms generated by $\xi = -\cA$. This also provides a  non-trivial consistency check for the 
correct identification of $G$. 
It is easy to check using \eqref{div-A-full} that these diffeomorphisms  satisfy the constraint
\begin{align}
  \nabla_\a \xi^\a = -\frac{3}{x_4^2}\, x \cdot \xi   
 \label{gauge-diffeo-constraint-1}
\end{align}
%% fixed V3
or equivalently 
\begin{align}
 \boxed{ \ \nabla_\a (\b^{3} \xi^\a) = 0 \ . \ } 
 \label{gauge-diffeo-constraint}
\end{align}
%% fixed V3
Hence they are essentially volume-preserving diffeos up to the factor $\b^{3}$,
leaving only 3 rather than 4 diffeomorphism d.o.f., unlike in GR.
This reflects the presence of a dynamical scalar metric degree of freedom, which we will study in detail
below.

\subsection{Generalities for the $\cA^{(-)}$ metric modes}
Among the $\cA^{(-)}[\phi^{(s)}]$ modes, only  the ones with spin $s=2$ can 
contribute to the metric, and these are in fact physical degrees of freedom as shown in 
\eqref{on-shell-B2pm}. The corresponding linearized metric fluctuation is \cite{Sperling:2019xar}
\begin{subequations}
\label{eq:prop_A2-mode}
\begin{align}
  h_{(-)}^{\mu\nu}[\phi] &\coloneqq
  h^{\mu\nu}[\cA^{(-)}[\phi]] 
 % = - \{x^\mu,\{x^\nu,\phi^{(2)}\}_-\}_- +  (\mu\leftrightarrow\nu)  \, , \\
 = -2 \{x^\mu,\{x^\nu,\phi\}_-\}_- = -2 \{x^\nu,\{x^\mu,\phi\}_-\}_- \nn\\
h_{(-)}[\phi] &\coloneqq
 \eta_{\mu\nu}  h_{(-)}^{\mu\nu}
  = -2\{x^\mu,\{x_\mu,\phi\}_-\}_-  \
  =   2 D_- D_-\phi  \, 
    \label{Hmunu-tr}  
% % 
\end{align}
\end{subequations}
for $\phi = \phi^{(2)}$.
It is not hard to derive the following formulas 
\begin{subequations}
\label{eq:prop_A2-prop}
\begin{align}
\{t_\mu, h_{(-)}^{\mu\nu}\} 
  &= -\frac{2}{R} \{x^\nu, D^-\phi\}_- 
  \label{Hmunu-div}  \\
\{t_\mu,\{t^\a, h^{(-)}_{\a\nu}\}\}  
 + (\mu\leftrightarrow \nu)
   &=\frac{2}{R^2} \Big(h^{(g)}_{\mu\nu}  - \frac 13\eta_{\mu\nu}h^{(g)} \Big)[D^-\phi] \\
x_\mu h^{\mu\nu}_{(-)}
% &=  - 2 x_\mu\{x^\mu,\{x^\nu,\phi^{(2)}\}_-\}_-  =  2 x_4 D^-(\{x^\nu,\phi^{(2)}\}_-) \nn\\
&=  2 x_4 \{x^\nu,D^-\phi\}_-
 \label{x-h-contract-id}
\end{align}
\end{subequations}
since $\{t^\nu,\phi^{(2)}\}_0 = 0$. Comparing \eqref{x-h-contract-id} and  \eqref{Hmunu-div}, we obtain
\begin{align}
% \{t_\mu,h^{\mu\nu}_{(-)}\} &= -\frac{2}{R} \{x^\nu,D^-\phi\}_-
%  = - \frac 1{R^2}\sinh^{-1}(\eta) x_\mu h^{\mu\nu} \nn\\
  \del_\mu h^{\mu\nu}_{(-)} &= - \frac 1{x_4^2} x_\mu h^{\mu\nu}_{(-)}
  \label{del-x-hminus-rel}
\end{align}
or equivalently 
\begin{align}
 \boxed{ \
 \del_\mu(\b h^{\mu\nu}_{(-)}) = 0 \ .
  \ }
  \label{A-h-nice-id}
\end{align}
This looks like a gauge-fixing condition. We can write it in covariant form 
using the explicit form of the Christoffel symbols \eqref{christoffels}, \eqref{christoffels-2}, which  gives 
\begin{align}
 \nabla_\mu h^{\mu\nu} &= \del_\mu h^{\mu\nu} - \frac 3{x_4^2} x_\mu h^{\mu\nu} + \frac{1}{2x_4^2}x^\nu h \ .
 \end{align}
 %%fixed V3
Since the $\cA^{(-)}[\phi^{(2,0)}]$ and the $\cA^{(-)}[\phi^{(2,1)}]$ modes  satisfy $h=0$, this can be written
using \eqref{del-x-hminus-rel} as
\begin{align}
 \boxed{
 \nabla_\mu (\b^{4} h^{\mu\nu}_{(-)}[\phi^{(2,j)}]) = 0 \qquad\mbox{for} \ \ j=0,1
 \label{h-covar-cons-21}
 }
\end{align}
Since this condition \eqref{del-x-hminus-rel} is not quite the same as
\eqref{div-puregauge-spin1} for the on-shell pure gauge gravitons, 
it follows that the extra 2 on-shell metric fluctuations 
$h^{\mu\nu}[\cA^{(-)}[\phi^{(2,1)}]]$ are in fact physical.

\paragraph{Linearized Ricci tensor.}

Using the constraint \eqref{A-h-nice-id} for 
$h^{\mu\nu}[\cA^{(-)}[\phi^{(2)}]]$, it follows from \eqref{Ricci-tensor-onshell} that all these 
on-shell (would-be massive) spin 2 modes are Ricci-flat up to cosmic scales,
\begin{align}
 2R_{(\rm lin)}^{\mu\nu} \
 &= \  0 +  \  O(\frac{\del G^{\mu\nu}}{x_4}) \ .
   \label{Ricci-tensor-onshell-massive}
\end{align}
This seems to suggest that these modes are exactly massless with only 
2 physical degrees of freedom, but this is not true,
as pointed out above. The point is that the $h^{\mu\nu}$ contributions from the 
would-be helicity 1 and 0 modes are typically dominated by diffeos,
which are trivially flat. However, we will see in the next section that the 
linearized Schwarzschild solution which arises from $h^{\mu\nu}[\cA^{(-)}[D^+D\phi]]$
 is {\em not} dominated by diffeos, but a genuine
non-trivial  Ricci-flat metric.

\subsection{Scalar modes $\cA^{(-)}[D^+D\phi]$ and the 
Schwarzschild metric}
\label{sec:A-DD-metric}

Now we work out the explicit metric perturbation arising from the 
on-shell $\cA^{(-)}[D^+D\phi]$ mode, which is
part of the would-be massless spin 2 multiplet $\cA^{(-)}[\phi^{(2)}]$.
We will see that this includes  a quasi-static
Schwarzschild metric, as well as other solutions which might be related to 
dark matter.
% The on-shell condition is $(\Box-\frac{4}{R^2})D^+D^+ \phi = 0$, which is 
% equivalent to 
% \begin{align}
% 0 = (\Box-\frac{4}{R^2})D^+D^+ \phi = D^+(\Box +\frac{4}{R^2} -\frac{4}{R^2})D^+ \phi
%  = D^+D^+ (\Box +\frac{2}{R^2})\phi
% \end{align}
We will use the on-shell condition 
$\Box \phi = - \frac{2}{R^2}\phi$ \eqref{onshell-0} throughout,
and focus on the late-time limit $\eta\to\infty$.
Starting with the explicit form \eqref{A-DDphi-explicit} for $\cA^{(-)}[D^+D^+\phi]$,
dropping the pure gauge contribution $\{t^\mu,\L\}$ 
and using the results of section \ref{sec:metric-contrib}, we obtain 
\begin{align}
 \frac{5}{2r^2} h^{\mu\nu}[\cA^{(2)}[D^+D\phi]]
  \ &= \  h^{\mu\nu}\big[ \big(r^2 \b t^\mu 
   +  r^2  \b x^\mu t^\a  \del_\a 
   - \frac 1{3} \theta^{\m\g}\del_\g (\t+4) \big)(\t+2) \phi\big] \nn\\
 %
%  &= r^2\Big(  \frac 23(2 + \t -\b^2)\eta^{\mu\nu}- \frac 23 \frac{\b^2}{R^2} x^\mu x^\nu
%    - \frac 13 (x^\mu \del^\nu +  x^\nu \del^\mu)
%   \Big)(\t+2) \phi  \nn\\
%   &\quad  +r^2\Big(- \frac 43\frac{\b^2}{R^2} x^\mu x^\nu  (1 + \t) 
%   + \frac{1}{3}(1 +\t-\b^2) (x^\nu\del^\mu  + x^\mu\del^\nu )\Big)(\t+2) \phi\nn\\
%   &\quad -\frac {r^2}{3}  \Big( 
%   - \frac{2(\t + 2)(\t+\b^2)}{3}\eta^{\mu\nu}
%   - \frac{2}{3R^2} \b^2 x^\mu x^\nu(2+\t) \nn\\
%  &\quad  + \frac{2\t + 1}3  (x^\nu  \del^\mu  +  x^\mu \del^\nu)  \
%   +\frac{2R^2}{3} \del^\nu  \del^\mu 
%   \Big)(\t+4)(\t+2) \phi  \nn\\
%   %%%
%    &= \frac 23\eta^{\mu\nu} r^2\Big( (2 + \t-\b^2) 
%      + \frac 13 (\t+4)(\t+2)(\t+\b^2) \Big)(\t+2) \phi  \nn\\
%    &\quad + \frac 13 r^2 x^\mu x^\nu \frac{\b^2}{R^2}
%    \Big(- 2 - 4 (1 + \t) + \frac 23(\t+4)(\t+2)\Big)(\t+2)\phi \nn\\
%    &\quad + \frac 13 r^2\Big(- 1  + (1 +\t-\b^2)  - \frac 13(\t+4)(2\t+1) \Big) 
%    (x^\nu\del^\mu + x^\mu\del^\nu)(\t+2)\phi \nn\\
%    &\quad  -\frac 29 R^2r^2\del^\nu  \del^\mu (\t+4)(\t+2)\phi \nn\\
%   %%%
 \!\! &\stackrel{\eta\to\infty}{=} \frac 29 r^2 \Big(\eta^{\mu\nu}(2 + \t)(3 + \t^2+4\t)
   +  \frac{\b^2}{R^2} x^\mu x^\nu (\t^2 -1)\nn\\
   &\qquad   - (x^\nu\del^\mu + x^\mu\del^\nu)(\t^2+3\t+2) 
    -  R^2\del^\nu  \del^\mu (\t+4)\Big)(\t+2)\phi \ .
\end{align}
Therefore
\begin{align}
 h^{\mu\nu} 
%  &= \frac{4r^4}{45}  
%  \Big((2 + \t)(\t+1)(\t + 3) \eta^{\mu\nu}
%     + \frac{\b^2}{R^2} x^\mu x^\nu (\t^2 -1) \nn\\
%    &\quad - (\t+1)(\t+2) (x^\nu\del^\mu + x^\mu\del^\nu)
%     -  R^2\del^\nu  \del^\mu (\t+4) \Big) (2 + \t)\phi \nn\\
   &= \frac{4r^4}{45} 
 \Big((2 + \t)(\t + 3)\eta^{\mu\nu}  
    + \frac{\b^2}{R^2} x^\mu x^\nu (\t -1)
    - (x^\nu\del^\mu + x^\mu\del^\nu) (\t+2) \Big)(\t+1)(\t+2)\phi\nn\\
   &\quad  -\frac{4r^4}{45}   R^2\del^\nu  \del^\mu  (\t+4)(\t+2)\phi  
\end{align}
%noting that $(3 + \t^2+4\t) = (\t+1)(\t + 3)$.
with trace
\begin{align}
  h 
%   &= \frac{4r^4}{45} (\t+2)(\t+1)
%  \Big(4(2 + \t)(\t + 3) - (\t -1) - 2(\t+2)\t \Big)\phi\nn\\
%    &\quad  -\frac{4r^4}{45} \b^2 (\t+2) (\t+4)(2+\t) \phi  \nn\\
%   &\stackrel{\eta\to\infty}{=}
%    \frac{4r^4}{45} (\t+2)(\t+1)
%  \Big(4(2 + \t)(\t + 3) - (\t -1) - 2(\t+2)\t \Big)\phi \nn\\
 &\stackrel{\eta\to\infty}{=} \frac{4r^4}{45} (\t+1) (2\t+5)(\t+5)(\t+2)\phi \ .
 \end{align} 
%  
% As a further check, we also compute the time-component
% \begin{align}
%  x_\mu x_\nu  h^{\mu\nu} &= 
%   \frac{4r^4}{45} (\t+2)(\t+1)
%  \Big(-R^2(2 + \t)(\t + 3)\cosh^2(\eta)
%     + (\t -1) \cosh^2(\eta) R^2 
%     +2 R^2  \cosh^2(\eta) (\t+2) \t \Big)\phi\nn\\
%    &\quad  -\frac{4r^4}{45}  (\t+2) (\t+4) R^2 \t (\t-1) \phi   \nn\\
%   &\stackrel{\eta\to\infty}{=}
%     \frac{4r^4R^2}{45} (\t+2)(\t+1)\cosh^2(\eta)
%  \Big(-(\t^2+5\t+6) + (\t -1) +2 (\t^2+2\t) \Big)\phi\nn\\
%   &=
%    \frac{4r^4R^2}{45} (\t+2)(\t+1) (\t^2  -7)\cosh^2(\eta)
% \phi
% \end{align}
Then the trace-reversed metric fluctuation $\tilde h^{\mu\nu}$ is 
\begin{align}
 \tilde h^{\mu\nu} &= h^{\mu\nu} - \frac 12 h \eta^{\mu\nu}  \nn\\
%   &= \frac{4r^4}{45} (\t+2)(\t+1)
%    \Big(\big((2 + \t)(\t + 3) - \frac 12 (2\t+5)(\t+5)  \big) \eta^{\mu\nu}  \nn\\
%    &\quad 
%     + \frac{\b^2}{R^2} x^\mu x^\nu (\t -1)
%     - (x^\nu\del^\mu + x^\mu\del^\nu)(\t+2)  \Big)\phi
%     -\frac{4r^4}{45}  R^2\del^\nu  \del^\mu(\t+4) (\t+2)  \phi  \nn\\
  &= \frac{4r^4}{45}
   \Big(-\frac 12 (5\t + 13) \eta^{\mu\nu} 
    + \frac{\b^2}{R^2} x^\mu x^\nu (\t -1)
    - (x^\nu\del^\mu + x^\mu\del^\nu) (\t+2) \Big) (\t+1)(\t+2)\phi \nn\\
 &\quad   -\frac{4r^4}{45}  R^2\del^\nu  \del^\mu   (\t+4)(\t+2)\phi \ .
    \label{tilde-h-nogauge}
\end{align}
Observe that for $\t\neq -2$  the  term $(x^\nu\del^\mu + x^\mu\del^\nu)\phi$ is dominant  at late times, since $x^0\sim R\cosh(\eta)$.
However this is essentially a large diffeomorphism contribution, which
can be removed from the effective metric fluctuation 
using \eqref{FRW-puregauge}, with the result
\begin{align}
 \tilde h^{\mu\nu} 
%   &\sim \frac{4r^4}{45} 
%    \Big(-\frac 12 (5\t + 13) \eta^{\mu\nu} 
%     + \frac{\b^2}{R^2} x^\mu x^\nu (\t -1)
%     +  \big(3\eta^{\mu\nu} + 2x^\nu x^\mu \frac{\b^2}{R^2}\big) (\t+2)\Big)(\t+1)(\t+2)\phi \nn\\
%    &\quad  -\frac{4r^4}{45}   R^2\b^2\del^\nu  \del^\mu (\t+4)(\t+2)\phi \nn\\
 %%
 &\sim \frac{4r^4}{45} 
   \Big( \frac 12(\t - 1)\eta^{\mu\nu}  
    + 3\frac{\b^2}{R^2} x^\mu x^\nu (\t +1)\Big)(\t+1)(\t+2)\phi 
    \label{eff-metric-after-diffeo}
\end{align}
for large $\eta$. Hence 
\begin{align}
 \tilde h_{\mu\nu}\, \dd x^\mu \dd x^\nu 
 \ \ &=  \ \frac{2r^4R^2}{45} 
   \sinh^2(\eta) \big(d\eta^2(5\t +7) + d\Sigma^2(\t - 1)\big)(\t+1)(\t+2)\phi \nn\\
 &\stackrel{\t\to-2}{=} \ \frac{2r^4R^2}{15} 
    \sinh^2(\eta)(\t+2)\phi\, \big(d\eta^2 + d\Sigma^2\big) \nn\\
 &= - 4 \phi' (d t^2 + a(t)^2 d\Sigma^2)
\end{align}
using \eqref{metric-sphericalcoords-trafo}
where $\tilde h_{\mu\nu}  = \eta_{\mu\mu'}\eta_{\nu\nu'} \tilde h^{\mu'\nu'}$,
and using the explicit form \eqref{eff-metric-FRW} of the scale parameter $a(t)$ for large $\eta$.
Here we define
\begin{align}
 \phi' := -\frac{r^4}{30} \b (\t+2)\phi \ 
 \label{phi-prime-new}
\end{align}
as in \eqref{phi-prime} (up to rescaling), which allows to take $\t\to -2$.
We will see that
this reduces to the linearized Schwarzschild metric for $\t\to -2$, while for $\t\neq -2$ it 
is a distinct metric which is not Ricci-flat.
However for $\t \neq-2$  the diffeo contribution in \eqref{tilde-h-nogauge}
grows very large at late times, which may invalidate the 
linearized approximation as discussed below. 
Therefore we focus on $\t\approx -2$, which is the most interesting and most reliable case.
Then the full perturbed metric can be written 
in the form \eqref{perturbed-metric-cosm}
\begin{align}
 \label{perturbed-metric-cosm-expl}  
\boxed{\
\begin{aligned}
 ds^2 = (G_{\mu\nu} - \d G_{\mu\nu})  \dd x^\mu \dd x^\nu 
  &= (\sinh(\eta)\eta_{\mu\nu} 
    - \tilde h_{\mu\nu})\, \dd x^\mu \dd x^\nu \\
%    &= -\dd t^2 + a(t)^2\dd\Sigma^2 
%     + 4\phi'R^2\sinh^2(\eta)(d\eta^2 + d\Sigma^2) \nn\\
  &= - \dd t^2 + a(t)^2\dd\Sigma^2 
      \ + 4 \phi' (d t^2 + a(t)^2 d\Sigma^2) \ .
 \end{aligned}
 \ }
\end{align}
The on-shell condition reduces to
$\Delta^{(3)}\phi = 0$ for $\t=-2$ due to \eqref{Box-Laplace-tau}, 
and in the spherically symmetric case the Newton potential 
on a $k=-1$ geometry is recovered  \eqref{harmonic-soln}, with
\begin{align}
 \phi \ &= \  \frac{e^{- \chi}}{\sinh(\chi)} \frac{1}{\cosh^2(\eta)}
 \ \sim \ \frac{1}{\r} e^{-\chi -2\eta} \ , \qquad \r = \sinh(\chi) \ .
 \label{phi-tis-2}
\end{align}
Strictly speaking we should  use $\phi'$ rather than $\phi$ in the $(\t+2)\phi = 0$ case. Then
the quasi-static condition becomes $(\t+3+\b^2)\phi' = 0$, and the on-shell condition \eqref{eom-phi'} is
$(\Delta^{(3)} - 4\b^4)\phi' = 0$. However
the $\b^2$ contributions can be dropped in the large $\eta$ limit giving again $\Delta^{(3)}\phi' = 0$, so that 
\begin{align}
  \phi' &\sim  \frac{1}{\r} e^{-\chi-3\eta} \ \sim  \frac{e^{-\chi}}{\r} \frac{1}{a(t)^2}
  \label{phiprime-solution}
\end{align}
 for large $\eta$, 
using \eqref{tau-eta} and recalling $a(t) \sim e^{-\frac 32 \eta}$ \eqref{a-eta}.
This metric is very close to the Vittie solution \cite{mcvittie1933mass} for the Schwarzschild metric 
for a point mass $M$ in a 
FRW spacetime, whose linearization for $k=-1$ is given by
\begin{align}
 ds^2 
   %&= -\Big(\frac{1-\mu}{1+\mu}\Big)^2 dt^2 + (1+\mu)^4 a(t)^2 d\Sigma^2  \nn\\
   &= -dt^2 + a(t)^2 d\Sigma^2  \ + \  4\mu(dt^2 + a(t)^2 d\Sigma^2)  \ + O(\mu^2) \ .
   \label{vittie}
 \end{align}  
Here 
\begin{align}  
  \mu = \mu(t,\chi) &= \frac{M}{2 \r} \frac{1}{a(t)} 
  \label{mu-vittie}
\end{align}
is the mass parameter, which is not constant but decays during the cosmic expansion; 
this is as it should be, because local gravitational systems do not participate in the 
expansion of the universe.
Comparing with  \eqref{phiprime-solution} we have
\begin{align}
  \phi' &\sim \  \mu(t,\chi)\, \frac{e^{-\chi}}{a(t)}  \ .
\end{align}
Since $\mu$ \eqref{mu-vittie}
looks like  a constant mass for a comoving observer  \cite{mcvittie1933mass}, 
the effective mass parameter in our solution effectively decreases like 
$a(t)^{-1}$ during the cosmic evolution.
This might be interpreted in terms of a time-dependent
Newton constant, although this is a bit premature since
we have not properly investigated the 
coupling to matter, and quantum effects may modify the result.
Nevertheless, the result is suggestive.
Also, while both metrics have the characteristic $\frac 1\r$ dependence of the 
Newton potential, our solution has an extra $e^{-\chi}$ factor, 
which reduces its range at space-like curvature scales.
Both effects are irrelevant at solar system scales, 
but they will be important for cosmological considerations,  reducing the gravitational attraction at long scales.

For completeness, we also recall the 
 linearized Schwarzschild solution in isotropic coordinates 
\begin{align}
 ds^2 &= -\frac{(1- \frac M {2r})^2}{(1+\frac M {2r})^2} \ d t^2
 + (1+ \frac M {2r})^4 (dx^2+dx^2+dz^2)  \nn\\
  &= \eta_{\mu\nu} dx^\mu dx^\nu + \frac {2M} {r} d x_0^2
  + \frac {2M} {r} (dx^2+dx^2+dz^2) \ + O(\frac{1}{r^2}) \ .
\end{align}
\eqref{vittie} reduces to this metric for a local comoving observer 
for large $a(t)$, 
while we obtained an extra factor $\frac 1{a(t)}$ in the effective mass.

Let us discuss the consistency and significance of these results.
The most striking point is that even though the metric \eqref{eff-metric-after-diffeo}
is Ricci-flat for $\t=-2$, for other values of $\t$ it is not. 
This seems to contradict the 
general result \eqref{Ricci-tensor-onshell-massive} for the linearized Ricci tensor, which should always vanish at scales shorter than the background curvature i.e. for large $\eta$. 
The resolution of this puzzle lies in the  diffeo contributions 
$(x\del+x\del)$ in \eqref{tilde-h-nogauge}, which were eliminated by a change of 
coordinates in \eqref{eff-metric-after-diffeo}. 
The point is that  for $\t\neq -2$, this term becomes very large for large $\eta$
as $x^0\sim R\cosh(\eta)$, and  completely dominates 
the other, non-trivial contributions to $\tilde h^{\mu\nu}$.
But the Ricci-tensor for a diffeo contribution vanishes trivially,
leading to \eqref{Ricci-tensor-onshell-massive}.
In other words, if the first terms in \eqref{tilde-h-nogauge} are non-trivial, they are 
dominated by the $(x\del+x\del)$ term, so that for large $\eta$
the linearized approximation becomes invalid\footnote{Recall that \eqref{A-DDphi-explicit} 
also contains large pure gauge contributions $\{t^\mu,\L\}$ which were 
discarded. This was justified, because these are exact gauge symmetries of the 
matrix model. In contrast, the above $(x\del+x\del)$ terms are diffeos
which are presumably not part of the 3 diffeo modes in the present 
framework, due to the constraint \eqref{gauge-diffeo-constraint}. Therefore this is an approximation 
whose validity needs to be checked carefully.}, unless $\t\approx -2$. 
Then the effective metric fluctuation 
\eqref{gravitons-H1} must be completed by the non-linear contribution, 
which will be discussed briefly below.

In contrast for $\t\approx -2$,
the pure gauge contribution in \eqref{tilde-h-nogauge}  vanishes,  
hence our Schwarzschild-like solution is fully justified.

A similar issue may arise for the would-be helicity 1 modes 
$\cA^{(-)}[\phi^{(2,1)}]$, but not for 
the helicity-2 gravitons $\cA^{(-)}[\phi^{(2,0)}]$, because there are no 
helicity 2 pure gauge contributions. Therefore these are indeed
Ricci-flat and non-trivial, as stated in \cite{Sperling:2019xar}.

One might worry that the restriction to $\t=-2$ of the Schwarzschild solution 
is too rigid for real physical systems such as the solar system.
However, systems with non-uniform motion lead to dynamical metric perturbations 
corresponding to physical spin 2 gravitons, which are realized here by the 
$\cA^{(-)}[\phi^{(2,0)}]$ modes. Therefore there should not be an obstacle 
to obtain  dynamical Ricci-flat metric perturbations 
as a combination of $\cA^{(-)}[D^+D\phi^{(0)}]$ and $\cA^{(-)}[\phi^{(2,0)}]$ modes.

\paragraph{Interpretation and physical significance.}

We  found that the scalar on-shell modes provide a Ricci-flat metric perturbation only for the
 specific quasi-static time-dependence $\t\approx -2$. 
Indeed it should be expected that a dynamical scalar metric mode, which does not 
exist in GR, is not Ricci-flat in general.
From a GR point of view, such non-Ricci-flat perturbations would be 
interpreted as dark matter.
Nevertheless, there better be a reason why in typical situations
such as the solar system, such non-Ricci-flat
deformations are suppressed. 
Strictly speaking 
this question can only be settled once the coupling of matter to the 
various modes is properly taken into account.
Quantum effects may also be important here, because they typically 
lead to an  induced Einstein-Hilbert term\footnote{As explained in \cite{Sperling:2019xar}, there is
a priori no matrix analog of the cosmological constant. However, it remains 
to be understood whether or not an analogous  term 
is induced through quantum effects, and what its consequences would be.} 
\cite{Sperling:2019xar},
which would  distinguish  Ricci-flat and -non-flat solutions.

However heuristically, we can give a classical mechanism which achieves that effect at the non-linear level
as follows.
For scalar modes with large diffeo contribution in \eqref{tilde-h-nogauge}, the 
linearized metric \eqref{gravitons-H1} must be replaced by the full non-linear expression \eqref{gamma-nonlinear}.
Then the large would-be diffeo contribution no longer decouples from a conserved $T^{\mu\nu}$, 
but strongly couples to matter\footnote{These are not expected to be exact gauge symmetries of the 
model, since there are only 3 pure gauge d.o.f.}. But if this large contribution governs the dynamics,
the first two terms in \eqref{tilde-h-nogauge} are effectively suppressed, and this suppression 
is stronger for shorter wavelengths due to the derivatives.
On the other hand  $\cA$ becomes small sufficiently far from matter, so that the  linearized 
treatment will suffice. Then the large contribution is indeed a flat diffeo, while 
the sub-leading non-Ricci-flat contribution  in \eqref{eff-metric-after-diffeo} 
is strongly suppressed.
This  does not apply to the Ricci-flat $\t=-2$ contribution since the diffeo vanishes, 
and we conclude that the non-Ricci-flat contributions are strongly suppressed, as desired.

For very long wavelengths, this suppression mechanism becomes weak, 
so that some non-Ricci-flat perturbations with very long - possibly galactic - wavelengths are 
expected. This would then be interpreted as dark matter from a GR point of view.
Moreover, the suppression mechanism is weaker in the earlier Universe, which 
might explain why dark matter seems to be more abundant in older galaxies
\cite{genzel2017strongly}.

This non-linear effect is somewhat reminiscent of the  vDVZ discontinuity in massive gravity
and its resolution through the
Vainshtein mechanism \cite{Vainshtein:1972sx}.
Indeed, the present modes arise precisely from  would-be massive 
spin 2 modes, albeit the details are different.

A time dependence $\sim a(t)^{-1}$ of the Newton constant seems to be somewhat large 
in view of recent estimates \cite{Mould:2014iga,Zhao:2018gwk}.
However, we have not properly taken into account the coupling to matter, and the 
underlying FLRW cosmology is non-standard.
Including an induced Einstein-Hilbert action 
in the quantum effective action could also affect the result.
These issues need to be understood before solid predictions can be made.

Finally, we note that 
the case $\t=-1$ is also special.
After suitable rescaling this leads to $\tilde h^{\mu\nu} \sim \eta^{\mu\nu}$, 
hence to a modification of the cosmological evolution $a(t)$ for 
$\phi \sim e^{-\eta}$. 
A similar modification may arise from $f \neq 0$ in \eqref{gaugefix-t-relation}.
This shows that modifications of the cosmic evolution are possible, 
but again this needs to be studied in more detail.

\paragraph{Pure gauge contribution and checks.} 
 
 An instructive check can be obtained by computing
 the metric fluctuation arising from the pure gauge term \eqref{A-puregauge-L}:
 \begin{align}
  h^{\mu\nu} &= \frac 25  r^2 h^{\mu\nu}[\theta^{\mu\a} \del_\a(\t+3)\phi] \ 
   -\frac 25  r^4 R^2 h^{\mu\nu}[ \sinh(\eta) t^\a \del_\a\del_\mu(\t+3)\phi]  \nn\\
   %%%
%    &=  \frac 2{15} r^4 \Big(- 2(\t + 2)\t\eta^{\mu\nu}
%   - \frac{2}{R^2} \b^2 x^\mu x^\nu(2+\t)
%    + (2\t + 1) (x^\nu  \del^\mu  +  x^\mu \del^\nu)   \
%   + 2R^2 \del^\nu  \del^\mu \Big)(\t+3) \phi  \nn\\
%   &\quad  -\frac 2{15}  r^4 \Big(2 R^2\sinh^2(\eta)\del_\mu\del_\nu(2 +\t)
%   - 3(x^\mu\del_\nu +  x^\nu\del_\mu )
%     - \frac 4{R^2} \b^2 x^\mu x^\mu (2+\t) \Big) (\t+3)\phi  \nn\\
%  %%%
%  &=  \frac 2{15} r^4 \Big(- 2(\t + 2)\t\eta^{\mu\nu}
%   + \frac{2}{R^2} \b^2 x^\mu x^\nu(2+\t)   \
%   + 2R^2 \del^\nu  \del^\mu 
%   -2 R^2\sinh^2(\eta)\del_\mu\del_\nu(2 +\t)\ \nn\\
%   &\quad  
%    + (2\t + 4) (x^\nu  \del^\mu  +  x^\mu \del^\nu) \Big) (\t+3)\phi\nn\\
%    %%
 &\stackrel{\eta\to\infty}{=}
   \frac 4{15} r^4 \Big(-\t\eta^{\mu\nu}
  + \frac{1}{R^2} \b^2 x^\mu x^\nu  \ 
  - R^2\sinh^2(\eta)\del_\mu\del_\nu 
  + (x^\nu  \del^\mu  +  x^\mu \del^\nu) \Big)(\t+3) (\t + 2)\phi  \nn
 \end{align}
 using \eqref{h-tdeldel-phi},
with trace
\begin{align}
 h 
 %&= \frac 4{15} (\t + 2)(\t+3) r^4 \Big(-4\t\phi
 % -\phi    \ 
 % - (2+\t)\phi  
 % + 2\t \phi \Big) \nn\\
 &= -\frac {12}{15} (\t+1)(\t + 2)(\t+3) r^4 \phi
\end{align}
consistent with \eqref{pure-gauge-H}.
%%% keep %%%
% \begin{align}
%  h_{(g)} &= \frac{6}{R}D^-\L = -\frac{2}{5}r^2 R\frac{6}{R}D^-(D(\t+3)\phi  ) \nn\\
%  &= -\frac{12}{5}r^2 (\t+3)D^-D\phi   \nn\\
%  &= \frac{12}{5}\frac{r^2 R^2}3 r^2 (\t+3) \cosh^2(\eta) \Delta^{(3)}\phi \nn\\
%  &\sim -\frac{12}{15}r^4 (\t+3)(\t+1)(\t+2) \phi 
%  \label{trace-puregauge-1}
% \end{align}
%%% keep %%%
Then the trace-reversed  pure gauge metric fluctuation  is 
\begin{align}
 \tilde h^{\mu\nu} &= h^{\mu\nu} - \frac 12 h \eta^{\mu\nu}  \nn\\
%  &= \frac 4{15} r^4(\t + 2)(\t+3)  \Big(-\t\eta^{\mu\nu}\phi
%   + \frac{1}{R^2} \b^2 x^\mu x^\nu\phi    \ 
%   - R^2\sinh^2(\eta)\del_\mu\del_\nu\phi  
%   + (x^\nu  \del^\mu  +  x^\mu \del^\nu) \phi \Big) \nn\\
%  & \quad +\frac{6}{15}r^4 (\t+3)(\t+2)(\t+1)\eta^{\mu\nu} \phi \nn\\
 &= \frac{4r^4}{15}  \Big(
  \frac 12 (\t+3) \eta^{\mu\nu}
   + \frac{1}{R^2} \b^2 x^\mu x^\nu  \ 
  - R^2\sinh^2(\eta)\del_\mu\del_\nu
  + (x^\nu  \del^\mu  +  x^\mu \del^\nu)  \Big)(\t+3) (\t + 2)\phi
  \label{pure-gauge-contrib-DD}
\end{align}  
and one can check using the results in section \ref{sec:diffeo-FRW} that 
this is indeed a diffeomorphism on the FLRW background.

Various other non-trivial checks were performed for the combined and the 
pure gauge contributions, comparing the trace $h$ and the time component
$x_\mu x_\nu h^{\mu\nu}$ with the general formulas 
\eqref{xxh-puregauge-id} and \eqref{Hmunu-tr} and \eqref{h-phi-relation}. All tests 
work out, so that we can be very confident that the above expressions for
the metric fluctuations are correct.

\subsection{Unphysical scalar  $\cA^{(+)}$ modes.}

Among  the  $\cA^{(+)}[\phi^{(s)}]$ modes, only the 
scalar mode $\cA^{(+)}[\phi^{(0)}]$  contributes to the linearized metric.
Even though it is  unphysical  because it does not satisfy the gauge-fixing 
constraint, we give its metric contribution
for completeness:
\begin{subequations}
\label{eq:prop_A2+mode}
\begin{align}
 h_{(+)}^{\mu\nu}[\phi^{(0)}] &= - \{x^\mu,\{x^\nu,\phi^{(0)}\}_+\}_- + 
(\mu\leftrightarrow\nu) \ \nn\\
  &= -2[\theta^{\mu\a}\theta^{\nu\b}]_0\del_\a\del_\b\phi - 
\left(\{x^\mu,\theta^{\nu\b}\}\del_\b +  \{x^\nu,\theta^{\mu\b}\}\del_\b\right) 
\phi^{(0)} \nn\\
  &= -\frac{2r^2R^2}3\left(P^{\mu\nu}P^{\a\b} - 
P^{\mu\b}P^{\nu\a}\right)\del_\a\del_\b\phi^{(0)} 
  - (\{x^\mu,\theta^{\nu\b}\}\del_\b +  \{x^\nu,\theta^{\mu\b}\}\del_\b) 
\phi^{(0)}\nn\\
 &= \frac{2r^2R^2}3\left(\del^\mu \del^\nu 
   - (\eta^{\mu\nu} + R^{-2}x^\mu x^\nu)\b^2(-\Box+\frac{1}{R^2}\t)\right)\phi^{(0)} \nn\\
 &\quad  
  - \frac 23 r^2 \eta^{\mu\nu}(\t  + 2)\t\phi^{(0)}
  + \frac 13 r^2(x^\nu \del^\mu + x^\mu \del^\nu)(1+ 2\t) \phi^{(0)} \, .
% 
% h_{(+)} &= -2\{x_\mu,\{x^\mu,\phi^{(0)}\}\}_- 
%    = -2 \{x_\mu,\{x^\mu,\phi^{(0)}\}\}_0 \,,\\
%  \{t_\mu, h_{(+)}^{\mu\nu}\} 
%  &= \{\{t_\mu,\cA^{(+)\mu}\},x^\nu\}_- - \frac 2R\, D^-\cA^{(+)\nu}  \nn\\
%  &=  \frac{3}R\{D^+\phi^{(0)},x^\nu\}_- - \frac 2R\, D^-\cA^{(+)\nu}  \nn\\
%   &=  -2r^2 \{t^\nu,\phi^{(0)}\} - \frac{5}{R}\{x^\nu,D\phi^{(0)} \}_- 
 \label{th-rel-A2+} 
%  
%\{t_\mu,\{t^\a, h^{(+)}_{\a\nu}\}\}  
%   &= -2r^2\{t_\mu, \{t^\nu,\phi^{(0)}\}\} - 
%\frac{5}{R}\{t_\mu,\{x^\nu,D\phi^{(0)} \}_-\}  \nn\\
%%
\end{align}
\end{subequations}
This is part of the $D\tilde\cA^{(\t)}$ \eqref{D-tilde-A-t-1} mode.
As a check, we recover \eqref{dAlembertian-x} by taking the trace.

To summarize, the $\cA^{(-)}[\phi^{(2)}],\cA^{(g)}[\phi^{(1)}], \cA^{(+)}[\phi^{(0)}]$ and $D\tilde\cA_\mu^{(\t)}[\phi]$
modes provide all $5+3+1+1 = 10$ off-shell d.o.f. of the most general metric fluctuation.
They lead to 5  physical on-shell modes
comprising $2$ graviton modes from $\cA^{(-)}[\phi^{(2,0)}]$, 
one scalar mode $\cA^{(-)}[D^+D\phi^{(0)}]$, and 
presumably 2 helicity 1 modes $\cA^{(-)}[\phi^{(2,1)}]$.

% As a check, consider $\phi = x^\a$: then 
% \begin{align}
%  h_{(+)}^{\mu\nu}[x^\a] &= - \{x^\mu,\{x^\nu,x^\a\}_+\}_- + (\mu \leftrightarrow \nu) \nn\\
%   &= -\{x^\mu,\theta^{\nu\a}\} + (\mu \leftrightarrow \nu) \nn\\
%   &= -r^2 \{\cM^{\nu\a},x^\mu\} + (\mu \leftrightarrow \nu) \nn\\
%   &= -r^2 (\eta^{\nu\mu}x^\a - \eta^{\a\mu} x^\nu) + (\mu \leftrightarrow \nu) \nn\\
%   &= -r^2 (2\eta^{\nu\mu}x^\a - \eta^{\a\mu} x^\nu - \eta^{\a\nu} x^\mu ) \nn\\
%   &\stackrel{!?}{=} \frac{2r^2R^2}3\left(\del^\mu \del^\nu 
%    - (\eta^{\mu\nu} + R^{-2}x^\mu x^\nu)\del^\b\del_\b \right)x^\a \nn\\
%  &\quad  - 2 r^2 \eta^{\mu\nu} x^\a
%   +  r^2(x^\nu \del^\mu + x^\mu \del^\nu) x^\a \,  \nn\\
%   &=  - 2 r^2 \eta^{\mu\nu} x^\a
%   +  r^2(x^\nu \eta^{\mu\a} + x^\mu \eta^{\nu\a}) \,  
% %
% \end{align}
% \begin{align}
%  - 2\{x^\mu,\{x^\mu,\phi\}_+\}_-
%   &= - \frac{2r^2R^2}3(3 -\cosh^2(\eta))\b^2(-\Box+\frac{1}{R^2}\t)\phi 
%  - \frac 23 r^2 (2\t + 7)\t \phi\, 
%  \label{xx-bracket-id}
% \end{align}
% for $\phi\in\cC^0$

% 

\section{Summary and conclusions}

We have studied in detail the scalar
fluctuations of the FLRW quantum space-time solution $\cM^{3,1}$ 
of Yang-Mills matrix models, based on the general results in \cite{Sperling:2019xar}.
In particular, we recovered the quasi-static linearized Schwarzschild metric
as a solution, which arises from 
the scalar  sector of the physical would-be massive spin 2 modes.
Quasi-static indicates that the corresponding effective mass  is found to decrease slowly 
during the cosmic evolution.

It is very remarkable 
that the linearized Schwarzschild solution can be obtained within the framework of
Yang-Mills matrix models, as we have shown.
Along with the propagating spin 2 graviton modes found in \cite{Sperling:2019xar},
this strongly supports the claim that 3+1-dimensional gravity 
can emerge from  the matrix model framework without compactification, 
in particular for the IKKT or IIB model \cite{Ishibashi:1996xs}.
The mechanism is very simple  in the spirit of 
noncommutative but almost-local field theory, 
by considering fluctuations around a background solution.

The present result is tied to  the specific structure of the background 
solution, which is  a twisted $S^2$ bundle over space-time, leading to 
a tower of higher-spin modes. 
It does not seem to work e.g. on simpler Moyal-Weyl type backgrounds, where
the linearized modes only lead to restricted metric fluctuations,
which includes some Ricci-flat metrics  \cite{Rivelles:2002ez,Yang:2006dk,Steinacker:2010rh} but not enough.

An important  issue is (local) Lorentz invariance, which is only partially manifest in the present framework.
This leads to a  different organization of modes in terms of the 
space-like $SO(3,1)$ isometry group. For example, the 
5 modes of a generic spin 2 irrep decompose into $2+2+1$ modes 
of $\phi^{(2,i)}$ as in \eqref{C-sk-def}. This is best understood in space-like gauge,
somewhat reminiscent of helicity modes.
This structure is indicated by the name ``would-be massive'' modes. 
Nevertheless, Lorentz-invariance appears to be largely
respected, presumably due to the large underlying gauge invariance. 
In particular, the propagation of all physical modes is governed by the same effective metric.

Aside from the higher spin modes, the present model includes extra on-shell metric modes 
beyond those of GR. This is not surprising, since the gauge invariance of the metric sector 
is reduced to 3 rather than 4 diffeomorphism d.o.f.
We studied in some detail the extra scalar modes, which arise from the 
would-be helicity zero sector of $\cA^{(-)}[\phi^{(2)}]$. 
Those are in general not Ricci-flat, but their proper
treatment is quite subtle and require non-linear considerations,
except (!) for the quasi-static Schwarzschild case.
We propose a heuristic argument why the non-Ricci-flat modes should be 
suppressed at the non-linear level, 
somewhat reminiscent of the Vainshtein mechanism \cite{Vainshtein:1972sx}.
They may however play a role at very long wavelengths, in the guise of dark matter.
Similarly, there are presumably two more physical modes arising from the would-be helicity 
1 gravitons, which are not studied here, and may also require 
the non-linear theory.

This leads us to the list of open issues and questions which need to be addressed
in future work. 
One important step is the inclusion of matter, in order to clarify how matter acts as a source of metric deformations. 
This was briefly discussed in \cite{Sperling:2019xar}, 
but it needs to be studied in detail, and at the non-linear level in order to clarify the above mechanism.
Only then a reliable assessment can be made whether a satisfactory behavior arises at the classical 
level, or if quantum effects such as an induced Einstein-Hilbert action are essential.

Another obvious tasks is to extend the present Schwarzschild solution,
and more generally the full  higher spin theory,
to the non-linear regime as far as possible. Even though some computations in the 
present paper are quite involved, the basic structure of the underlying
$\cA^{(-)}[D^+D\phi^{(0)}]$ solution is very simple and based only on 
Lie-algebraic structures. This -- along with black hole 
solutions in higher spin theories \cite{Didenko:2009td,Iazeolla:2011cb,Iazeolla:2017vng} -- leads to the hope that an exact
analytic solution can be found, not only at the semi-classical level, 
but also at the fully non-commutative level.
These are only some of many open questions which can  be studied using the tools provided here and in \cite{Sperling:2019xar}.

\paragraph{Acknowledgements.} 
 
 I would like to thank Marcus Sperling for related collaboration, and 
 T. Damour, Carlo Iazeolla, H. Kawai,
 Jan Rosseel and A. Tsuchiya for valuable discussions,
 notably during workshops at the IHES in Bures-sur-Yvette and 
 the ESI Vienna. The hospitality and support of these institutes is gratefully acknowledged.
 This work was supported by the Austrian Science Fund (FWF) grant P32086. 
 The work also profited from the COST network QSPACE through various meetings.

\section{Appendix}

\subsection{Ladder operators and eigenmodes.}
\label{sec:ladder-ops}

We provide a simpler and more conceptual derivation of the 
eigenmodes $\cA^{(\pm)}$ \eqref{D2-A2p-eigenvalues}, \eqref{D2-A2m-eigenvalues} found in \cite{Sperling:2019xar}.
Starting from the observation  
$[\Theta^{\mu\nu},X^4]  \sim i\{\theta^{\mu\nu},x^4\} = 0$ we  obtain
% \begin{align}
%  \tilde\cI(D(\cA_\mu)\} = D( \tilde\cI(\cA_\mu)) ,
% \end{align}
% which implies 
\begin{align}
 \tilde\cI(D^\pm(\cA_\mu)\} = D^\pm( \tilde\cI(\cA_\mu) \ .
\end{align}
Together with the relations \cite{Sperling:2019xar}
\begin{align}
  \Box D^+\phi^{(s)}   &= D^+\left(\Box+\frac{2s+2}{R^2}\right)\phi^{(s)}  \,,\\
 \Box D^-\phi^{(s)}   &= D^-\left(\Box-\frac{2s}{R^2}\right)\phi^{(s)}  \,
 \label{Box-x4-relations}
\end{align}
we obtain
\begin{align}
\cD^2 D^+\cA^{(s)}
 &= (\Box + \frac{2}{r^2R^2}\tilde\cI)D^+\cA^{(s)}
 = (D^+(\Box + \frac{2s+2}{R^2}) + \frac{2}{r^2R^2}D^+\tilde\cI)\cA^{(s)} \nn\\
  &= D^+(\cD^2+\frac{2s+2}{R^2}) \cA^{(s)}, \qquad  \cA^{(s)} \in \cC^s
  \label{D2-D+-relation}
\end{align}
and similarly for $D^-$.
Therefore $D^\pm$ are intertwiners for $\cD^2$ which rise or lower the eigenvalues.
Now observe
\begin{align}
 \cD^2 D^+\cA^{(g)}[\phi^{(s)}] &= D^+(\cD^2+\frac{2s+2}{R^2}) \cA^{(g)}[\phi^{(s)}] = D^+\cA^{(g)}[(\Box +\frac{2s+5}{R^2})\phi^{(s)}] \nn\\
 \cD^2 \cA^{(g)}[D^+\phi^{(s)}] &= \cA^{(g)}[ (\Box +\frac{3}{R^2})D^+\phi^{(s)}] = \cA^{(g)}[D^+ (\Box +\frac{2s+5}{R^2})\phi^{(s)}] \ .
\end{align}
But this implies that $\cA^{(+)}[.] = D^+\cA^{(g)}[.] - \cA^{(g)}[D^+[.]]$  has the same intertwiner property, 
\begin{align}
 \cD^2 \cA^{(+)}[\phi^{(s)}] &=\cA^{(+)}[(\Box +\frac{2s+5}{R^2})\phi^{(s)}]\ ,
\end{align}
and a similar argument based on 
$D^-\cA^{(g)}[.] = \cA^{(g)}[D^-[.]] +  \cA^{(-)}[.]$ gives 
\begin{align}
 \cD^2 \cA^{(-)}[\phi^{(s)}] &=\cA^{(-)}[(\Box +\frac{-2s+3}{R^2})\phi^{(s)}] 
\end{align}
as desired.
These properties  originate from the underlying $\mso(4,2)$ Lie algebra structure,
and they should apply to the fully noncommutative case as well as the 
semi-classical Poisson limit.
In particular, the solution $\cA^{(-)}[D^+D\phi]$ 
underlying the Schwarzschild metric should easily generalize
to the noncommutative setting.

\subsection{Useful relations}
\label{sec:useful-formulas}

From the basic commutation relations \eqref{basic-CR-H4} it is easy to obtain
\begin{align}
 \b^{-1}\{x^ \mu, \b\} &= - \b\{x^ \mu, \b^{-1}\} 
 % = - \frac{1}{R}\b\{x^ \mu,x^4\} \nn\\ &= - \frac{1}{R}\b\theta^{\mu 4} 
  = r^2\b t^{\mu} \nn\\
 \b^{-1}\{t^ \mu, \b\} &= - \b\{t^ \mu, \b^{-1}\} 
%   =- \frac{1}{R}\b\{t^ \mu,x^4\}
   = \frac 1{R^2} \b x^\mu\label{x-t-beta-relations}  \\
 \b^{-1}\t\b &= - \b \t \b^{-1} 
 = -(\b^2+1) \ .
 \label{tau-beta-relation}
\end{align}
Furthermore, it is not hard to derive
\begin{align}
 \Box x^\a &=  \frac 1{R^2} x^\a \nn\\
  \Box x^4 &=  \frac 4{R^2} x^4
  \label{Box-x}
\end{align}
and 
\begin{align}
 \Box(\sinh(\eta)\phi) 
 % &= \sinh(\eta)\Box \phi - \frac 2R \{t^\a,x^4\}\{t_\a, \phi\} + (\Box\sinh(\eta)) \phi  \nn\\
 % &= \sinh(\eta)\Box \phi
 % + \frac 2{R^2} x^\a\{t_\a, \phi\} + \frac{4}{R^2}\sinh(\eta) \phi \nn\\
 % &= \sinh(\eta)\Box \phi + \frac 2{R^2} \sinh(\eta) (\t + 2) \phi   \nn\\
  &= \sinh(\eta)\big(\Box + \frac 2{R^2} (\t + 2)\big) \phi  \nn\\
 \Box\b \phi &= \b\Box\phi - \frac 2{R^2} (\t + 2)\b \phi  \label{Box-b2-id-1} \\
 \Box \t\phi &= \t\Box \phi + 2 \b^2(-\Box + \frac{1}{R^2}\t)  \phi  \ 
    \label{box-tau-relation-1}  
\end{align}
for scalar functions
$\phi\in\cC^0$.
%This can be checked for $\phi=x^4$.
% \begin{align}
%  4\frac{1}{R^2}(1+\b^2) x^4 &= \Box (1+\b^2) x^4 - 2 \b^2(-\frac 4{R^2} + \frac{1}{R^2}(1+\b^2)) x^4 \nn\\
%  &= \Box x^4 + R \Box\b - 2 \b^2(-\frac 4{R^2} + \frac{1}{R^2}(1+\b^2)) x^4 \nn\\
%  &= 4\frac{1}{R^2} x^4 -  \frac 2{R} (\t + 2)\b
%  - 2 \b^2(-\frac 3{R^2} + \frac{1}{R^2}\b^2) x^4 \nn\\
%  &= \frac{4}{R^2} x^4 - \frac 2{R} (-(1+\b^2) + 2)\b
%  - 2 \b(-\frac 3{R} + \frac{1}{R}\b^2)  \nn\\
%  &= \frac{4}{R^2} x^4 + \frac 4{R}\b
% \end{align}
% proof:
% \begin{align}
%  - \t\Box \phi &=  \t\b^{-2} \eta^{\a\b} \del_\a \del_\b \phi
%   -  \frac{1}{R^2} \t \t\phi   \nn\\
%   &= \b^{-2} \big(2(1+\b^2) +  \t) \eta^{\a\b} \del_\a \del_\b \phi
%   -  \frac{1}{R^2} \t \t \phi   \nn\\
%   &=  \b^{-2}  \eta^{\a\b} \del_\a \del_\b (\t-2)\phi 
%   -  \frac{1}{R^2} \t \t \phi  
%    + 2\b^{-2}(1+\b^2)\eta^{\a\b} \del_\a \del_\b \phi\nn\\
%   &= - \Box \t\phi 
%   -2\b^{-2}  \eta^{\a\b} \del_\a \del_\b \phi
%    + 2(1+\b^{-2})\eta^{\a\b} \del_\a \del_\b\phi  \nn\\
%   &= - \Box \t\phi  + 2\eta^{\a\b} \del_\a \del_\b\phi \nn\\
%   &= - \Box \t\phi  + 2 (-\b^2\Box + \frac{\b^2}{R^2}\t)  \phi \nn\\
%   &= - \Box \t\phi  + 2 \b^2(-\Box + \frac{1}{R^2}\t)  \phi 
%   \label{box-tau-relation}
% \end{align}
Finally, we note that \eqref{tau-relns}
gives 
\begin{align}
 D(\t+s)\phi &= (\t+s)D\phi,  \nn\\
  D^+D^- \t &= \t D^+ D^- \ .
 \label{D-tau-relation}
\end{align}

\subsection{Metric fluctuations from $\cA$ contributions}
\label{sec:metric-contrib}

In this section we obtain the metric fluctuations $h^{\mu\nu}[\cA]$ \eqref{tilde-H-def}
arising from the various 
terms in the tangential perturbations $\cA$. 
We will use the averaging formulas \eqref{averaging-relns} and 
the on-shell relation $\Box \phi = - \frac{2}{R^2}\phi$ throughout, as well as
\begin{align}
 \del^\a\del_\a\phi = \b^2(-\Box+\frac{1}{R^2}\t)\phi  = \frac{\b^2}{R^2}(2+\t)\phi 
 \label{deldel-Box-rel}
\end{align}
using \eqref{deldel-Box-relation}.

Consider first $\cA^\mu = \theta^{\mu\nu}\del_\nu \phi$. 
Then
\begin{align}
 h^{\mu\nu} &= -\{x^\mu,\cA^\nu\}_0 + (\mu\leftrightarrow \nu)  \
  = -\{x^\mu,\theta^{\nu\a}\del_\a \phi\} + (\mu\leftrightarrow \nu)  \nn\\
 % &= -[\theta^{\nu\a}\{x^\mu,\del_\a \phi\}]_0 
 %    -\{x^\mu,\theta^{\nu\a}\}\del_\a \phi  + (\mu\leftrightarrow \nu)  \nn\\
  &= -[\theta^{\nu\a} \theta^{\mu\b}]_0 \del_\a\del_\b \phi
     -\{x^\mu,\theta^{\nu\a}\}\del_\a \phi  + (\mu\leftrightarrow \nu)  \nn\\
%   &= -\frac{r^2 R^2}{3}
%   \big((\eta^{\mu\nu} + \frac{1}{R^2}x^\mu x^\nu)(\eta^{\a\b} + \frac{1}{R^2}x^\a x^\b)
%    - (\eta^{\mu\a} + \frac{1}{R^2}x^\mu x^\a)(\eta^{\nu\b} + \frac{1}{R^2}x^\nu x^\b)
%   \big) \del_\a\del_\b \phi \nn\\
%   &\quad   -r^2\{\cM^{\nu\a},x^\mu\}\del_\a \phi  + (\mu\leftrightarrow \nu)  \nn\\
%  &= -\frac{r^2 R^2}{3}
%   \big((\eta^{\mu\nu} + \frac{1}{R^2}x^\mu x^\nu)
%   (\del^\a\del_\a + \frac{1}{R^2}(\t^2-\t))\phi
%    - (\eta^{\mu\a} + \frac{1}{R^2}x^\mu x^\a)
%    (\del^\nu + \frac{1}{R^2}x^\nu \t) \del_\a \phi
%   \big) \nn\\ 
%   &\quad -r^2 (\eta^{\mu\nu} \t \phi - x^\nu\del^\mu \phi)
%    + (\mu\leftrightarrow \nu) \nn\\
%   &= -\frac{r^2}{3}
%   (\eta^{\mu\nu} + \frac{1}{R^2}x^\mu x^\nu)(\b^2(2+\t) + (\t^2-\t))\phi
%  +\frac{r^2}{3}R^2\del^\nu  \del^\mu \phi 
%   +\frac{r^2}{3}\big(x^\nu  \del^\mu  +  x^\mu \del^\nu + \frac{1}{R^2}x^\mu x^\nu \t \big)(\t-1) \phi  \nn\\ 
%   &\quad -r^2 (\eta^{\mu\nu} \t \phi - x^\nu\del^\mu \phi)  + (\mu\leftrightarrow \nu) \nn\\
%   %%
%   &= -\frac{2r^2}{3}(\eta^{\mu\nu} + \frac{1}{R^2}x^\mu x^\nu)(\b^2(2+\t) + (\t^2-\t))\phi
%   +\frac{2r^2R^2}{3} \del^\nu  \del^\mu \phi 
%   +\frac{2r^2}{3}\big(x^\nu  \del^\mu  +  x^\mu \del^\nu + \frac{1}{R^2}x^\mu x^\nu \t \big)(\t-1) \phi  \nn\\ 
%   &\quad -2r^2\eta^{\mu\nu} \t \phi +r^2 (x^\nu\del^\mu +  x^\mu\del^\nu) \phi  \nn\\ 
%    %%
  &= \frac{r^2}3\Big(- 2(\b^2 + \t)(\t + 2)\eta^{\mu\nu}
   - \frac{2\b^2}{R^2}  x^\mu x^\nu(2+\t)
   +  (x^\nu  \del^\mu  +  x^\mu \del^\nu)(2\t + 1)
   + 2R^2 \del^\nu  \del^\mu\Big) \phi 
   \label{h-theta-del-phi}
\end{align}
%since $x^\a x^\b \del_\a\del_\b \phi = (\t^2-\t)\phi$ and 
using \eqref{deldel-Box-relation} and the on-shell condition. 
Next consider $\cA^\mu = \b t^\mu \phi$. Then
\begin{align}
 h^{\mu\nu} &= -\{x^\mu,\cA^\nu\}_0 + (\mu\leftrightarrow \nu)  
  = - \{x^ \mu,\b t^\nu \phi\}_- + (\mu \leftrightarrow \nu)  \nn\\
   &= \sinh \eta^{\mu\nu} \b\phi - [t^\nu \theta^{\mu\a}]_0\del_\a(\b\phi) + (\mu \leftrightarrow \nu) \nn\\
%   &= \sinh \eta^{\mu\nu} \b \phi 
%     - \frac 13\big(\sinh(\eta)(\eta^{\nu\a} x^\mu -\eta^{\mu\nu}x^\a)
%         + \epsilon^{\nu\a\mu\b}x_\b\big)\del_\a(\b\phi) 
%        + (\mu \leftrightarrow \nu) \nn\\
%    &=  \eta^{\mu\nu}\phi 
%      - \frac 13\sinh(\eta)\big(\eta^{\nu\a} x^\mu -\eta^{\mu\nu}x^\a\big)\del_\a(\b\phi) 
%        + (\mu \leftrightarrow \nu) \nn\\
%    &= \eta^{\mu\nu} (1 + \frac 13 \t  +\frac 13(\b^{-1}\t\b)) \phi
%         -\frac 13 x^\mu \{t^\nu,\b\phi\} 
%        + (\mu \leftrightarrow \nu)  \nn\\
%    &= \eta^{\mu\nu} (1 + \frac 13 \t - \frac 13 \b^2 \cosh^2(\eta))\phi -\frac 13 (x^\mu \{t^\nu,\b\}\phi + x^\mu \del^\nu\phi)
%        + (\mu \leftrightarrow \nu)  \nn\\
%    &= \eta^{\mu\nu} (1 + \frac 13 \t - \frac 13 \b^2 (1+\sinh^2(\eta)))\phi -\frac 13 (\frac{\b^2}{R^2} x^\mu x^\nu \phi + x^\mu \del^\nu\phi)
%        + (\mu \leftrightarrow \nu)  \nn\\
   &= \frac 23\eta^{\mu\nu} (2 + \t - \b^2)\phi - \frac 23 \frac{\b^2}{R^2} x^\mu x^\nu \phi 
   - \frac 13 (x^\mu \del^\nu +  x^\nu \del^\mu)\phi \ .
   \label{h-t-phi}
 \end{align}
%since $\epsilon^{\nu\a\mu\b}$ drops out upon symmetrization in $\mu\nu$, and using 
%\eqref{tau-beta-relation}.
%This can be checked for the trace and for $\phi = x^4$.
% 
% The trace is 
% \begin{align}
%  h = -2\{x_ \mu,\b t^\mu \phi\}_- 
%  &\stackrel{!}{=} \frac 83(2 + \t) \phi + \frac 23 \b^2\cosh^2(\eta) \phi 
%    - \frac 23 \t\phi   \nn\\
%   8\phi -2t^\mu\{x_ \mu,\b  \}_- \phi - 2\b t^\mu\{x_ \mu, \phi\}_- 
%  &\stackrel{!}{=}  \frac 83(2 + \t) \phi + \frac 23 (1-\t)\phi = (6+2\t)\phi  \nn\\
%   2\phi + 2 \b^{-1} (\t\b) \phi 
%  &\stackrel{!}{=}  0
% \end{align}
% which is correct
% for large $\eta$, hence 
% \begin{align}
%  h =  (6+2\t)\phi 
% \end{align}
% 
% 
% 
% As a further check, consider $\phi = x^4$. Then $\t\phi \approx \phi$ for large $\eta$, and 
% \begin{align}
%   h^{\mu\nu} &= - \{x^ \mu, t^\nu \b x^4\}_- + (\mu \leftrightarrow \nu) 
%      = - R(\{x^ \mu, t^\nu\} + \{x^ \nu, t^\mu\}) 
%    = 2 R \eta^{\mu\nu}
% \end{align}
% works out.
For $\cA^\mu =  \b x^\mu t^\a  \del_\a \phi = \frac{\b}{r^2R} x^\mu D\phi$, we obtain
\begin{align}
 h^{\mu\nu} &= -\{x^\mu,\cA^\nu\}_0 + (\mu\leftrightarrow \nu)  
  = - \{x^ \mu,\b x^\nu t^\a  \del_\a \phi\}_0 + (\mu \leftrightarrow \nu)  \nn\\
 % &= - \theta^{\mu\nu} \b t^\a  \del_\a \phi
  % - x^\nu \{x^ \mu, t^\a\}  \b \del_\a \phi
  % - x^\nu  [t^\a \{x^ \mu, \b \del_\a \phi\}]_0 + (\mu \leftrightarrow \nu) \nn\\
  &=  x^\nu \del_\mu \phi
    - x^\nu  [t^\a \{x^ \mu, \b\}]_0 \del_\a \phi
   - \b x^\nu  [t^\a \theta^{\mu\s}\del_\s\del_\a \phi]_0 + (\mu \leftrightarrow \nu)  \nn\\
%   &=  x^\nu \del_\mu \phi
%     - r^2\b^2  x^\nu  [t^\a t^{\mu} ]_0 \del_\a \phi
%    - \b x^\nu  [t^\a \theta^{\mu\s}]_0 \del_\s\del_\a \phi + (\mu \leftrightarrow \nu)  \nn\\
%   &=  x^\nu \del_\mu \phi
%     - \b^2  \frac{\cosh^2(\eta)}{3} x^\nu  P_\perp^{\a\mu}  \del_\a \phi
%    - \frac 13 x^\nu  (x^\mu \del^\a\del_\a \phi - \t \del_\mu \phi ) + (\mu \leftrightarrow \nu)  \nn\\
%  &=  x^\nu \del_\mu \phi
%     - \b^2  \frac{\cosh^2(\eta)}{3} x^\nu \big(\eta^{\a\mu} + \frac{x^\mu x^\a}{R^2 \cosh^2(\eta)}\big) \del_\a \phi
%    - \frac 13 x^\nu  (x^\mu \frac{\b^2}{R^2}(2 + \t) \phi - \del^\mu(\t-1) \phi ) + (\mu \leftrightarrow \nu)  \nn\\
%   %
 &=  - \frac 43\frac{\b^2}{R^2} x^\nu x^\mu (1 + \t)\phi 
   + \frac 13(1+\t-\b^2)(x^\nu \del^\mu + x^\mu \del^\nu) 
     \label{h-xtdel-phi}
\end{align}
using \eqref{x-t-beta-relations}. Again the trace provides some check.
% as a check, the trace is 
% \begin{align}
%  h = - 2 \{x_\mu,\b x^\mu t^\a  \del_\a \phi\}_0 
%   &\stackrel{!}{=}  \frac 43 (1 + \t) \phi 
%   + \frac{2}{3}(1 +\t)\t \phi  \nn\\
%  - 2 \b x^\mu \{x_\mu, t^\a  \del_\a \phi\}_0 
%   &\stackrel{!}{=}  \frac 23(\t^2+3\t+2)\phi  \nn\\
%   \frac{2}{r^2} D^-D \phi
%   &\stackrel{!}{=}  \frac 23(\t^2+3\t+2)\phi \nn\\
%  - \frac{2R^2}{3} \cosh^2(\eta)\Delta^{(3)} \phi &\stackrel{!}{=}  \frac 23(\t^2+3\t+2)\phi  \nn\\
%   \frac{2}{3} (\t+1)(\t+2) \phi  &\stackrel{!}{=}  \frac 23(\t^2+3\t+2)\phi  \nn\\
% \end{align}
% using \eqref{onshell-Delta-tau}. Is consistent.
Finally,  for $\cA^\mu =  \sinh(\eta) t^\a \del_\a\del_\mu\phi$ we obtain
\begin{align}
 h^{\mu\nu} &= -\{x^\mu,\cA^\nu\}_0 + (\mu\leftrightarrow \nu)  
  = - \{x^ \mu,\sinh(\eta) t^\a \del_\a\del_\nu\phi\} + (\mu\leftrightarrow \nu) \nn\\
%   &= - \{x^ \mu,\sinh(\eta)\}  t^\a \del_\a\del_\nu\phi
%   -\sinh(\eta)  \{x^ \mu,t^\a\}  \del_\a\del_\nu\phi
%   - \sinh(\eta) t^\a \{x^ \mu,\del_\a\del_\nu\phi\} 
%   + (\mu\leftrightarrow \nu)  \nn\\
  &=  r^2 [t^\mu t^\a]_0 \del_\a\del_\nu\phi
   + \sinh^2(\eta) \del_\mu\del_\nu\phi
  - \sinh(\eta) [t^\a \theta^{ \mu\g}]_0\del_\g\del_\a\del_\nu\phi
  + (\mu\leftrightarrow \nu)  \nn\\
%   &=  \frac{\cosh^2(\eta)}3 P_\perp^{\mu\a} \del_\a\del_\nu\phi
%    + \sinh^2(\eta) \del_\mu\del_\nu\phi
%   - \frac{1}{3}\sinh^2(\eta)(\eta^{\a\g}x^\mu - \eta^{\a\mu}x^\g)\del_\g\del_\a\del_\nu\phi
%   + (\mu\leftrightarrow \nu)  \nn\\
%   &=  \frac{\cosh^2(\eta)}3 (\eta^{\mu\a} +\frac{x^\mu x^\a}{R^2\cosh^2(\eta)}) \del_\a\del_\nu\phi
%    + \sinh^2(\eta) \del_\mu\del_\nu\phi 
%    - \frac{1}{3}\sinh^2(\eta)(x^\mu \del_\nu\del^\a\del_\a\phi- \t \del^\mu\del_\nu\phi)
%   + (\mu\leftrightarrow \nu)  \nn\\
%   &=  \frac{1}3 \cosh^2(\eta)\del^\mu\del_\nu\phi + \sinh^2(\eta) \del_\mu\del_\nu\phi
%   +\frac{1}{3R^2} x^\mu\del_\nu(\t-1)\phi   \nn\\
%   &\quad - \frac 1{3R^2}\b^{-2} x^\mu \del_\nu\b^2(2+\t)\phi
%   +\frac 13 \sinh^2(\eta)\del^\mu\del_\nu(\t-2)\phi
%   + (\mu\leftrightarrow \nu) \nn\\
  &= \frac 23 \sinh^2(\eta)\del_\mu\del_\nu(2 +\t)\phi
  -\frac{1}{R^2} (x^\mu\del_\nu +  x^\nu\del_\mu )\phi
  - \frac 4{3R^4} \b^2 x^\mu x^\mu (2+\t)\phi
   \label{h-tdeldel-phi}
\end{align}
usng the on-shell relation.
We also note the relations
\begin{align}
 \{t^\mu,\b t_\mu(\t+2)\phi\} 
 %&= t_\mu\{t^\mu,\b (1+\b^2)(\t+2)\phi\}
  &= t^\mu\del_\mu(\t+2)\phi  \nn\\
 \{t^\mu,\b x_\mu t^\a  \del_\a (\t+2)\phi\} 
%   &= \{t^\mu, x_\mu\}\b t^\a  \del_\a \phi  + x_\mu\{t^\mu,\b  t^\a  \del_\a \phi\} \nn\\
%   &= 4 t^\a  \del_\a \phi
%    +x_\mu\{t^\mu, t^\a \}\b  \del_\a \phi
%    + t^\a   x_\mu\{t^\mu,\b   \del_\a \phi\} \nn\\  
%  &= 4 t^\a  \del_\a \phi
%    - \frac{\b}{r^2 R^2} x_\mu \theta^{\mu\a}   \del_\a \phi
%    +  \b^{-1} t^\a  \t(\b   \del_\a \phi) \nn\\  
%  &= 4 t^\a  \del_\a \phi
%    +  t^{\a}   \del_\a \phi
%    +  \b^{-1} t^\a  \t(\b)\del_\a \phi +  t^\a  \del_\a (\t-1)\phi  \nn\\  
%   &= 5 t^\a  \del_\a \phi
%    - (1+\b^2) t^\a \del_\a \phi +  t^\a  \del_\a (\t-1)\phi  \nn\\  
 &= t^\a  \del_\a (\t+3-\b^2)(\t+2)\phi  \nn\\
 \{t_\nu,\theta^{\nu\a} \del_\a  (\t+2)\phi\} 
 % &= \{t_\nu,\theta^{\nu\a}\} \del_\a  (\t+2)\phi + \theta^{\nu\a} \{t_\nu,\del_\a  (\t+2)\phi\} 
  % &= r^2\{\cM^{\nu\a}, t_\nu\} \del_\a  \phi 
  &= 3 r^2 t^\a \del_\a  (\t+2)\phi
  \label{gauge-fixing-relations-scalar}
\end{align}  
% hence
% \begin{align}
%  \{t_\nu,\theta^{\nu\a} \del_\a (\t+4+\b^2)(\t+2) \phi\} 
%   &=3 r^2 t^\a \del_\a (\t+4+\b^2)(\t+2) \phi 
% \end{align}
due to $t^\a\del_\a\b = 0$, which are used to check gauge invariance.

\subsection{$DD$ operator on scalar fields}

Let  $\phi \in \cC^0$. The  explicit formula \eqref{D-properties} for $D$ 
gives
\begin{align}
 D D \phi &= r^4 R^2 t^\mu t^\nu  \nabla^{(3)}_\mu  \nabla^{(3)}_\nu \phi   \nn\\
 D D D D \phi &= r^8 R^4 t^\mu t^\nu t^\r t^\s  \nabla^{(3)}_\mu  \nabla^{(3)}_\nu  \nabla^{(3)}_\r  \nabla^{(3)}_\s \phi  
\end{align}
where $\nabla^{(3)}$ is the covariant derivative along the space-like $H^3$.
In particular,
\begin{align}
 D^- D^+ \phi &= r^4 R^2 [t^\mu t^\nu]_0  \nabla^{(3)}_\mu  \nabla^{(3)}_\nu \phi 
 = \frac{r^2 R^2}3 \cosh^2(\eta) P_\perp^{\mu\nu} \nabla^{(3)}_\mu  \nabla^{(3)}_\nu \phi \nn\\
  &= -\frac{r^2 R^2}3 \cosh^2(\eta)  \Delta^{(3)}\phi
  \label{D-D+-explicit}
\end{align}
where $\Delta^{(3)} = -\nabla^{(3)\mu}  \nabla^{(3)}_\mu$ is the covariant Laplacian on $H^3$.
Note that both expressions are $SO(3,1)$-invariant second order differential operators. 
The averaging is given 
in terms of  the projector  $P_\perp$  on $H^3$ in \eqref{kappa-average}.
% Moreover, they agree
% on the function $\phi = x^\mu$, since
% \begin{align}
%  D D x^\mu &= r^2 R D t^\mu =  r^2 x^\mu \stackrel{?}{= } -\frac{r^2 R^2}3 \cosh^2(\eta)  \Delta^{(3)} x^\mu
% \end{align}
% and
% \begin{align}
%  \Delta^{(3)} x^\a &= - \del_\mu  (P_\perp^{\mu\nu}  \del_\nu x^\a) 
%  =  - \del_\mu  ((\eta^{\mu\nu} +\frac{1}{R^2\cosh^2(\eta)} x^\mu x^\nu) \del_\nu x^\a) \nn\\
%  &= - \frac 1{R^2}\del_\mu  (\frac{1}{\cosh^2(\eta)} x^\mu x^\a) \nn\\
%  &= - \frac 1{R^2}\Big((\frac{5}{\cosh^2(\eta)}  x^\a 
%   + x^\mu x^\a \del_\mu  (\frac{1}{\cosh^2(\eta)} \Big) \nn\\
%  &= - \frac 1{R^2}\Big(\frac{5}{\cosh^2(\eta)}  x^\a 
%   - \frac{1}{R^2\cosh^4(\eta)} x^\mu x^\a \del_\mu (-x^\b x_\b) \Big)\nn\\
%  &= - \frac 1{R^2}\Big(\frac{5}{\cosh^2(\eta)}  x^\a 
%   + \frac{2}{R^2\cosh^4(\eta)} x^\a x^\b x_\b \Big)\nn\\
%  &= - \frac 1{R^2}\frac{3}{\cosh^2(\eta)}  x^\a 
%  \label{Delta-3-x}
% \end{align}
Now we compute
\begin{align}
 [D D D D \phi]_0 &= r^8 R^4 [t^\mu t^\nu t^\r t^\s]_0  \nabla^{(3)}_\mu  \nabla^{(3)}_\nu  \nabla^{(3)}_\r  \nabla^{(3)}_\s \phi   \nn\\
  &= \frac 35 r^8 R^4 \big([t^{\mu}t^{\nu}][t^{\r} t^{\s}]_0 
   + [t^{\mu}t^{\r}][t^{\nu} t^{\s}]_0  + [t^{\mu}t^{\s}][t^{\nu} t^{\r}]_0\big)
  \nabla^{(3)}_\mu  \nabla^{(3)}_\nu  \nabla^{(3)}_\r  \nabla^{(3)}_\s \phi   \nn\\
  &= \frac {\cosh^4(\eta)}{15} R^4 r^4 \big(P^{\mu\nu}_H P_\perp^{\r\s}
   + P_\perp^{\mu\r}P_\perp^{\nu\s}  + P_\perp^{\mu\s}P_\perp^{\nu\r}\big)
  \nabla^{(3)}_\mu  \nabla^{(3)}_\nu  \nabla^{(3)}_\r  \nabla^{(3)}_\s \phi   \nn\\
 % &= \frac {\cosh^4(\eta)}{15} R^4 r^4 \big(3\Delta^{(3)}\Delta^{(3)} + 2\nabla^{(3)}_\mu (R_{(3)}^{\mu\a}\nabla^{(3)}_\a) \big) \phi \nn\\
 % &= \frac {\cosh^4(\eta)}{15} R^4 r^4 \big(3\Delta^{(3)}\Delta^{(3)} 
 %   - \frac 23 R_{(3)} \Delta^{(3)} \big) \phi    \nn\\
  &= R^4 r^4\frac {\cosh^4(\eta)}{5}  \big(\Delta^{(3)} + \frac 43\frac{1}{R^2\cosh^2(\eta)}  \big) \Delta^{(3)}\phi   
\end{align}
where $P^{\mu\nu}_\perp = g^{\mu\nu}_{(3)}$ is the tangential induced metric on $H^3$  which satisfies 
$\nabla^{(3)} P_\perp^{\mu\s} = 0$.
The individual terms are given by
\begin{align}
 P^{\mu\nu}_H P_\perp^{\r\s} \nabla^{(3)}_\mu  \nabla^{(3)}_\nu  \nabla^{(3)}_\r  \nabla^{(3)}_\s \phi   
  &=  \Delta^{(3)}\Delta^{(3)}  \phi \nn\\
 P_\perp^{\mu\r}P_\perp^{\nu\s}\nabla^{(3)}_\mu  \nabla^{(3)}_\nu  \nabla^{(3)}_\r  \nabla^{(3)}_\s \phi   
 % &=  \Delta^{(3)}\Delta^{(3)}  \phi 
 %  + P_\perp^{\mu\r}P_\perp^{\nu\s}\nabla^{(3)}_\mu {R_{\nu\r;\s}}^{\a}  \nabla^{(3)}_\a \phi \nn\\
   &=  \Delta^{(3)}\Delta^{(3)}  \phi 
   +   \nabla^{(3)}_\mu ( R^{\mu\a}_{(3)}\del^{(3)}_\a \phi) \nn\\
 &=  P_\perp^{\mu\s}P_\perp^{\nu\r}\nabla^{(3)}_\mu  \nabla^{(3)}_\nu  \nabla^{(3)}_\r  \nabla^{(3)}_\s \phi   
%  &= -P_\perp^{\mu\s}\nabla^{(3)}_\mu  \Delta^{(3)} \nabla^{(3)}_\s \phi   
%   = \Delta^{(3)} \Delta^{(3)}  \phi  
%    +  P_\perp^{\mu\s}\nabla^{(3)}_\mu (R_{\s\g}^{(3)} g^{\g\d}\del_\d \phi) \nn\\
 % &= \Delta^{(3)} \Delta^{(3)}  \phi  
 %    + \nabla^{(3)}_\mu (R^{\mu\d}_{(3)} \del_\d \phi)
\end{align} 
where 
\begin{align}
R_{(3)}^{\mu\a} = \frac 13 P_\perp^{\mu\nu} R_{(3)}, \qquad 
 R_{(3)} = -\frac{6}{R^2\cosh^2(\eta)} \ 
\end{align}
 are the Ricci tensor and scalar on $H^3$.
% and using
% \begin{align}
%  \nabla^{(3)}_\mu  [\nabla^{(3)}_\nu, \nabla^{(3)}_\r]  \nabla^{(3)}_\s \phi 
%   &= \nabla^{(3)}_\mu  {R_{\nu\r;\s}}^{\a} \nabla^{(3)}_\a \phi  
%  % =  {R_{\nu\r;\s}}^{\a} \nabla^{(3)}_\mu \nabla^{(3)}_\a \phi  \nn\\
% %   \nabla^{(3)}_\s \Delta^{(3)} \phi &=  \Delta^{(3)}\nabla^{(3)}_\s \phi 
% %     + R_{\s\g}^{(3)} g^{\g\d}_{(3)}\del_\d \phi
% \end{align}
%This can be checked e.g. for $\phi = x^\a$.
% To check the normalization, we apply this to $x^\a$, which gives indeed
% \begin{align}
%  D^4 x^\a = r^4 x^\a 
%  =  -R^4 r^4\frac {\cosh^4(\eta)}{5}  \big((- \frac 1{R^2}\frac{3}{\cosh^2(\eta)} )
%  + \frac 43\frac{1}{R^2\cosh^2(\eta)}  \big)  \frac 1{R^2}\frac{3}{\cosh^2(\eta)} x^\a
% \end{align}
Combining these, we obtain
\begin{align}
D^- D^- D^+ D^+ \phi &= 
 [D D D D \phi]_0 - D^- D^+ D^- D^+ \phi \nn\\
  %&= R^4 r^4\frac {\cosh^4(\eta)}{5}  \big(\Delta^{(3)} + \frac 43\frac{1}{R^2\cosh^2(\eta)}  \big) \Delta^{(3)}\phi   
  %     - \frac{r^4 R^4}9 \cosh^4(\eta) \Delta^{(3)} \Delta^{(3)}\phi  \nn\\
  %   &= R^4 r^4 \cosh^4(\eta) \Big((\frac{1}{5}-\frac 19)\Delta^{(3)} 
  %   +\frac{4}{15R^2\cosh^2(\eta)}  \Big)\Delta^{(3)}\phi \nn\\
     &= \frac 4{15} R^4 r^4 \cosh^4(\eta) \Big(\frac 13 \Delta^{(3)} 
     +\frac{1}{R^2\cosh^2(\eta)}  \Big)\Delta^{(3)}\phi  \ .
\end{align}
%noting that $ \Delta^{(3)}(f(\eta)  \phi) =  f(\eta) \Delta^{(3)} \phi$. 
Note that this vanishes for $x^\mu$, 
%since  $\Delta^{(3)}x^\mu = -\frac{3}{R^2\cosh^2(\eta)}x^\mu$ \eqref{Delta-3-x}, 
consistent with 
$D^- D^- D^+ D^+ x^\mu = 0$.
%As a further check, consider $\phi = x^4$. Then $\Delta^{(3)} x^4 = 0 = D x^4$. 
Now we  apply this to on-shell solution with 
$\big(\Box + \frac 2{R^2}\big) \phi=0$. 
Then \eqref{Box-Laplace-tau} gives 
\begin{align}
 \cosh^2(\eta)\Delta_g^{(3)} \phi
  &= -\frac{1}{R^2}(1+ \t  + \b^2)(2 +\t)\phi 
 \label{Delta3-phi-static}
\end{align}
so that 
\begin{align}
 \frac 12 h =  D^- D^- D^+ D^+ \phi
 %&\sim - \frac 4{15} R^4 r^4 \cosh^4(\eta) \Big(\frac 13 \Delta^{(3)} 
 %    +\frac{1}{R^2\cosh^2(\eta)}  \Big)\frac{\b^2}{R^2}(\t+1)(\t+2) \phi   \nn\\ 
 %&\sim - \frac 4{15} R^2 r^4 \cosh^2(\eta) \Big(\frac 13 \Delta^{(3)} 
 %    +\frac{1}{R^2\cosh^2(\eta)} \Big)(\t+1)(\t+2) \phi   \nn\\ 
% &= -\frac 4{15} r^4  
% \Big(-\frac 13 (2 + 3\t + \t^2)  +1 \Big)(\t+1)(\t+2)\phi \nn\\
  &\stackrel{\eta\to\infty}{\sim} -\frac 4{45} r^4  (1 - 3\t - \t^2)(\t+1)(\t+2) \phi \ .
 \label{h-phi-relation}
\end{align}
This can be used as a consistency check for the computation of the trace $h$ in section \ref{sec:A-DD-metric}.

\subsection{Evaluation of $\cA^{(-)}[D^+D\phi]$}
\label{sec:eval-A-DD}

To find the corresponding metric fluctuation mode, we need to
elaborate the fluctuation mode $\cA^{(-)}_\mu$ explicitly.
For $\phi\in \cC^0$, we have  
\begin{align}
 DD\phi &= r^4 R^2 t^\a t^\b  \del_\a \del_\b \phi 
  - r^2 R \frac{1}{x_4} t_\a\theta^{\a\b}\del_\b\phi \nn\\
   &= r^4 R^2 t^\a t^\b  \del_\a \del_\b \phi + r^2 \t\phi 
\end{align}
hence
\begin{align}
 \{x^\mu,DD\phi\}_1 &= r^4 R^2 \{x^\mu,t^\a t^\b  \del_\a \del_\b \phi\}_1   
   + r^2 \{x^\mu, \t\phi \}  \nn\\
  &= 2 r^4 R^2 t^\a \{x^\mu, t^\b\} \del_\a  \del_\b \phi 
    + r^4 R^2 [t^\a t^\b \theta^{\m\g}]_1 \del_\g (\del_\a \del_\b \phi) 
      + r^2 \theta^{\mu\nu}\del_\nu \t\phi   \nn\\
  &= - 2 r^4 R x_4 t^\a \del_\a  \del_\mu \phi 
    +  \frac 35 r^4 R^2\Big(2 t^\a [t^\b \theta^{\m\g}]_1  
     + [t^\a t^\b]_0 \theta^{\m\g}  \Big)\del_\g (\del_\a \del_\b \phi)
      + r^2 \theta^{\mu\nu}\del_\nu \t\phi \nn\\
%   &= - 2 r^4  R^2 \sinh(\eta)  t^\a \del_\a  \del_\mu \phi 
%      + \frac 15 r^2 R^2\cosh^2(\eta)  \theta^{\m\g}   P_\perp^{\a\b}\del_\g (\del_\a \del_\b \phi)  \nn\\
%   &\quad  + \frac 25 r^4 R^2 t^\a 
%   \Big(\sinh(\eta) ( \eta^{\b\g} x^\mu - \eta^{\b\mu} x^\g) 
%   +  x_\d \varepsilon^{\d 4\b\mu \g}  \Big)
%   \del_\g (\del_\a \del_\b \phi)
%    + r^2 \theta^{\mu\nu}\del_\nu \t\phi \nn\\
%   &= - 2 r^4 R^2 t^\a \{t_\mu,  \del_\a \phi\} 
%      + \frac 15 r^2 R^2\cosh^2(\eta)  \theta^{\m\g}P_\perp^{\a\b}\del_\g (\del_\a \del_\b \phi) 
%     + r^2 \theta^{\mu\nu}\del_\nu \t\phi  \nn\\
%   &\quad + \frac 25 r^4 R^2 
%     t^\a \sinh(\eta)  \Big( x^\mu \del^\b (\del_\b \del_\a \phi) -  x^\g\del_\g (\del_\mu \del_\a \phi) \Big)  \nn\\
%   &= - 2 r^4 R^2 ( \{t_\mu, t^\a  \del_\a \phi\} +\frac{1}{r^2 R^2} \theta^{\mu\a}  \del_\a \phi)
%      + \frac 15 r^2 R^2\cosh^2(\eta)  \theta^{\m\g}\big(\eta^{\a\b} - \frac{x^\a x^\b}{x_\g x^\g}\big) \del_\g (\del_\a \del_\b \phi) \nn\\
%   &\quad  + r^2 \theta^{\mu\nu}\del_\nu \t\phi  \nn\\
%   &\quad + \frac 25 r^4 R^2 
%  t^\a \sinh(\eta)  \Big( x^\mu \del_\a \big(-\b^2\Box + \frac{R^2}{\sinh^2(\eta)}\t\big)  \phi -  x^\g\del_\g (\del_\mu \del_\a \phi) \Big)   \nn\\
  &= - 2 r^4 R^2 \{t_\mu, t^\a  \del_\a \phi\} - 2 r^2 \theta^{\mu\a}  \del_\a \phi
     + \frac 15 r^2 R^2\cosh^2(\eta)  \theta^{\m\g}\del_\g (\del^\a \del_\a \phi) 
     + \frac 15 r^2 \theta^{\m\g} x^\a x^\b  \del_\g (\del_\a \del_\b \phi)  \nn\\
  &\quad  + r^2 \theta^{\mu\nu}\del_\nu \t\phi  
   + \frac 25 r^4 R^2 
 t^\a \sinh(\eta)  \Big( x^\mu \del_\a \big(-\b^2\Box + \frac{R^2}{\sinh^2(\eta)}\t\big)  \phi -  x^\g\del_\g (\del_\mu \del_\a \phi) \Big)  
\end{align}
using the averaging formulas \eqref{averaging-relns}, \eqref{average-3} and \eqref{deldel-Box-relation}.
The first term is pure gauge, and the last term can be rewritten as 
\begin{align}
  t^\a \sinh(\eta) x^\g\del_\g \del_\mu \del_\a \phi
 % &= t^\a \sinh(\eta)\del_\a  \del_\mu (\t-2)\phi \nn\\
 % &= t^\a  \{t_\mu,\del_\a(\t-2) \phi\}  \nn\\
 &=  \{t_\mu, t^\a \del_\a(\t-2) \phi\} +\frac{1}{r^2 R^2} \theta^{\mu\a}\del_\a((\t-2) \phi) 
\end{align}
using $\t\del = \del(\t-1)$.
Further, 
\begin{align}
  x^\a x^\b  \del_\g (\del_\a \del_\b \phi) 
  % &=  x^\b  \del_\g (x^\a \del_\a \del_\b \phi) - x^\b \del_\g \del_\b \phi  \nn\\
  % &=   \del_\g (x^\b x^\a \del_\a \del_\b \phi) - 2 x^\a \del_\a \del_\g \phi\nn\\
   &=   \del_\g((\t-1)(\t-2) \phi) \ .
\end{align}
Therefore 
\begin{align}
 \{x^\mu,DD\phi\}_1 &= - 2 r^2 \theta^{\mu\a}  \del_\a \phi
     + \frac 15 r^2 R^2\cosh^2(\eta)  \theta^{\m\g}\del_\g (\del^\a \del_\a \phi) 
     + \frac 15 r^2 \theta^{\m\g} x^\a x^\b  \del_\g (\del_\a \del_\b \phi)  \nn\\
  &\quad  + r^2 \theta^{\mu\nu}\del_\nu \t\phi + \{t_\mu,\L\} \nn\\
  &\quad  + \frac 25 r^4 R^2 x^\mu \sinh(\eta) t^\a  \del_\a \big(-\b^2\Box + \frac{1}{R^2\sinh^2(\eta)}\t\big)  \phi 
 - \frac 25 r^4 R^2 \frac{1}{r^2 R^2} \theta^{\mu\a}\del_\a((\t-2) \phi)   \nn\\
% %  &=  \frac 15 r^2 R^2\cosh^2(\eta)  \theta^{\m\g}\del_\g (\del^\a \del_\a \phi) 
% %      + \frac 15 r^2 \theta^{\m\g} \del_\g( (\t^2 - 3\t + 2) \phi)  \nn\\
% %   &\quad  + r^2 \theta^{\mu\nu}\del_\nu((\t-2)\phi) + \{t_\mu,\L\} \nn\\
% %   &\quad  - \frac 25 r^4 R^2 x^\mu \frac 1{\sinh(\eta)} t^\a  \del_\a (\Box  \phi - \frac 1{R^2}\t \phi) 
% %  - \frac 25 r^2 \theta^{\mu\a}\del_\a((\t-2) \phi)     \nn\\
% %   &= - \frac 15 r^2 R^2\cosh^2(\eta)  \theta^{\m\g}\del_\g \b^2\big(\Box - \frac 1{R^2}\t\big) \phi \nn\\
% %   &\quad - \frac 25 r^4 R^2 x^\mu \frac 1{\sinh(\eta)} t^\a  \del_\a (\Box - \frac 1{R^2}\t) \phi
% %    \nn\\
% %  &\quad   +r^2 \theta^{\m\g} \del_\g\Big(\frac 15 (\t^2 - 3\t + 2)  + (\t-2) 
% %  - \frac 25 (\t-2)\Big)   \phi 
% %  + \{t_\mu,\L\}   \nn\\
  &= - \frac 25 r^4 R^2\b^3 \cosh^2(\eta)  t^\mu \big(\Box - \frac 1{R^2}\t\big) \phi 
  - \frac 15 r^2 R^2\frac{\cosh^2(\eta)}{\sinh^2(\eta)}  \theta^{\m\g}\del_\g \big(\Box - \frac 1{R^2}\t\big) \phi 
  \nn\\
  &\quad  - \frac 25 r^4 R^2 x^\mu \frac 1{\sinh(\eta)} t^\a  \del_\a (\Box - \frac 1{R^2}\t) \phi
     + \frac{r^2}5 \theta^{\m\g} \del_\g(\t^2-4)  \phi 
 + \{t_\mu,\L\}  
 \label{A-DDphi-1}
\end{align}
using 
 \begin{align}
  \theta^{\m\g}\del_\g \b^2
  % &=  2\b^4 \frac{1}{R^2}\theta^{\m\g} x_\g 
   =  2\b^3 r^2 t^\mu
 \end{align}
where 
\begin{align}
 \L 
   %&=  2 r^4 R^2 (- t^\a  \del_\a \phi - \frac 15 t^\a \del_\a(\t-2) \phi) \nn\\
   %&= \frac 25  r^4 R^2 t^\a \del_\a(-\t-3) \phi  \nn\\
   &= -\frac 25  r^2 R  D((\t+3) \phi) \ .
 \label{gaugeparam-schwarzschild}
\end{align}

\paragraph{Degenerate case.}

In the  special case $\big(\Box - \frac 1{R^2}\t\big)\phi = 0$, we obtain 
\begin{align}
 \{x^\mu,DD\phi\}_1 &= \frac{r^2}5 \theta^{\m\g} \del_\g(\t^2-4)  \phi 
     + \{t_\mu,\L\}  
  = \frac{r^2}5 \{x^\mu,(\t^2-4)  \phi\}  + \{t_\mu,\L\} 
  \label{degeneracy-Apm}
\end{align}
i.e. there is a linear dependence between the $\cA^{(\pm)}$ modes.
Imposing also the on-shell condition would imply 
$(2 + \t)\phi = 0$. We will see that  then $\cA^{(-)}[D^+D\phi]$ vanishes,
but a non-trivial mode can be extracted by taking a suitable limit,
which corresponds precisely to the Schwarzschild solution.

\paragraph{On-shell condition.}

Now consider on-shell solutions, so that $\Box\phi = -\frac{2}{R^2}\phi$. Then
\eqref{A-DDphi-1} becomes
\begin{align}
 \{x^\mu,DD\phi\}_1 
%  &= \frac 25 r^4 \b^3 \cosh^2(\eta)  t^\mu \big(2 + \t\big) \phi 
%   + \frac 15 r^2\frac{\cosh^2(\eta)}{\sinh^2(\eta)} \theta^{\m\g}\del_\g \big(2 + \t\big) \phi 
%   \nn\\
%   &\quad + \frac 25 r^4 x^\mu \frac 1{\sinh(\eta)} t^\a  \del_\a (2 + \t) \phi
%      + \frac{r^2}5 \theta^{\m\g} \del_\g(\t^2-4)  \phi 
%  + \{t_\mu,\L\}  \nn\\
 &=  \frac 25 r^4 \b(1+\b^2) t^\mu \big(2 + \t\big) \phi 
  + \frac 15 r^2 (1+\b^2)\theta^{\m\g}\del_\g \big(2 + \t\big) \phi  \nn\\
  &\quad + \frac 25 r^4  \b x^\mu t^\a  \del_\a (2 + \t) \phi
     + \frac{r^2}5 \theta^{\m\g} \del_\g(\t^2-4)  \phi 
 + \{t_\mu,\L\} \ .
\end{align}
But in fact we need 
\begin{align}
 \cA_\mu^{(-)}[D^+D^+\phi] = \{x_\mu,DD\phi\}_1 - \cA_\mu^{(+)}[D^-D\phi]
\end{align}
where 
\begin{align}
  \cA_\mu^{(+)}[D^-D^+\phi] &= \theta^{\mu\nu}\del_\nu(D^-D^+ \phi) 
  % = -\frac{r^2 R^2}{3} \theta^{\mu\nu}\del_\nu(\cosh^2(\eta)\Delta^{(3)} \phi) \nn\\
   = \frac{r^2}{3} \theta^{\mu\nu}\del_\nu(\b^2+ \t+1)(2 +\t)\phi 
 \end{align}
 on-shell, using \eqref{D-D+-explicit} and  \eqref{Delta3-phi-static}.
 Combining with the above and using
\begin{align}
  -  r^2 \theta^{\m\g} (\del_\g\b^2)(\t+2)\phi
  % =  - \frac 2{R^2} r^2 \theta^{\m\g}x_\g  \b^4 (\t+2)\phi
    =  - 2 r^4 t^{\m}  \b^3 (\t+2)\phi
\end{align}
one finds the on-shell form
\begin{align}
  \cA_\mu^{(-)}[D^+D^+\phi] 
%  &= \frac 25 r^4 \b  (1+\b^2) t^\mu \big(2 + \t\big) \phi 
%    + \frac 25 r^4  \b x^\mu t^\a  \del_\a (2 + \t) \phi + \{t_\mu,\L\} \nn\\
%   &\quad + \frac 15 r^2  (1+\b^2)\theta^{\m\g}\del_\g \big(2 + \t\big) \phi 
%      + \frac{r^2}5 \theta^{\m\g} \del_\g(\t^2-4)  \phi 
%  -  \frac{1}{3} r^2\theta^{\mu\nu}\del_\nu (\t+ 1 + \b^2)(2 + \t)\phi  \nn\\
%  %
%   &= \frac 25 r^4 \b  (1+\b^2) t^\mu \big(2 + \t\big) \phi 
%    + \frac 25 r^4  \b x^\mu t^\a  \del_\a (2 + \t) \phi + \{t_\mu,\L\}  \nn\\
%   &\quad +  r^2 \theta^{\m\g}\del_\g \Big(\frac 15(1+\b^2) + \frac{1}5 (\t-2)  
%  -  \frac{1}{3} (\t+ 1+\b^2)\Big)(\t+2)\phi  
%  -  \frac{r^2}{5} \theta^{\m\g} \del_\g\b^2(\t+2)\phi \nn\\
 %%
%   &= \frac 25 r^4 \b (1+\b^2) t^\mu (2 + \t) \phi 
%    + \frac 25 r^4  \b x^\mu t^\a  \del_\a (2 + \t) \phi \nn\\
%   &\quad - \frac 2{15} r^2 \theta^{\m\g}\del_\g (\t+4+\b^2)(\t+2)\phi  + \{t_\mu,\L\}  \nn\\
%  %
  = \frac {2r^4}5  \Big(\b (t^\mu  + x^\mu t^\a  \del_\a) 
   - \frac 1{3r^2} \theta^{\m\g}\del_\g (\t+4 + \b^2)\Big)(\t+2)\phi  + \{t_\mu,\L\} \ .
   \label{A-DDphi-explicit-app}
\end{align}

\subsection{Background FLRW geometry and covariant derivatives}
\label{sec:diffeo-FRW}

The effective FLWR metric \eqref{eff-metric-G} is conformally flat, 
\begin{align}
 G^{\mu\nu} = \b \eta^{\mu\nu}, \qquad \b  = \frac 1{\sinh(\eta)} \ .
\end{align}
Then the Christoffel symbols in the Cartesian coordinates $x^\mu$ are 
\begin{align}
 \Gamma_{\mu\nu}^\r 
 %&= \frac 12 G^{\r\s}(\del_\mu G_{\nu\s} + \del_\nu G_{\mu\s}
 %  - \del_\s G_{\mu\nu})  \nn\\
%     &=  \frac 12 \eta^{\r\s} \b(\del_\mu \b^{-1} \eta_{\nu\s} + \del_\nu\b^{-1} \eta_{\mu\s}
%    - \del_\s \b^{-1} \eta_{\mu\nu})   \nn\\
%     &=  \frac 12  \b(\d^\r_{\nu}\del_\mu \b^{-1}  + \d^{\r}_{\mu}\del_\nu\b^{-1}
%    - \eta_{\mu\nu} \eta^{\r\s}\del_\s \b^{-1})   \nn\\
    &= - \frac{1}{2x_4^2} (\d^\r_{\nu}\eta_{\mu\a} x^\a   + \d^{\r}_{\mu}\eta_{\nu\a} x^\a 
    - \eta_{\mu\nu} x^\r )   \nn\\
  %  &= - \frac{R}{2x_4^3} (\d^\r_{\nu}G_{\mu\a} x^\a   + \d^{\r}_{\mu}G_{\nu\a} x^\a 
  %  - G_{\mu\nu} x^\r )   \nn\\
    &= - \frac 1{2R^2}\b^3 (\d^\r_{\nu}G_{\mu\a} x^\a   + \d^{\r}_{\mu}G_{\nu\a} x^\a 
    - G_{\mu\nu} x^\r ) 
    \label{christoffels}
 \end{align}   
so that
\begin{align}
 \Gamma^\r &= G^{\mu\nu}\Gamma_{\mu\nu}^\r =  \frac{R}{x_4^3} x^\r, \qquad
 \Gamma_{\mu\nu}^\mu = - \frac{2}{x_4^2}\eta_{\nu\a} x^\a  
 \label{christoffels-2}
\end{align}
%% fixed! V3
using \eqref{tau-beta-relation}.
% \begin{align}
%   \del_\mu x_4^2 &= -2 \eta_{\mu\nu} x^\nu , \qquad \del_\mu x_4 = - \frac{1}{x_4}\eta_{\mu\nu} x^\nu , \qquad
%  % \del_\mu \cosh^2(\eta) = -\frac{2}{R^2}\eta_{\mu\nu} x^\nu  
%  \nn\\
%   \b\del_\mu\b^{-1} &= - \b^{-1}\del_\mu\b = -\frac{R}{x_4} (\frac{1}{R x^4}) \eta_{\mu\nu}  x^\nu
%   = -\frac{1}{x_4^2}  \eta_{\mu\nu}  x^\nu 
%   \label{del-x4-formulas} 
%  \end{align}
Note that $\G_{\mu\nu}^\r$ is suppressed by the cosmic 
curvature scale.
For example, the pure gauge metric perturbations arising from diffeomorphisms 
generated by $\xi^\mu$ are given by
\begin{align}
\d_\xi G^{\mu\nu} &= \nabla^\mu\xi^\nu + \nabla^\nu \xi^\mu 
%  = \del^\mu\xi^\nu + \del^\nu \xi^\mu 
%   + G^{\mu\mu'}\Gamma_{\mu'\a}^\nu \xi^\a +  G^{\nu\nu'}\Gamma_{\nu'\a}^\mu \xi^\a   \nn\\
%  &= \del^\mu\xi^\nu + \del^\nu \xi^\mu 
%   - \frac{1}{2x_4^2}G^{\mu\mu'}(\d^\nu_{\mu'}\eta_{\a\g} x^\g 
%           + \d^{\nu}_{\a} \eta_{\mu'\g} x^\g - \eta_{\mu'\a}x^\nu ) \xi^\a \nn\\
%   &\qquad - \frac{1}{2x_4^2}G^{\nu\nu'}(\d^\mu_{\nu'} \eta_{\a\g} x^\g 
%    + \d^{\mu}_{\a} \eta_{\nu'\g} x^\g - \eta_{\nu'\a}x^\mu ) \xi^\a  \nn\\
  = \del^\mu\xi^\nu + \del^\nu \xi^\mu 
  - \frac{1}{x_4^2}G^{\nu\mu} x\cdot\xi \ .
 \label{puregauge-grav-covar}
\end{align}
As an application, the divergence of a vector field can be expressed as follows
\begin{align}
 \nabla_\mu \cA^\mu &= \del_\mu \cA^\mu + \Gamma_{\mu\nu}^\mu \cA^\nu 
  = \del_\mu \cA^\mu - \frac{2}{x_4^2}x^\a \eta_{\a\nu}  \cA^\nu \ .
  \label{div-A-full}
\end{align}
%%fixed V3

\paragraph{Diffeomorphisms and standard form on the FRW background.}

The terms $(x^\mu \del^\nu + x^\nu \del^\mu)\phi$ and $\del^\mu\del^\nu \phi$
in the  expression \eqref{tilde-h-nogauge} for $\tilde h^{\mu\nu}$ can be eliminated by a suitable 
diffeomorphism. Since $(x^\mu \del^\nu + x^\nu \del^\mu)\phi$  
becomes large at late times, one must be careful 
to use the proper covariant derivatives.
For example, consider the following vector fields on the FRW background
\begin{align}
 \xi^\mu = x^\mu \b \phi \ .
\end{align}
Then 
\begin{align}
 \nabla^\mu\xi^\nu + \nabla^\nu \xi^\mu 
 &=  G^{\mu\mu'}\del_{\mu'}(x^\nu \b\phi) + (\mu\leftrightarrow\nu) 
  - \frac{1}{x_4^2}\b G^{\nu\mu} x\cdot x \phi \nn\\
 &= (2 + \frac{\cosh^2}{\sinh^2}) \b\phi G^{\mu\nu}
  + (x^\nu G^{\mu\mu'}\del_{\mu'} \b + ...)\phi +  \b (x^\nu G^{\mu\mu'}\del_{\mu'} + ...)\phi  \nn\\
 &\stackrel{\eta\to\infty}{=} \b^2\Big(3\phi \eta^{\mu\nu}
  + 2 x^\nu x^\mu \frac{\b^2}{R^2}\phi 
     + (x^\nu \eta^{\mu\a}\del_\a + x^\mu \eta^{\nu\a}\del_\a )\phi\Big) \ .
  \label{puregauge-FRT-special-1}
\end{align}
Hence 
\begin{align}
 \boxed{\
 \b^2(x^\nu \eta^{\mu\a}\del_\a + x^\mu \eta^{\nu\a}\del_\a )\phi 
 \sim  - \b^2\big(3\eta^{\mu\nu} + 2x^\nu x^\mu \frac{\b^2}{R^2}\big) \phi
 \ }
 \label{FRW-puregauge}
\end{align}
where $\sim$ indicates equivalence up to diffeos.

Next, consider the following vector fields
\begin{align}
 \xi^\mu = x^\mu \phi \ .
\end{align}
Then 
\begin{align}
 \nabla^\mu\xi^\nu + \nabla^\nu \xi^\mu 
 &=  \del^\mu(x^\nu \phi) + (\mu\leftrightarrow\nu) 
  - \frac{1}{x_4^2} G^{\nu\mu} x\cdot x \phi \nn\\
 %&= (2 + \frac{\cosh^2}{\sinh^2}) \phi G^{\mu\nu}
 % +  (x^\nu \del^\nu + ...)\phi  \nn\\
 &\stackrel{\eta\to\infty}{=} \b\Big(3\phi \eta^{\mu\nu}
     + (x^\nu \eta^{\mu\a}\del_\a + x^\mu \eta^{\nu\a}\del_\a )\phi\Big)
  \label{puregauge-FRT-special}
\end{align}
hence 
\begin{align}
 \boxed{
 \b(x^\mu \del^\nu + x^\nu \del^\mu)\phi \sim - 3 \b\eta^{\mu\nu} \phi \ .
 }
 \label{FRW-puregauge-2}
\end{align}
Finally, consider 
\begin{align}
 \xi^\mu = \b^{-1}\eta^{\mu\nu}\del_\nu \phi \ .
\end{align}
Then 
\begin{align}
 \nabla^\mu\xi^\nu + \nabla^\nu \xi^\mu 
 &= \del^\mu(\b^{-1}\eta^{\nu\a}\del_\a \phi ) + (\mu\leftrightarrow\nu)  -\frac{1}{x_4^2} \b^{-1} G^{\nu\mu}\t\phi  \nn\\
 %&= \eta^{\mu\mu'}\eta^{\nu\nu'}\del_{\mu'}\del_{\nu'} \phi 
 % + \eta^{\mu\mu'}\b\del_{\mu'}\b^{-1}\eta^{\nu\nu'}\del_{\nu'} \phi 
 %  + (\mu\leftrightarrow\nu) 
 %  -\frac{1}{x_4^2} \eta^{\nu\mu} \t\phi  \nn\\
 &= 2\eta^{\mu\mu'}\eta^{\nu\nu'}\del_{\mu'}\del_{\nu'} \phi 
   - \frac{1}{R^2}\b^2 (x^{\mu}\eta^{\nu\nu'}\del_{\nu'}  + x^{\nu}\eta^{\mu\mu'}\del_{\mu'}) \phi 
   -\frac{1}{x_4^2} \eta^{\nu\mu} \t\phi \ .
 \end{align}
 The second term can be rewritten using \eqref{FRW-puregauge},
 and therefore 
\begin{align}
 \boxed{
 R^2\eta^{\mu\mu'}\eta^{\nu\nu'}\del_{\mu'}\del_{\nu'} \phi 
  % \sim \frac 12\b^2 (x^{\mu}\eta^{\nu\nu'}\del_{\nu'} +  x^{\nu}\eta^{\mu\mu'}\del_{\mu'} )\phi  
  %    + \frac 12\b^2\eta^{\nu\mu} \t\phi 
   \sim  - \b^2\big(\frac 12(3-\t)\eta^{\mu\nu} + x^\nu x^\mu \frac{\b^2}{R^2}\big) \phi \ .
 }
 \label{FRW-puregauge-3}
\end{align}
One can check with these results that the pure gauge contribution \eqref{pure-gauge-contrib-DD} is 
indeed a diffeomorphism.
% \begin{align}
%  \b^2\tilde h^{\mu\nu} 
%  &\propto  \frac 12 \b^2(\t+3) \eta^{\mu\nu}\phi
%    + \frac{1}{R^2} \b^2\b^2 x^\mu x^\nu\phi    \ 
%   - R^2\del_\mu\del_\nu\phi  
%   + \b^2(x^\nu  \del^\mu  +  x^\mu \del^\nu) \phi \nn\\
%   &\sim \frac 12 \b^2(\t+3) \eta^{\mu\nu}\phi
%    + \frac{1}{R^2} \b^2\b^2 x^\mu x^\nu\phi    \ 
%   + \b^2\big(\frac 12(3-\t)\eta^{\mu\nu} + x^\nu x^\mu \frac{\b^2}{R^2}\big) \phi
%   + \b^2(x^\nu  \del^\mu  +  x^\mu \del^\nu) \phi \nn\\
%   &\sim \frac 12 \b^2(\t+3) \eta^{\mu\nu}\phi
%    + \frac{1}{R^2} \b^2\b^2 x^\mu x^\nu\phi    \ 
%   + \b^2\big(\frac 12(3-\t)\eta^{\mu\nu} + x^\nu x^\mu \frac{\b^2}{R^2}\big) \phi
%    - \b^2\big(3\eta^{\mu\nu} + 2x^\nu x^\mu \frac{\b^2}{R^2}\big) \phi \nn\\
%   &= 0
% \end{align} 
% excellent!

\subsection{Massless scalar fields $(\Box+\frac{2}{R^2})\phi = 0$}
\label{sec:Box}

Using \eqref{G-Box-relation}, the on-shell relation can be written 
for rotationally invariant $\phi(\eta,\chi)$ in the form 
\begin{align}
 0 &= (\Box+\frac{2}{R^2})\phi = \sinh(\eta)^3 \Box_G \phi + \frac{2}{R^2}\phi \nn\\
 % &= \frac{1}{R^2}\Big(\frac{1}{\cosh^3(\eta)} 
 %\del_\eta\big(\cosh^3(\eta)\del_\eta\big)
 %-\frac{\sinh(\eta)^2}{\sinh^2(\chi)\cosh^2(\eta)}
 % \del_\chi\big(\sinh^2(\chi)\del_\chi\big) + 2 \Big)\phi \nn\\
  &=  \frac{\tanh^2(\eta)}{R^2} \Big(\frac{1}{\sinh^2(\eta)\cosh(\eta)} \del_\eta\big(\cosh^3(\eta)\del_\eta\big)
  + 2\frac{\cosh^2(\eta)}{\sinh^2(\eta)}
  + R^2 \cosh^2(\eta)\Delta^{(3)} \Big)\phi \ . 
%   &=  \frac{\tanh^2(\eta)}{R^2} \Big(\frac{1}{\sinh^2(\eta)\cosh(\eta)} \del_\eta\big(\cosh^3(\eta)\del_\eta\big)
%   + 2\frac{\cosh^2(\eta)}{\sinh^2(\eta)}
%  - \frac{1}{\sinh^2(\chi)}
%   \del_\chi\big(\sinh^2(\chi)\del_\chi\big)  \Big)\phi 
 \label{eom-phi-1}
\end{align}
We make a separation ansatz 
\begin{align}
 \phi(\eta,\chi) = f(\eta) g(\chi) \ .
\end{align}
Then the eom becomes
\begin{align}
 \frac{1}{\sinh(\eta)^2\cosh(\eta)}\frac 1f \del_\eta\big(\cosh^3(\eta)\del_\eta f\big)
  + 2\frac{\cosh^2(\eta)}{\sinh(\eta)^2} 
 \ = \  -R^2\cosh^2(\eta)\Delta_g^{(3)} g \ .
\end{align}
The factor $\cosh^2(\eta)$ in front of $\Delta_g^{(3)}$ drops out, see \eqref{Delta-3-H},
which leads to two equations
\begin{align}
  -R^2\cosh^2(\eta)\Delta_g^{(3)} g &= c\, g
  \label{eom-radial-H3-laplace}  \\[1ex]
   \frac{1}{\sinh(\eta)^2\cosh(\eta)} \del_\eta\big(\cosh^3(\eta)\del_\eta f\big)
  +2\frac{\cosh^2(\eta)}{\sinh^2(\eta)} f
 &=  c\, f 
 \label{separation-eq-eta}
\end{align}
where $c=const$.

\paragraph{Space-like harmonics.}

Consider first the space-like equation \eqref{eom-radial-H3-laplace}.
For rotationally invariant functions $\phi(\chi)$, this reduces using \eqref{Delta-3-H} to
\begin{align}
  \frac{1}{\sinh^2(\chi)}\del_\chi\big(\sinh^2(\chi)\del_\chi g \big)
   &= c\, g   \ .
   \label{separation-eq-chi}
\end{align}
The general solution is 
\begin{align}
 g(\chi) = \Big(c_1 e^{-\sqrt{1+c} \chi} 
    +c_2 e^{\sqrt{1+c} \chi}  \Big)\frac 1{\sinh(\chi)} \ .
    \label{g-chi-solution}
\end{align}
For $(1+c)>0$, there is at least one solution 
which is decreasing for $\chi\to \infty$. 
For $(1+c)<0$, the solutions are oscillating in radial direction.

\paragraph{Time dependence.}

The second equation \eqref{separation-eq-eta} is 
\begin{align}
% \frac{1}{\sinh(\eta)^2\cosh(\eta)} \del_\eta\big(\cosh^3(\eta)\del_\eta f\big)
%   +2\frac{\cosh^2(\eta)}{\sinh(\eta)^2} f
%  &=  c\ f \nn\\
 \frac{\cosh^2(\eta)}{\sinh(\eta)^2} f'' + 3 \frac{\cosh(\eta)}{\sinh(\eta)} f'
  + 2\frac{\cosh^2(\eta)}{\sinh(\eta)^2} f- c f &= 0 \ .
  \label{eq-f-eta-time}
\end{align}
Asymptotically, this is 
\begin{align}
 e^{-3\eta}\del_\eta(e^{3\eta}\del_\eta f)  &= (-2+c)  f  \nn\\
 (\del_\eta^2 + 3 \del_\eta + 2-c) f &= 0
 \label{lambda-eq-timedep}
\end{align}
which is solved by $f= e^{\l\eta}$ with 
\begin{align}
 \l^2 + 3\l +2 -c &= 0  \nn\\
 \l_{1,2} &= \frac 12 (-3 \pm \sqrt{1 + 4 c}) \ .
  \label{lambda-eq-timedep-2}
\end{align}
The most interesting quasi-static Schwarzschild solution arises for $\t\phi=-2\phi$,
which corresponds to $\Delta^{(3)}\phi = 0$ via \eqref{Box-Laplace-tau} hence to $c=0$.
Then \eqref{g-chi-solution} and \eqref{eq-f-eta-time} have the exact solutions  
\begin{align}
 g(\chi) =  \frac{e^{- \chi}}{\sinh(\chi)}, 
 \qquad f(\eta) = \frac{1}{\cosh^2(\eta)}   \sim e^{-2\eta} \ ,
 \label{harmonic-soln}
 \end{align}
where $\rho = \sinh(\chi)$ \eqref{ds-induced} is the appropriate distance variable on $H^3$.
Thus $\phi = f(\eta) g(\chi)$ exhibits  the typical $\frac 1\r$  behavior of the harmonic Newton potential in 3 dimensions,  
with time dependence  given by 
$f(\eta)  \sim e^{-2\eta}$.
Note that $\phi(\eta)$ remains finite for $\eta\to 0$, so that the Schwarzschild solution does not blow up at any time.
For $c<-\frac 14$, this will lead to propagating scalar modes.

\bibliographystyle{JHEP}
\bibliography{papers}

\end{document}